\documentclass{aa}  

\usepackage{graphicx}
\usepackage{txfonts}
\usepackage{xcolor}

\newcommand{\maestro}{Maestro}
\newcommand{\solar}{\ensuremath{_{\odot}}}
\newcommand{\avgsph}[1]{\ensuremath{\left<#1\right>}}
\newcommand{\avgtime}[1]{\ensuremath{\left<#1\right>_t}}
\newcommand{\avgvol}[1]{\ensuremath{\overline{#1}}}

\begin{document}

   \title{Calibrating Core Overshooting Parameters With Two-dimensional Hydrodynamical Simulations}
   \titlerunning{Calibrating Core Overshooting With Two-dimensional Hydrodynamical Simulations}


   \author{J. Higl
          \inst{1,2,3}
          E. M\"uller \inst{1}
          \and
          A. Weiss\inst{1}
          }

   \institute{Max-Planck-Institut f\"ur Astrophysik,
     Karl-Schwarzschild-Str.~1, 85748 Garching, Germany
     \and Technische Universit\"at M\"unchen, Physik Department,
     James~Franck~Str.~1, 85748 Garching, Germany
     \and Heidelberg Institute for Theoretical Studies, 
     Schloss-Wolfsbrunnenweg~35, 69118 Heidelberg, Germany\\
            \email{johann.higl@h-its.org}
             }

   \date{Received ; accepted}

 
   \abstract{The extent of mixed regions around convective zones is
     one of the biggest uncertainties in stellar evolution.  1D
     overshooting descriptions introduce a free parameter ($f_{ov}$)
     that is in general not well constrained from observations.
     Especially in small central convective regions the value is
     highly uncertain due to its tight connection to the pressure
     scale height.
     \\
     Long-term multi-dimensional hydrodynamic simulations can be used
     to study the size of the overshooting region as well as the
     involved mixing processes.  Here we show how one can
     calibrate an overshooting parameter by performing 2D
     \maestro{} simulations of Zero-Age-Main-Sequence stars ranging
     from $1.3$ to $3.5 \,M\solar$.  The simulations cover the
     convective cores of the stars and a large fraction of the
     surrounding radiative envelope.  We follow the convective flow
     for at least 20 convective turnover times, while the longest
     simulation covers 430 turnover time scales.  This allows us to
     study how the mixing as well as the convective boundary itself
     evolve with time, and how the resulting entrainment can be
     interpreted in terms of overshooting parameters.
     \\
     We find that increasing the overshooting parameter $f_{ov}$
     beyond a certain value in the initial model of our simulations,
     changes the mixing behaviour completely.  This result can be used
     to put limits on the overshooting parameter. We find
     $0.010 < f_{ov} < 0.017$ to be in good agreement with our simulations
     of a $3.5M\solar$ mass star.  We also identify a
     diffusive mixing component due to internal gravity waves (IGW)
     that is active throughout the convectively stable layer, but most
     likely overestimated in our simulations.  Furthermore,
     applying our calibration method to simulations of less
     massive stars suggests a need for a mass-dependent
     overshooting description where the mixing in terms of the
     pressure scale height is reduced for small convective cores.
     \\
   }

  
   \maketitle
%

\section{Introduction}\label{s:intro}
The treatment of turbulent convection is still the biggest uncertainty
in the calculation of stellar models. Modern stellar evolution codes
usually apply the classical Mixing-Length-Theory
\citep[MLT;][]{boehmvitense58} to allow for a treatment of this
intrinsically three-dimensional problem in one-dimensional stellar
evolutionary calculations.  MLT provides an estimate for the
convective velocities inside convective zones. By construction, this
velocity drops instantly to zero at a convective boundary, i.e., a
mass element moving towards such a boundary is assumed to stop
instantly, although physically only the acceleration vanishes. A more
natural way would be to assume that any mass element that crosses the
boundary is slowed down, eventually stopped, and finally returned to
the convective zone (CZ) by the buoyancy force.  Consequently, matter
beyond the CZ boundary is mixed into the CZ.  Simple energy estimates
indicate that in the case of convective stellar cores this is a
negligible, spatially unresolvable effect
\citep{lattanzio17}. Nevertheless there is evidence from eclipsing
binaries \citep[e.g.][]{ribas00, valle16}, globular clusters
\citep[e.g.][]{aparicio90,bertelli92}, and asteroseismology
\citep[e.g.][]{deheuvels16} to suggest that a region of
significant spatial extent is well mixed around convective cores. 

Therefore 1D stellar evolution codes consider well mixed overshooting
regions around CZ. The width of the overshooting region is usually
parametrized by a single parameter $f_{ov}$, defined as a fraction of
the pressure scale height $H_p$.  Attempts to calibrate $f_{ov}$ from
observations often involve uncertainties that are of the same order of
magnitude as the parameter itself. Moreover, stellar parameters of
eclipsing binaries can usually be explained by stellar models without
overshooting \citep[e.g.][]{pols97, higl17}.  A more accurate estimate
of $f_{ov}$ is provided by space born observations of stellar
pulsations, which allow one to probe the deep interior of stars by
asteroseismology \citep{moravveji16}. \citet{pedersen18} have shown
that the imprints of different mixing descriptions at convective cores
have a potentially observable influence on asteroseismic properties.
However, only a small sample of stars with sufficiently accurate
observations exist so far.

Calibrating $f_{ov}$ with observations is always model dependent,
while multidimensional hydrodynamic simulations can give first
principle estimates of mixing processes near a convective boundary and
about the extent of the overshooting layer.  Simulations of convective
envelopes by \citet{freytag96} using the "star in a box" idea found
that convective boundary mixing in thin envelopes of A stars can be
described by a diffusive process, where the diffusion constant decays
exponentially with distance from the CZ. This form of overshooting is
now applied in most codes for all convective boundaries.

Envelope convection covers several pressure scale heights, which makes
pressure fluctuations an important driving mechanism for the transport
of kinetic energy and enthalpy \citep{viallet13}. Interior CZ, on the
other hand, are more shallow.  Hence, the contribution of pressure
fluctuations is reduced and buoyancy fluxes dominate the convective
motions.  Whether the same overshooting description can be applied in
both cases is unclear, but least one would expect that shallow
convection should be described using a different overshooting
parameter.  Furthermore, stellar interior convection zones are
dominated by low Mach number flows, which require a special treatment
in hydrodynamic simulations in order to resolve all relevant
timescales \citep{miczek15}.  Some groups circumvent this problem by
increasing the energy input into their simulations to increase the
Mach number \citep[e.g.][]{meakin07,cristini17,horst20}, while others
modify the hydrodynamic equations such that sound waves are prohibited
\citep[e.g.][]{rogers13,almgren06a}. 

One goal of multi-D simulations is to provide a more accurate
description of convection that can be used in 1D codes. In the case of
convective envelopes it was recently shown
\citep{jorgensen18,mosumgaard20} that combining the mean
stratification of 3D models \citep{magic13} with the interior
structure of 1D models improves the agreement with asteroseismic
observations.

Terrestrial atmospheric sciences often describe the mixing at 
convective boundaries by entrainment models, where the convective
region grows constantly with time \citep[e.g.][]{mellado17,stevens02}.
\citet{meakin07} compared simulations of interior convection zones during oxygen burning
with such entrainment models and found that the mixing across the convective 
boundary is also well described by entrainment in the stellar context.   

This was confirmed for carbon
burning shells in \citet{cristini17}, and for core hydrogen burning in
\citet{gilet13}. \citet{staritsin13} tested the entrainment model in
1D stellar evolution models and found that it significantly improves
the consistency of 1D models with observations. However, they had to
use entrainment rates that are orders of magnitude smaller than what
is found in hydrodynamic simulations in order to prevent the whole
star from becoming convective.
 
In this work we use the low Mach number hydrodynamic code \maestro{}
\citep{almgren06a,almgren06b,almgren08,nonaka10} to demonstrate that one can
calibrate the overshooting parameter of a 1D mixing description on the
main-sequence with the help of long-term hydrodynamic simulations in
combination with consistent 1D models. The paper is organized as
follows: In Sect.~\ref{s:setup} we describe the properties of
\maestro{} and of our initial models, followed by an analysis of an
intermediate mass star in Sect.~\ref{s:results}.  This section also
contains a description of our calibration method. In
Sect.~\ref{s:mass} we extend the analysis to stars of smaller masses,
and give some conclusions in Sect.~\ref{s:conclusion}.

\section{Numerical Setup}\label{s:setup}
In order to make predictions about the mixing at convective
boundaries it is necessary to follow convective
motions over several convective turnover times in a quasi steady state.
Getting to such a steady state might take another few
turnover timescales. MLT predicts that on the main-sequence the sound
crossing timescale is roughly four orders of magnitude smaller than
the convective timescale, which is limiting the usage of explicit
hydrodynamic codes in this regime, since numerical stability requires
one to resolve the sound crossing time.  \maestro{} overcomes this
problem by removing sound waves from the Euler equations, thereby
allowing for timestep sizes that resolve the advection timescale. 
For the simulations discussed in this work the resulting timesteps 
are two orders of magnitude larger than in fully compressible explicit codes.
With \maestro{} we could therefore cover several hundred convective 
turnovers in this low Mach number regime with our 2D simulations. 

\subsection{\maestro{}}\label{s:maestro}
\maestro{} was introduced in \citet{almgren06a,almgren06b,almgren08}
and later extended by an adaptive mesh refinement in \citet{nonaka10}.
It uses a generalized version of the pseudo incompressible
approximation \citep{durran89} to remove sound waves from its
simulations.  This method has the advantage that it allows one to
follow the evolution of large scale density and temperature
perturbations, which is not the case when using the anelastic
approximation. Only pressure fluctuations are assumed to be small, and
the velocity field $\vec{U}$ has to fulfil the constraint
\begin{equation}
\label{eq:constraint}
\nabla \cdot (\beta_{0} \vec{U}) = \beta_0 S ,
\end{equation}
where $\beta_{0}$ depends on the background density $\rho_0$ and
pressure $P_0$, and is given by
\begin{equation}
  \beta_{0}(r,t) = \rho_0(0,t)\exp{\left( \int_0^r
      \frac{1}{\avgsph{\Gamma_1} P_0} \frac{\partial P_0}{\partial r'} dr'\right) } .
\end{equation}
$\avgsph{\Gamma_1}$ is the angularly averaged value of
$\Gamma_1 = d(\log P)/d(\log \rho)$ at constant entropy, where $\rho$
and $P$ are the density and pressure, respectively.

The quantity $S$ in Eq.\,(\eqref{eq:constraint}) represents the source
terms of the system. As we do not follow any compositional changes due
to nuclear reactions in our simulations, the source term reduces to
\begin{equation}
  S = \chi H_{\mathrm{ext}} 
\end{equation}
where $H_{\mathrm{ext}}$ is the energy released by nuclear reactions,
and $\chi = p_T / (\rho c_p p_{\rho})$ with
$p_T     = \partial P / \partial T |_{\rho}$,
$p_{\rho} = \partial P / \partial \rho |_T$, and
the specific heat at constant pressure
$c_p = \partial h /\partial T |_{p}$. Here $h$ is the specific enthalpy.

Incorporating Eq.\,(\eqref{eq:constraint}) into the Euler equations one
can define a set of equations for density $\rho$, velocity $\vec{U}$
and specific enthalpy $h$ (for details of this derivation see
\citet{almgren06b})
\begin{align}
\frac{\partial \rho}{\partial t} + \nabla \rho \vec{U} = & \, 0 
\\ 
\rho\frac{\partial (\vec{U})}{\partial t} + \rho \vec{U}\nabla \vec{U}
  = & \frac{\beta_0}{\rho}\nabla \frac{\pi}{\beta_0} -
      \frac{(\rho-\rho_0)}{\rho}g\vec{e_r} 
\label{eq:maestro_momentum} 
\\
\rho \frac{Dh}{Dt} - \frac{DP}{Dt} = & \nabla (\kappa \nabla T) + \rho
                                       H_{ext} 
\label{eq:maestro_enthalpy} 
\\
\nabla \beta_{0} \vec{U} = & \beta_{0} S ,
\end{align} 
where $D/Dt = \partial t + \vec{U} \cdot \nabla$ is the Lagrangian
derivative, $\pi = P - P_0$ the deviation of the pressure $P$ from the
background pressure $P_0$, and $g$ the gravitational acceleration
acting in radial direction defined by the unit vector $\vec{e_r}$.

Different from the equation set derived in \citet{almgren06b} we
include an additional contribution to the enthalpy equation
(Eq.\,\eqref{eq:maestro_enthalpy}) from radiative diffusion determined
by the conductivity $\kappa$ and the gradient of temperature
$T$. Moreover, as \citet{vasil13} showed that the first term on the
right hand side of Eq.\,(\eqref{eq:maestro_momentum}) as given in
\citet{almgren06b} does not conserve the energy of the system, we use
here the corrected term as proposed by \citet{vasil13} and included
into \maestro{} by \citet{jacobs16}.

This set of equations does not allow for sound waves to
propagate. Hence, we choose a timestep size $\Delta t$ according to
the advection velocity $u$ and the cell width $\Delta x$
\begin{equation}
\label{eq:dt_advect}
\Delta t < \frac{\Delta x}{u} .
\end{equation}

\maestro{} uses a fractional step method, where first density,
enthalpy, and velocity are advected, without taking the velocity
constraint into account.  Then one enforces
Eq.\,(\eqref{eq:constraint}), which also sets the updated pressure
\citep[see][]{bell02}.  The advection step is done using a two step
predictor-corrector (PC) scheme. Details of this scheme, including a
flow chart, can be found in \citet{nonaka10}.  For our simulations we
found that this scheme causes unrealistically large velocities in
stably stratified regions.  These velocities are caused by an
insufficient time resolution of internal gravity waves (IGW), which
evolve on a timescale of the Brunt-Vaisala frequency $N$ defined as
\begin{equation}
  N^2 = - \frac{g \xi_{T}}{\xi_{\rho}H_p} \left( \nabla_{ad} - \nabla
        - \frac{\xi_{\mu}}{\xi_{T}} \nabla_{\mu} \right)
\end{equation}
where $\nabla_{ad}$ and $\nabla$ are the adiabatic and actual
temperature gradient, and where $\nabla_{\mu}=d\ln{\mu}/d\ln{P}$ is
the molecular weight gradient.  The quantities
$\xi_{\rho}=\partial \ln{P}/\partial \ln{\rho} |_{T,\mu}$,
$\xi_{\mu}=\partial \ln{P}/\partial \ln{\mu} |_{\rho,T}$, and
$\xi_{T}=\partial \ln{P}/\partial \ln{T} |_{\rho,\mu}$ are obtained
from the equation of state (EOS).

The frequency of an IGW is given as
$\omega = N
\left|\vec{k_\bot}\right|/\left(\left|\vec{k_\parallel}+\vec{k_\bot}\right|\right)$
\citep[e.g.][]{sutherland10}, where $\vec{k_\bot}$ and
$\vec{k_\parallel}$ are the wave vectors perpendicular and parallel to
the direction of gravity, respectively.  The corresponding phase
velocity of an IGW
$\vec{v}_{ph}=(\omega/\left|\vec{k}\right|^2) \cdot \vec{k}$ is larger
than its group velocity $\vec{v}_{gr}=\nabla_{\vec{k}}\, \omega$,
where $\vec{k} = \vec{k}_\parallel + \vec{k}_\bot$ and
$\nabla_{\vec{k}}$ is the gradient operator with respect to $\vec{k}$.
Accordingly, we can use the maximum phase velocity
$v_{ph,max} = \lambda N / 2\pi$ of an IGW of wavelength $\lambda$ as
an upper limit for its propagation speed, and hence substitute the
timestep criterion Eq.\,(\eqref{eq:dt_advect}) by
\begin{equation}
\label{eq:brunt_timestep}
\Delta t < \frac{\Delta x  2 \pi} {\lambda N } .
\end{equation}
to numerically resolve the time evolution of IGW.

Assuming that IGW have typical wavelengths of a pressure scale height,
one ends up with timesteps that are only a factor $\sim 5$ to 10
larger than the usual CFL timestep which is required to resolve sound
waves.  Fully resolving IGW in time therefore requires  
one order of magnitude more computing time with \maestro{}. This 
prevents us from performing simulations up to hundreds of convective
turnover timescales with our computational resources. However, such simulations are 
needed to estimate the extent of the mixed region.  

The maximum frequency of IGW allowed by $N$ is of the order of several
hundred $\mu \mathrm{Hz}$, while observations show us that the
dominant component of IGW is in the low frequency regime
\citep[e.g.][]{bowman19} up to several tens of $\mu \mathrm{Hz}$.
Angular momentum transport by IGWs is also dominated by low frequency
waves \citep{Aerts18}. \citet{Aerts10} also observed g-mode periods
between 0.5 - 3 days, corresponding to $4 - 20 \,\mu \mathrm{Hz}$.
Theoretical wave spectra of IGWs \citep{lecoanet14} also predict IGW
frequencies predominantly below the convective turnover frequency
(which is well resolved in our simulations). While the last statement
is challenged by the simulations of \citet{edelmann19}, who found a
non-negligible contribution from frequencies above the turnover
frequency, the wave frequencies still remain far below the mHz limit.
We therefore expect that high frequency IGW do not play a significant
role in the mixing and that the benefit of long simulations would
outweigh the downside of unresolved high frequency IGW.

\begin{figure}
\resizebox{\hsize}{!}{\includegraphics{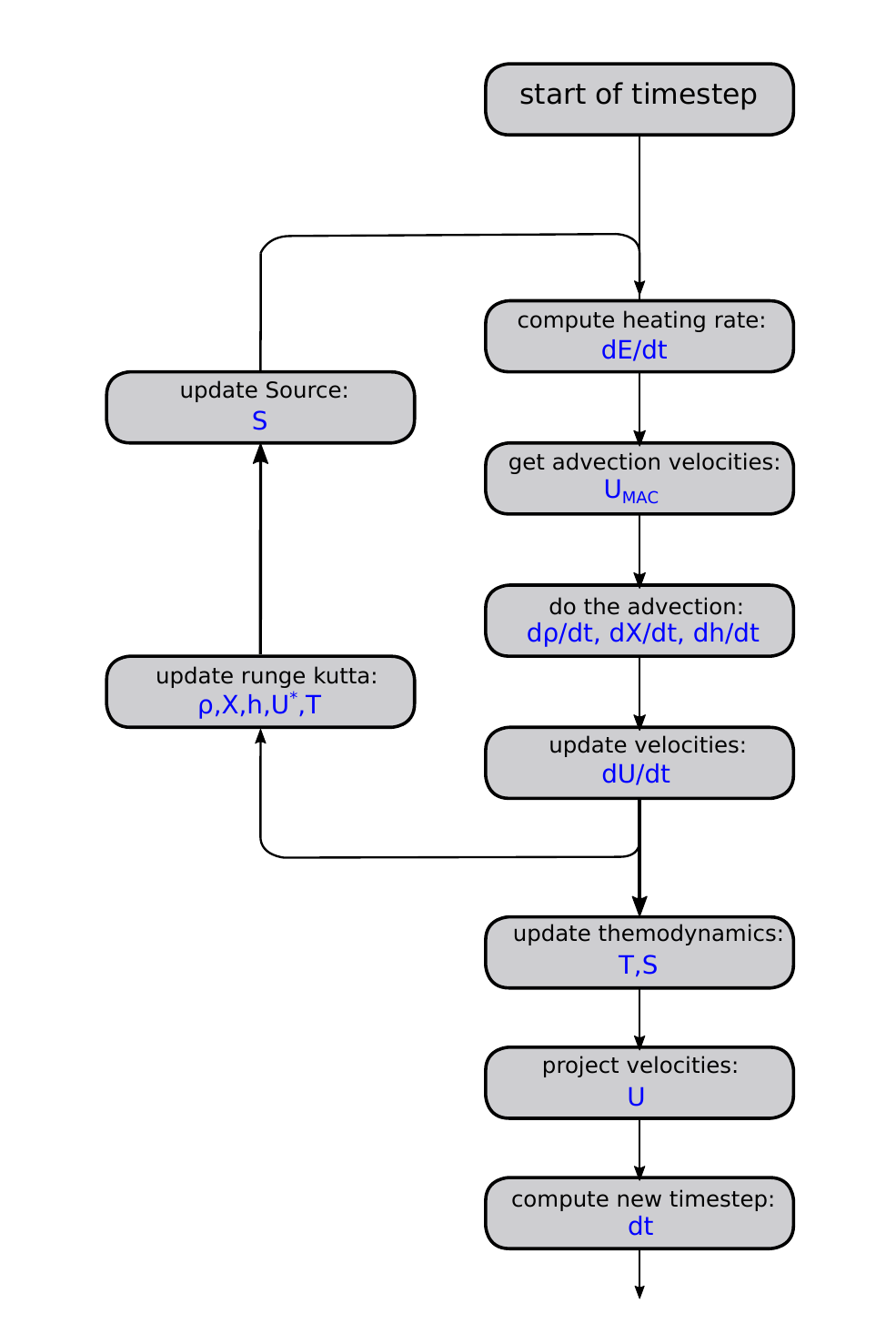}}
\caption{\label{p:flow_chart}Flow chart of the modified time
  advancement algorithm. The computed/updated quantities of each sub
  step are given in blue.}
\end{figure}

We therefore decided to perform our simulations with timesteps
according to Eq.\,(\eqref{eq:dt_advect}) and to mitigate the problem of
spurious velocities in the stable layer by a higher order, multi-step
scheme for the time integration.  \citet{bell02} showed that it is
possible to exchange the time advancement before the final projection
by any other method.  We replaced the PC method with a 4-step
Runge-Kutta (RK) integrator.  A flow chart of this new method is
depicted in Fig.\,\ref{p:flow_chart}, where we give in blue the
updated quantity in each sub-step.

During the RK loop we need to introduce two additional velocities in
the scheme.  $U^*$ is the updated velocity in each RK step, computed
using the reconstructed velocity at cell interfaces
$U_{\mathrm{MAC}}$, which is forced to fulfil the velocity constraint
(Eq.\,\eqref{eq:constraint}), while $U^*$ does not do this in general.

\begin{figure}
\resizebox{\hsize}{!}{\includegraphics{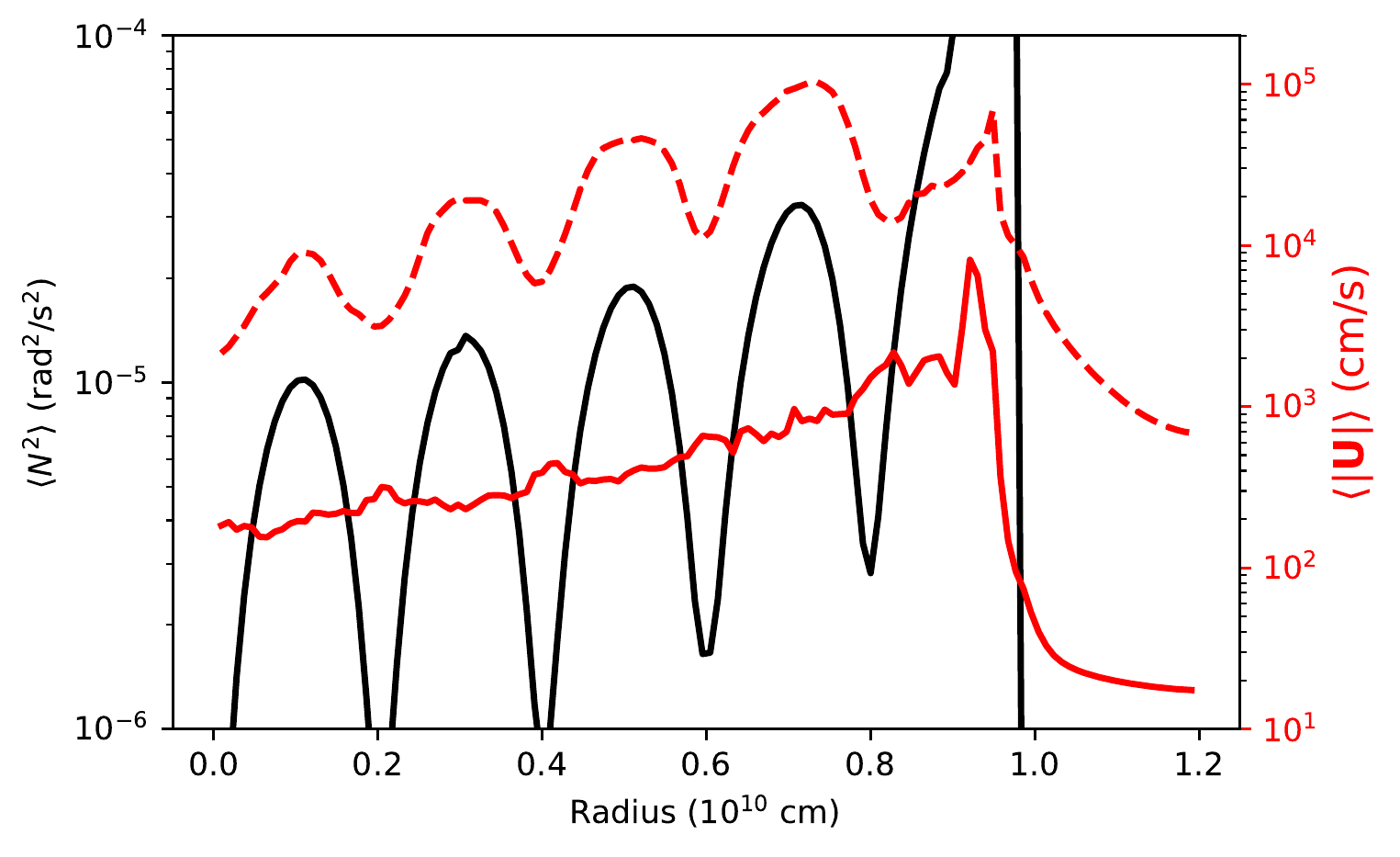}}
\caption{\label{p:pc_rk_demonstration}The black solid line shows the
  initial $N^2$ profile of a constructed stable atmosphere. Red lines
  give the velocity magnitude profile after $3\, 10^5\,\mathrm{s}$ for
  the predictor-corrector scheme (dashed) and the Runge-Kutta
  integrator (solid), respectively.}
\end{figure}

In order to demonstrate the capability of the new time integration
scheme, we set up a stable atmosphere with several g-mode cavities.
We use a constant gravity and a linearly declining density, modulated
by a sine function (avoiding density inversions).  After integrating
the hydrostatic equilibrium (HSE), we end up with a stratification
that has several peaks in the Brunt-Vaisala-Frequency
(see black line in Fig.\,\ref{p:pc_rk_demonstration}). 

A stable atmosphere should not develop significant velocities, yet
Fig.\,\ref{p:pc_rk_demonstration} shows that the predictor-corrector
method produces large velocities.  After $3\, 10^5\,\mathrm{s}$ the
profile exhibits a peaked velocity profile. The peaks coincide with
the $N^2$ cavities and they increase in amplitude as $N^2$
increases. The reason for the peaks is the presence of numerical
artefacts that act as high frequency gravity waves. Such artefacts
will be trapped in the g-mode cavities and pile up over time until the
waves break.  Breaking gravity waves can develop a mean flow as has
been shown in previous studies \citep[e.g.][]{couston18}.  The drop in
velocity magnitude after the last peak in $N^2$ is due to a velocity
damping that is applied to reduce boundary effects.

In test simulations of convective boundary mixing, we found that this
numerical phenomenon can create mean flows of a velocity magnitude
similar to that of the convective motions.  The RK integrator, on the
other hand, smoothes the numerical artefacts such that we do not see
this phenomenon any longer.  The velocities get reduced by more than
one order of magnitude (see red line in
Fig.\,\ref{p:pc_rk_demonstration}), and are now significantly smaller
than the expected convective velocities.

\maestro{} is a purely Cartesian code with a one-dimensional
background state. In order to compute spherical stars it is therefore
necessary to adjust the scheme, such that the background state is
evaluated consistently with the domain centered
dataset. \citet{nonaka10} proposed the necessary algorithm to
guarantee this in three-dimensional simulations. We extended their
scheme to allow also for 2D simulations of spherical datasets.  The
resulting domain is a planar slice through the star containing its
center.

\subsection{Microphysics}
In the \maestro{} simulations we use the Helmholtz equation of state,
including effects of radiation, ionization, degeneracy of electrons
and Coulomb corrections \citep{timmes00}.  

The nuclear heating term is based on hydrogen burning equilibrium
rates for the PP and the CNO cycle \citep{kippenhahn12}. We followed
the evolution of three independent species (H, He, CNO) during the
simulation, CNO species representing the total abundance of elements
involved in the CNO cycle.  Its atomic weight and charge is calculated
according to the ratio of solar abundances of C, N, and O.  The
equilibrium rates reproduce the nuclear energy generation of the 1D
models within a factor of two.

In contrast to most other simulations of convective boundary mixing
\citep[e.g.][]{meakin07,cristini17} we do \emph{not} increase our
energy production by an additional boosting factor in any of our
simulations. This is possible because the timestep size of \maestro{}
increases as the flow velocity decreases, and hence the number of
convective turnover times that one can simulate does not depend on the
absolute value of the flow velocity.

We also include energy transport by radiative diffusion using the
analytic stellar opacities provided by \citet{timmes00b}, which
combine analytic expressions for hydrogen-free and hydrogen-containing
compositions by \citet{iben75} and \citet{christy66},
respectively. The opacities also contain contributions from
Compton-scattering based on \citet{weaver78}.

\subsection{Initial Models}
\label{s:initial_model}
We obtain our initial models with the 1D stellar evolution code
Garstec \citep{weiss08}, and evolve them until the beginning of core
hydrogen burning when less than 1\% of hydrogen has been consumed.
This leads to a very shallow jump in the hydrogen profile at the
boundary of the convectively mixed core.  Therefore, changes in the
hydrogen profile can be seen quickly.
Consequently, this quick reaction time of the conditions at the
boundary also allows us to study the time evolution of the mixing.
 
For our calibration method (see \ref{s:calibration}) we need initial
1D models computed with and without convective overshooting.  Garstec
implements overshooting as a diffusive process according to
\citet{freytag96}, where a diffusion constant $D$ is computed based on
the pressure scale height $H_p$ and the distance to the convective
boundary $c_z$ as
\begin{equation}
\label{eq:overshooting}
D = D_0 \exp{\left(-\frac{2c_z}{f_{ov}H_p}\right)}\,,
\end{equation}
with $f_{ov}$ being the overshooting parameter, and $D_0$ the
diffusion constant at the convective boundary. 
In Garstec $D_0$ is evaluated inside the CZ at some distance to the boundary, 
since MLT predicts that the velocity as well as the diffusivity vanish  
right at the convective boundary.

Our models that include overshooting start from the same initial
conditions as the non-overshooting ones and are self-consistently
evolved until they reach a similar central hydrogen content as the
models without overshooting.  As a consequence of our self-consistent
approach and the fact that we assume a radiative temperature
stratification in the overshooting region, we find that the size of
the convective core according to the Schwarzschild criterion in the 1D
models increases slightly as we increase the overshooting
parameter. The size of the mixed core, on the other hand, increases at
a much large rate.
 
Garstec provides us with thermally relaxed models on a Lagrangian
grid. In order to map these into \maestro{}, we need to interpolate
the model onto an Eulerian grid.  The interpolation introduces slight
deviations from HSE in the mapped models.  Since \maestro{} expects
its background state to be in perfect HSE, it is necessary to
reintegrate the equation of HSE:
\begin{equation}
\label{eq:hse}
\frac{\partial P}{\partial r} = - g \rho
\end{equation}
During this reintegration, we also switch from the OPAL equation of
state used in Garstec to the Helmholtz EOS in \maestro{}, which
slightly modifies the thermal structure of the models, and thus no
longer guarantees thermal equilibrium.  MLT predicts the temperature
gradient in the CZ of our models to correspond to a small
overadiabaticity of the order of $10^{-8}$.  Because overadiabacity
acts as an energy reservoir for the convective flow in a hydrodynamic
simulation, a small change in the temperature stratification can
increase the internally stored energy significantly or remove the
convective core entirely.  To keep the simulations energetically as
close as possible to the initial models, we take the MLT temperature
gradient $\nabla_{\mathrm{mlt}}$ into account while 
recalculating HSE, i.e., we substitute the temperature gradient
$\nabla$ by $\nabla_{mlt}$, so that
\begin{equation}
\label{eq:nabla}
  \nabla - \nabla_{\mathrm{ad}} \, \rightarrow \, 
  \nabla_{\mathrm{mlt}} - \nabla_{\mathrm{ad}} := \nabla_{\mathrm{ex}}\,,
\end{equation}
where $\nabla_{\mathrm{ex}}$ is the superadiabaticity.

We achieve this requirement by simultaneously solving equations
(\eqref{eq:hse}) and (\eqref{eq:nabla}), which also ensures that the
location of the convective boundary does not change during the
reintegration as has been shown by \citet{edelmann17}.

In Fig.\,\ref{p:initial_stratification} we demonstrate the effect of
this integration method. Models that are reintegrated while keeping
the temperature constant (red dotted line) lead to a temperature
stratification that is stable in the very centre of the convective
core and largely superadiabatic towards the convective
boundary. Keeping the overadiabaticity constant instead (red solid
line) we achieve a much more consistent temperature stratification.
With this procedure we get a HSE with a relative accuracy of $10^{-5}$
and a temperature gradient that is only $5\,10^{-5}$ larger than the
adiabatic temperature gradient in the CZ.  While this value is still
three orders of magnitude larger than that predicted by MLT (red
dashed line), it is sufficiently small for the purpose of our
simulations. In Fig.\,\ref{p:initial_stratification} we also display
the density profile before (dashed black line) and after reintegration
(solid black line).  The change in the density profile introduced by
the reintegration is within the thickness of the plotted line, and
hence can be neglected.

\section{An Intermediate Mass Star}\label{s:results}
An intermediate mass star on the main-sequence has a mass between
$\approx 1.2 M\solar$ and $\approx 8 M\solar$, a convective core, and
a radiative envelope. Such stars are not massive enough to evolve all
the way to core collapse and will end their life as a carbon-oxygen
white dwarf.  Here we will discuss 2D simulations of the convective
core in a $3.5 M\solar$ star with a solar like composition of
$\mathrm{X}=0.710$, $\mathrm{Y}=0.276$, and $\mathrm{CNO}=0.014$.  We
chose this mass, because it has a convective core large enough to
avoid problems with our overshooting description (see
\ref{s:initial_model}).  Intermediate mass stars are also preferred
targets for observers to study IGW.  Using the observed frequencies of
IGW, originating at the boundary of the convective core, it is
possible to estimate the sizes of the mixed cores
\citep{deheuvels16,moravveji16}.

\begin{figure}
\resizebox{\hsize}{!}{\includegraphics{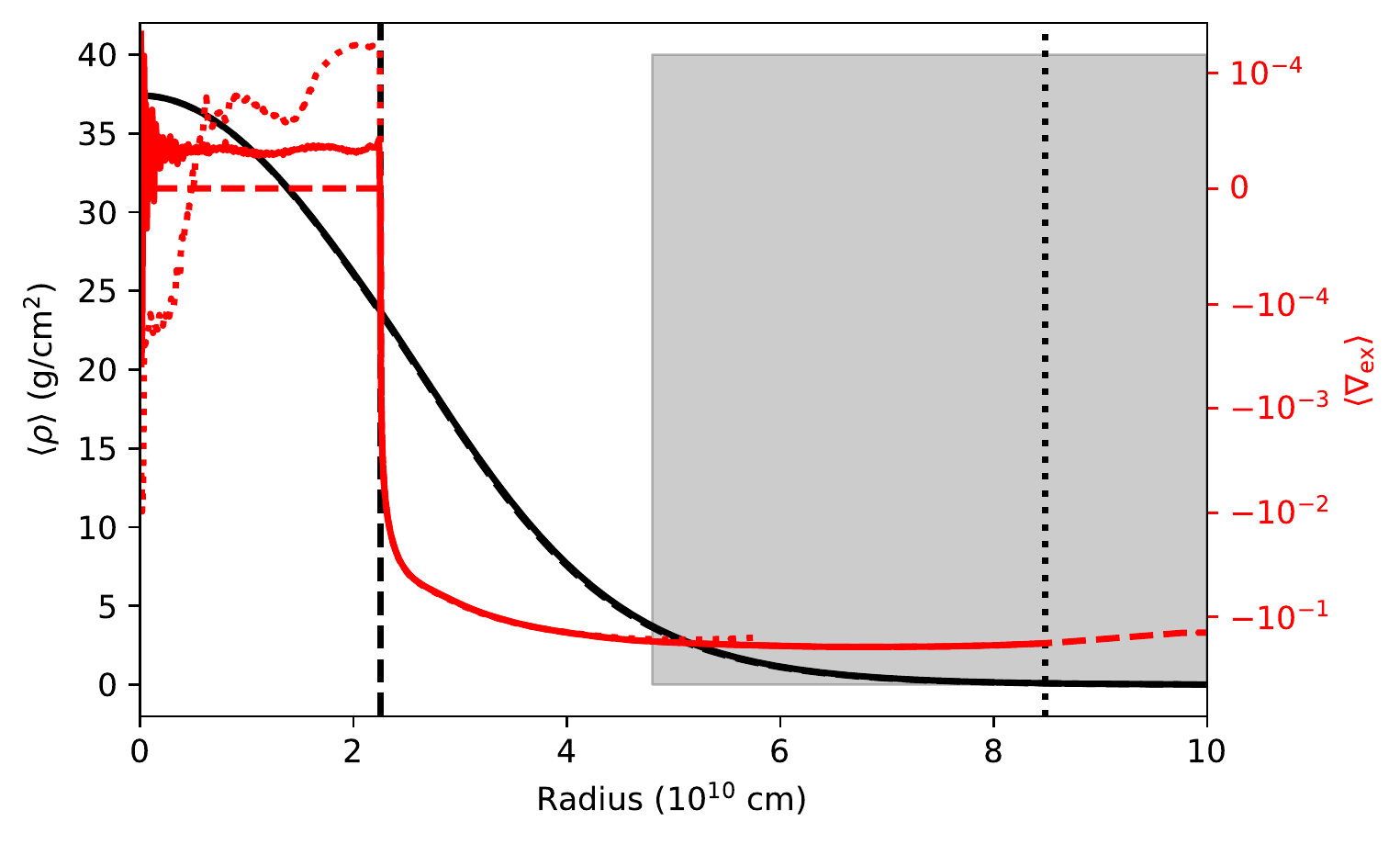}}
\caption{\label{p:initial_stratification} Initial density (black) and
  superadiabaticity (red) of the model H3.5 (solid; see Table \ref{t:intermediate}) and H3.5-T
  (dotted), respectively. Dashed lines give the stratification as
  predicted by Garstec. The vertical dashed and dotted lines indicate
  the boundary of the convective core and of the computational domain,
  respectively. In the shaded region we damp the velocities. }
\end{figure}

Our simulations cover the complete convective core on a Cartesian grid
as well as the star's stable layer up to roughly $50\%$ of the stellar
radius. This corresponds to a total of 6 pressure scale heights and 5
density scale heights (see Fig.\,\ref{p:initial_stratification}).
\citet{gilet12} showed that mapping and averaging errors in the
corners of the Cartesian grid can lead to spurious velocities that
quickly grow in amplitude.  To reduce this influence we apply a
velocity damping in the outer parts of the computational domain
following \citet{almgren08}. The shaded region in
Fig\,\ref{p:initial_stratification} shows to the damping region, whose
inner boundary is located at a radius of $4.8\,10^{10} \,\mathrm{cm}$.

\citet{Robinson03} found that velocities and temperatures are strongly
influenced by domain boundaries up to a distance of at least two
pressure scale heights. Hence, our velocity damping starts only
$\approx 2.5$ pressure scale heights beyond the convective boundary,
thereby reducing such boundary effects.

We performed 10 simulations in total, exploring the effects of
overshooting, time integration, initial model preparation, and
resolution in time and space (see Table \ref{t:intermediate}). Our
longest running simulation spans more than 6 years of physical time
and covers roughly 340 convective turnover timescales.

\begin{table*}
\begin{tabular}{l|rclccccccccc}
Name & Grid & 1D Model & $f_{ov}$ & $M_{\mathrm{CZ},i}$ & $M_{\mathrm{mixed},i}$ & Int. & dt & Mixing Rate & $Ri_{b}$  & $t_{\mathrm{max}}$ & $t_{\mathrm{max}}/\tau_{\mathrm{conv}}$ \\
\hline
\hline
H3.5  & $1024^2$ & $\nabla_{\mathrm{ex}}$ & 0 & $0.69$ & $0.69$& RK & \vec{U} &$4.0 \cdot 10^{-6}$ & $627$  & $2\cdot 10^8$ & 340 \\
M3.5  & $512^2$ & $\nabla_{\mathrm{ex}}$ & 0  & $0.69$ & $0.69$& RK & \vec{U} &$6.5 \cdot 10^{-6}$ & $592$ & $1\cdot 10^8 $ & 200 \\
E3.5  & $2048^2$ & $\nabla_{\mathrm{ex}}$ & 0  & $0.69$ & $0.69$& RK & \vec{U} &$1.0 \cdot 10^{-5}$ & $3031$ & $1\cdot 10^7 $ & 20 \\
H3.5-pc & $1024^2$ & $\nabla_{\mathrm{ex}}$ & 0  & $0.69$ & $0.69$& PC & \vec{U} &$6.0 \cdot 10^{-6}$ & $2141$ & $6\cdot 10^7 $ & 100 \\
H3.5-igw & $1024^2$ & $\nabla_{\mathrm{ex}}$ & 0  & $0.69$ & $0.69$& PC & IGW &$1.3 \cdot 10^{-5}$ & $766$ & $1\cdot 10^7 $ & 20 \\
H3.5-T & $1024^2$ & T & 0  & $0.69$ & $0.69$& PC & \vec{U} &$4.2 \cdot 10^{-6}$ & $1771$ & $2\cdot 10^8$ & 430 \\
H3.5-ov1.0 & $1024^2$ & $\nabla_{\mathrm{ex}}$ & 0.01  & $0.72$ & $0.83$ & RK & \vec{U} &$2.5 \cdot 10^{-6}$ & $936$  & $8\cdot 10^7 $ & 140 \\
H3.5-ov1.7 & $1024^2$ & $\nabla_{\mathrm{ex}}$ & 0.017 & $0.78$ & $1.00$ & RK & \vec{U} &$1.5 \cdot 10^{-7}$ & $1193$  & $3\cdot 10^7 $ & 50 \\
H3.5-ov2.0 & $1024^2$ & $\nabla_{\mathrm{ex}}$ & 0.02 & $0.78$ & $1.02$ & RK & \vec{U} &$6.4 \cdot 10^{-8}$ & $1580$ & $7\cdot 10^7 $ & 130 \\
H3.5-ov3.0 & $1024^2$ & $\nabla_{\mathrm{ex}}$ & 0.03 & $0.78$ & $1.14$ & RK & \vec{U} &$2.5 \cdot 10^{-16}$ & $2236$  & $4\cdot 10^7$ & 70 \\
\end{tabular}
\caption{\label{t:intermediate} Overview of our 2D simulations of a 
  $3.5\,M\solar$ star. The model name, which is given in the first 
  column, indicates the grid size of the simulation
  (M: medium; H: high; E: extremely high), which is listed in the 
  second column. It also indicates the mass of the star in solar units 
  (3.5). The model name of simulations which differ otherwise from the 
  reference model H3.5 includes further characters (see text). 
  The third column gives the quantity that is kept identical in the 
  2D model to that in the 1D one (see Sect.\,\ref{s:initial_model}). 
  The fourth column gives the overshooting parameter used in the 1D
  stellar evolution calculations. The mass (in solar units) of the 
  initial CZ and that of the homogeneously mixed core can be found 
  in columns five and six, respectively. 
  The seventh column shows the time integration method, the timestep  
  being calculated according to the criterion given in column eight 
  (see text for details).
  The ninth and tenth column give the time-averaged mixing rates 
  for $t > 10^7\,\mathrm{s}$ in units of $M\solar/\mathrm{yr}$. and 
  the bulk Richardson number $Ri_{b}$ of the convective boundary, 
  respectively.
  The last two columns show the final physical time and the number 
  of convective turnovers covered by the simulation. }
\end{table*}

\subsection{Onset of Convection and Steady State Properties}
\label{s:transient}

\begin{figure} 
\resizebox{\hsize}{!}{\includegraphics{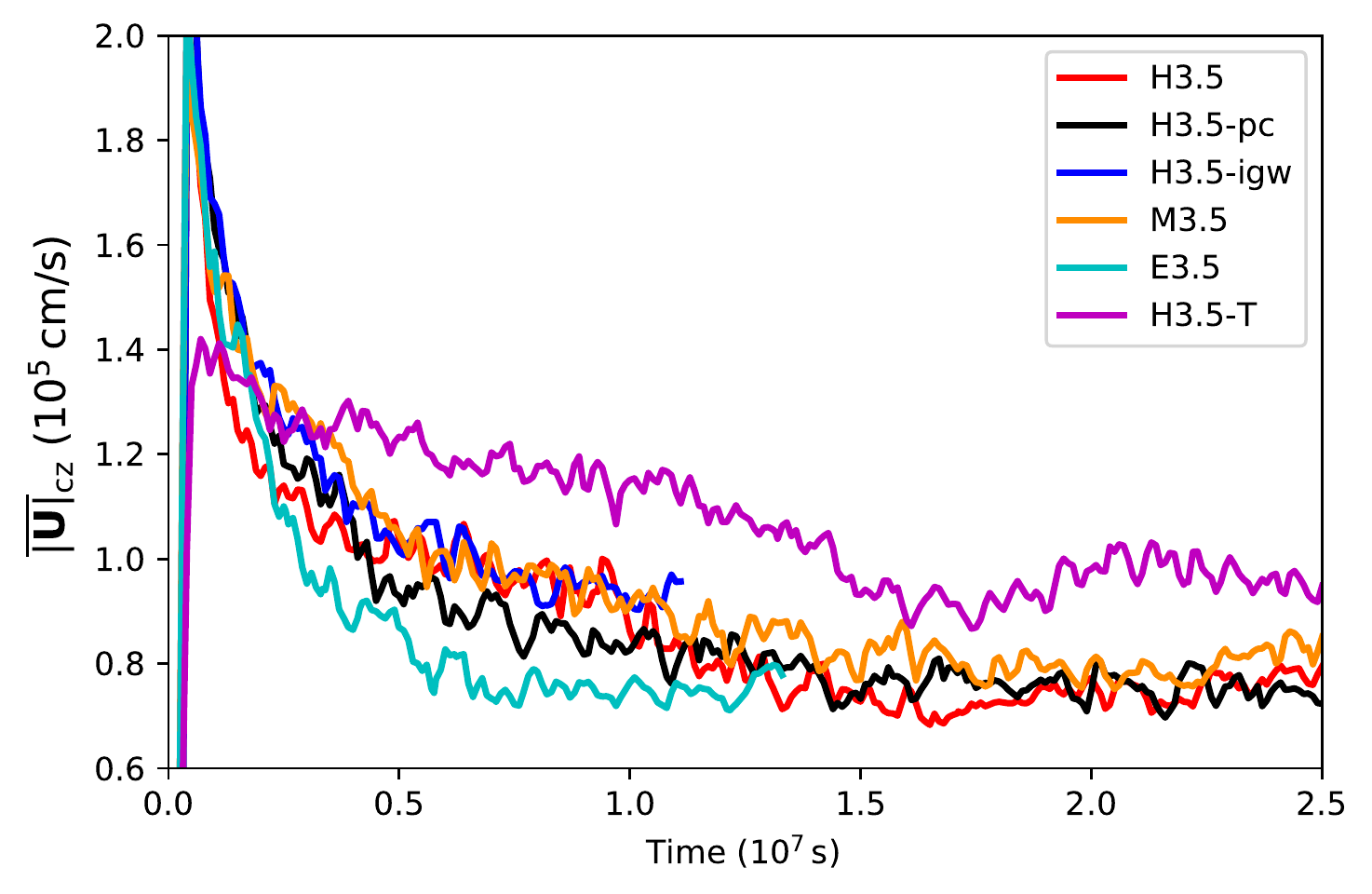}}
\caption{\label{p:vel_evolution} Evolution of the density-averaged
  convective velocity magnitude
  $\overline{\left|U\right|}_{\mathrm{CZ}}$ in the CZ for different 2D
  simulations of a $3.5M\solar$ mass star.}
\end{figure}

We initialize all of our simulations by imposing a small velocity
perturbation in the inner part of the CZ as in \citet{zingale09}. The
perturbation quickly grows and causes a pronounced peak in the
density-averaged convective velocity amplitude
$\overline{\left|U\right|}_{\mathrm{CZ}}$ after
$\approx 10^5\,\mathrm{s}$ (see Fig. \ref{p:vel_evolution}).  Similar
transients have been seen in other simulations
\citep[e.g.][]{meakin07,jones17,gilet13}.  The transient is connected
to a release of thermal energy in the CZ, which brings the slightly
overadiabatic temperature gradient closer to the adiabatic
one. Subsequently, the density-averaged convective velocities slowly
dissipate away over a few convective turnover times until they reach a
quasi steady state after around $10^7 \,\mathrm{s}$.  Their values are
then quite similar in simulations of different resolution
(Fig. \ref{p:vel_evolution}), indicating convergence.  We also
performed simulations with the original predictor-corrector scheme
(denoted with PC in column 7 of Table \ref{t:intermediate}) of
\maestro{} and found that we reach a similar quasi steady state as
with our 4th order Runge-Kutta time integrator (denoted with RK in
column 7 of Table \ref{t:intermediate}).

The simulation H3.5-T uses an initial profile where the temperature of
the 1D model was preserved during reintegration of the HSE. This
simulation produces during the whole simulation $\approx 30\%$ larger
convective velocities than the models initialized with a preserved
$\nabla_{\mathrm{ex}}$, which indicates that the initial thermal
energy reservoir acts as an additional non negligible heating source
in H3.5-T.

\begin{figure*}
\resizebox{\hsize}{!}{\includegraphics{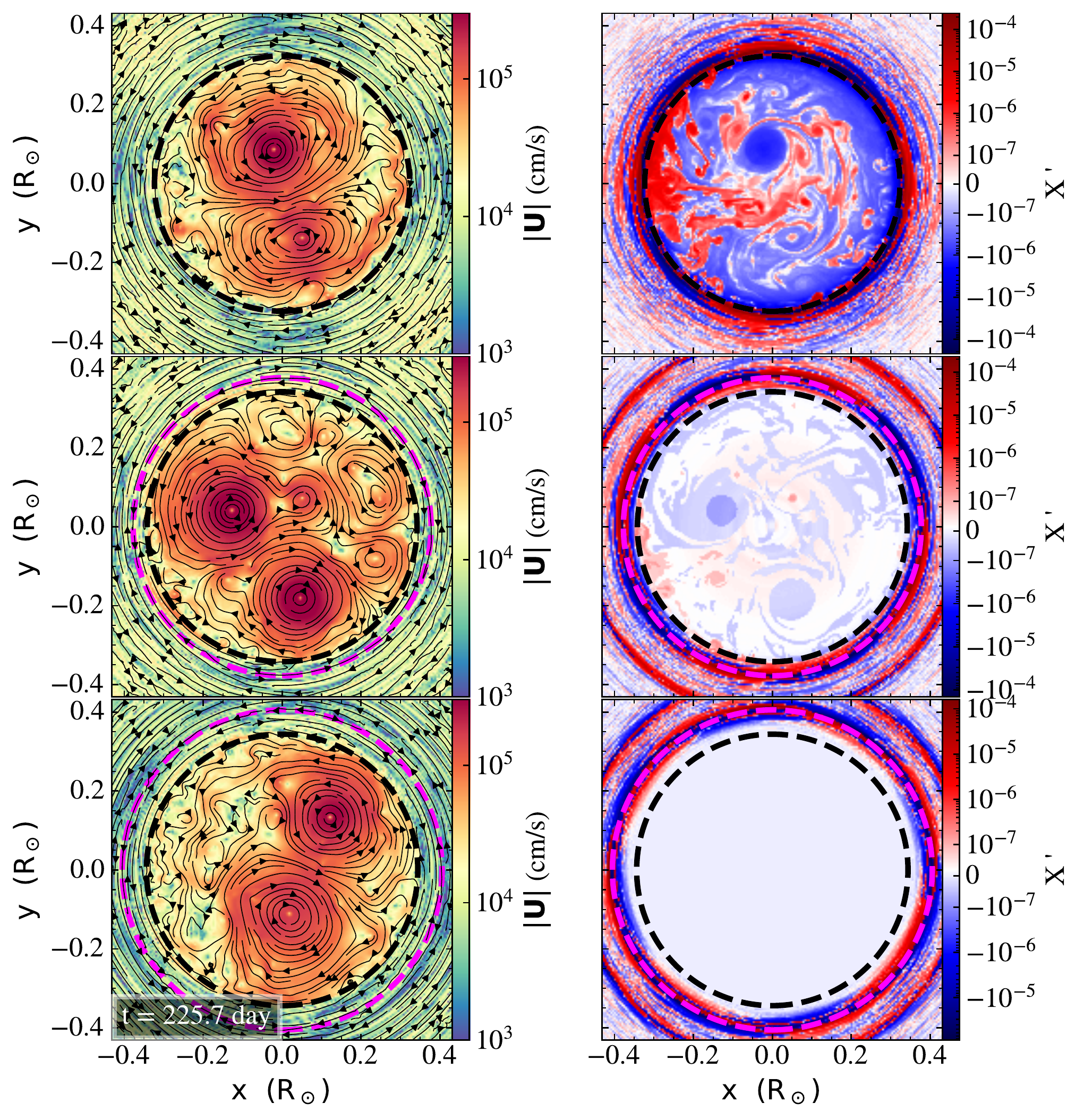}}
\caption{\label{p:flow}Velocity magnitude (left panels) and
  deviations of the hydrogen mass fraction
  from its angulary averaged value 
  $X'(x,y) = X(x,y) - \avgsph{X}(r)$ (right panels) of simulation
  H3.5, H3.5-ov1.7, and H3.5-ov3.0 (from top to bottom) after
  $2\, 10^7\,\mathrm{s}$. The magenta dashed circles mark the initial
  size of the mixed core, while the black dashed circles show the size
  of the convective core according to the Schwarzschild criterion. The
  streamlines in the left panels indicate the flow direction. }
\end{figure*}

\begin{figure}
\resizebox{\hsize}{!}{\includegraphics{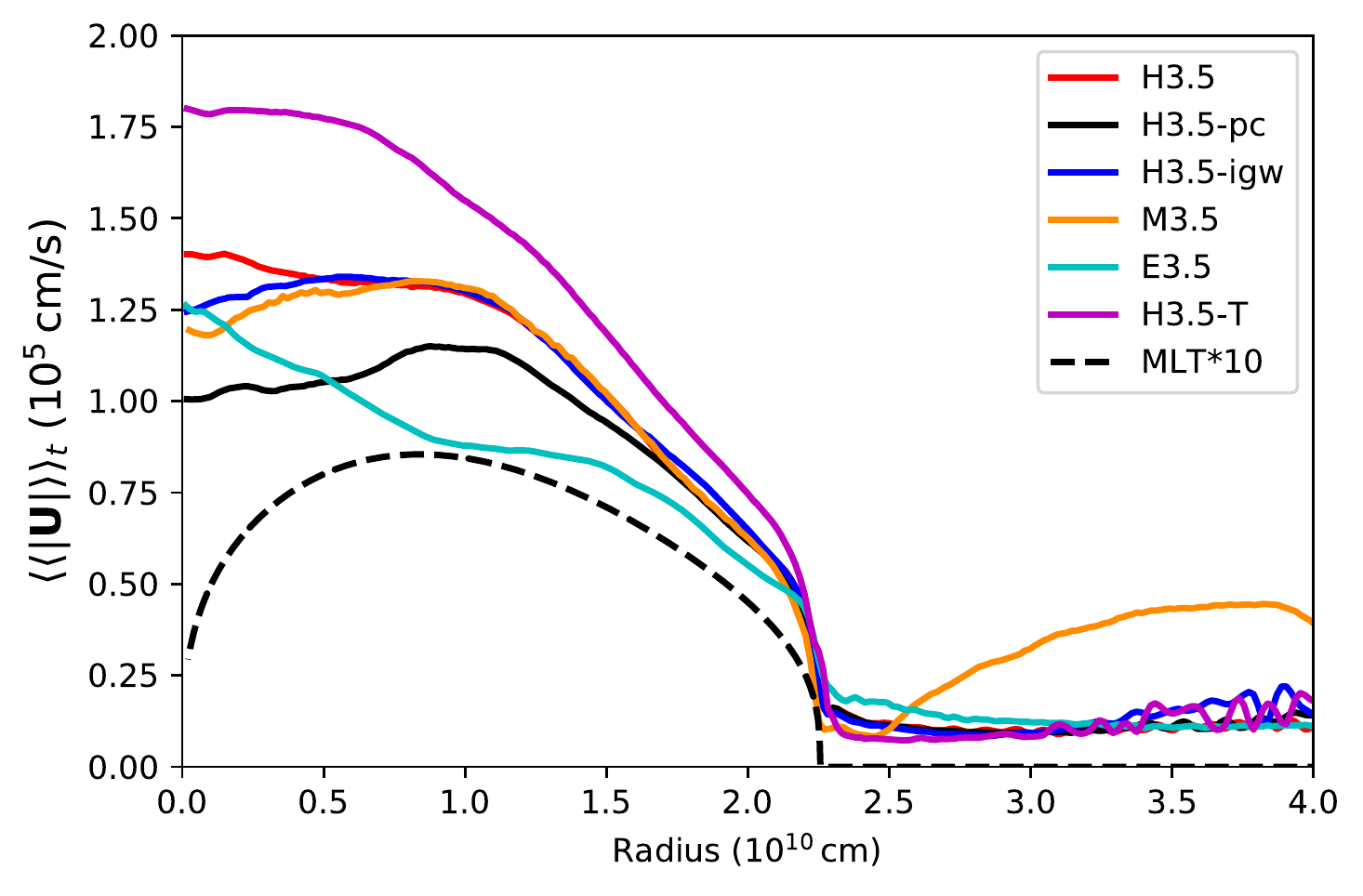}}
\caption{\label{p:vel_profiles} Time-averaged velocity profiles for
  different 2D simulations of a $3.5M\solar$ mass star. The averages
  extend from $5\,10^6\,\mathrm{s}$ to $10^7\,\mathrm{s}$. The dashed
  black line shows, scaled up by a factor of 10, the velocity profile 
  predicted by MLT.}
\end{figure}

The velocity field in the CZ is dominated by two counter-rotating
vortices (see left panels in Fig. \ref{p:flow}), which is a well known
effect due to vorticity conservation in two-dimensional simulations
\citep[e.g.][]{kercek98, meakin07}.  As expected from the inverse
energy cascade in 2D simulations \citep{kraichnan67, batchelor69}, the
vortices fill as much space as is available in the CZ.

Fig.\,\ref{p:vel_profiles} shows velocity magnitude profiles
$\avgtime{\avgsph{\left| U \right|}}$, which are first angularly
averaged (indicated by \avgsph{.}), and then averaged in time
(\avgtime{.}).  Angularly averages are performed using the yt python
package \citep{turk11}, which sorts the Cartesian cells into radial
bins based on the central point of each cell and then computes a mass
weighted average for each bin.  The time average is performed over 50
output files between $5\,10^6\,\mathrm{s}$ and $10^7\,\mathrm{s}$,
which corresponds to roughly 10 convective turnover timescales.
Overall, the shape of the velocity profile within the CZ follows the
predictions by MLT (black dashed line in Fig.\,\ref{p:vel_profiles})
in all simulations.  However, the velocities are larger by more than
one order of magnitude. Similar discrepancies were also found in other
2D simulations \citep{meakin07,pratt16}. \citet{meakin07} and
\citet{pratt20} confirmed that the higher velocities are an artefact
of the reduced dimensionality of 2D simulations.

Besides the shift in amplitude we also find that contrary to MLT
predictions the velocity peaks in our simulations in the centre of the
star. This demonstrates a shortcoming of MLT, which 
treats the centre of the star as a convective boundary and 
therefore predicts zero velocity at that point. Our
simulations do not suffer from this shortcoming due the usage of a
Cartesian grid, which allows flows across the centre of the star. 
 MLT also predicts that the velocities should drop to
zero at the boundary of the CZ. Instead, we find a noticeable velocity
magnitude throughout the stable layer, which is approximately one
order of magnitude smaller than in the CZ. The latter behaviour does
not hold, however, for the medium resolution model M3.5, which suffers
from the problem of unresolved gravity waves as discussed in
\ref{s:maestro}.  Nevertheless, in the CZ the velocity profile of
model M3.5 agrees perfectly with that of model H3.5 indicating that
the time-averaged flow in the CZ is numerically converged.  The small
discrepancies present between model H3.5 and the extremely highly
resolved model E3.5 can be explained by the different transient
behaviours of these simulations (see Fig.\,\ref{p:vel_evolution}),
because model E3.5 reaches the quasi steady state at an earlier time
than model H3.5, which reduces the average velocities in model E3.5.

In simulation H3.5-igw we used the timestep criterion according to
Eq.\,(\eqref{eq:brunt_timestep}), which resolves the timescale of IGW up
to a wavelength of a pressure scale height. For efficiency reasons we
used the PC integrator in this case.  According to linear theory IGW
are exponentially damped inside of a CZ, which means that we do not
expect any influence of unresolved IGW on the convective flow itself.
Fig.\,\ref{p:vel_evolution} and Fig.\,\ref{p:vel_profiles} show that
models H3.5-igw and H3.5 indeed produce almost identical results in
the CZ.

Simulation H3.5-igw also confirms the need for a higher order time
integration. Fig.\,\ref{p:vel_profiles} shows that model H3.5-pc,
which uses the original PC time integration, has consistently smaller
velocities than those found in model H3.5. However, the results of
simulation H3.5-igw, which were obtained with the same timestep
integrator but smaller timesteps than model H3.5, agree perfectly with
those of model H3.5.

\subsection{Convective Boundary}
\label{s:conv_boundary}

\begin{figure} 
\resizebox{\hsize}{!}{\includegraphics{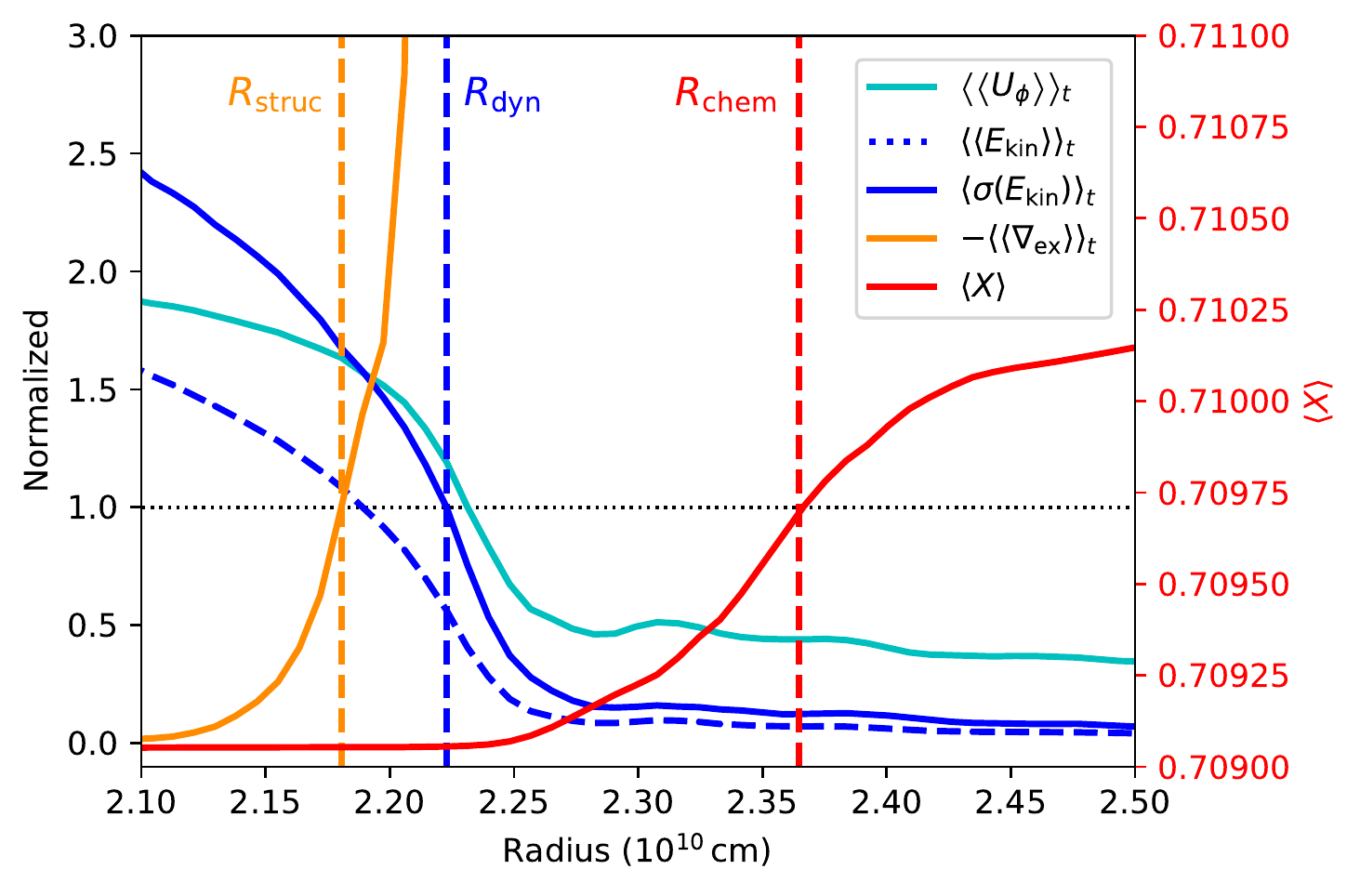}}
\caption{\label{p:boundaries} Various radial profiles in the
  neigbourhood of the convective boundary.
  $\avgtime{\avgsph{U_\phi}}$, $\avgtime{\avgsph{E_{\mathrm{kin}}}}$,
  and $\avgtime{\avgsph{\nabla_{\mathrm{ex}}}}$ are the combined
  spatial (indicated by \avgsph{.}) and time (\avgtime{.}) averages of
  the angular velocity $U_\phi$, the kinetic energy
  $E_{\mathrm{kin}}$, and the superadiabaticity
  $\nabla_{\mathrm{ex}} = \nabla_{\mathrm{mlt}} - \nabla_{\mathrm{ad}}$,
  respectively.  The time averages are taken from $1.82\,10^8\,\mathrm{s}$ to
  $2.06\,10^8\,\mathrm{s}$ using 100 output files from the H3.5
  simulation.
  $\avgtime{\sigma(E_{\mathrm{kin}})}$ is the time-averaged standard
  deviation of the kinetic energy (see Eq.\,\eqref{eq:variation}). 
  The angularly averaged hydrogen mass fraction profile $\avgsph{X}$ is
  taken at $2.06\,10^8\,\mathrm{s}$.
  The curves of $\avgtime{\avgsph{U_\phi}}$, $\avgtime{\avgsph{E_{\mathrm{kin}}}}$,
  $\avgtime{\avgsph{\nabla_{\mathrm{ex}}}}$, and $\avgtime{\sigma(E_{\mathrm{kin}})}$ 
  are normalized to the values corresponding to their respective
  boundary definitions (see text), i.e., the black dotted line
  indicates the boundary values of each line. The dashed vertical lines mark
  the radial locations of our favored boundaries. 
}
\end{figure}

Before we can discuss the mixing across convective boundaries in more
detail we first need to define the position of the convective
boundary.  In a 1D stellar evolution context the most commonly used
definition is the Schwarzschild boundary, i.e., the point where the
acceleration of mass elements by buoyancy is no longer positive. A
proxy for the Schwarzschild criterion is a negative superadiabaticity
$\nabla_{\mathrm{ex}}$ as defined in Eq.\,\eqref{eq:nabla}. We denote
the corresponding radial position by $R_{\mathrm{struc}}$.

In hydrodynamic simulations, however, the position of a convective
boundary can also be determined by the velocity field.
\citet{rogers15} and \citet{brummell02} account for this by defining
the position of the convective boundary as the point where the kinetic
energy $E_{\mathrm{kin}}$ drops to $1\%$ or $5\%$ of its peak value,
respectively.  We propose instead to use the standard deviation of a
dynamical quantity as an indicator for the boundary, because
time-averaged profiles smooth out all rare mixing events that might
happen during the averaging period. \citet{pratt17} argue that such
rare mixing events will set the final depth of the mixed region around
CZ. Using the standard deviation of the velocity field (or kinetic
energy field) does include a larger contribution from those rare
events into the time average.

To this end we define the time-averaged standard deviation
$\avgtime{\sigma(E_{\mathrm{kin}})}$ by combining the standard
deviations $\sigma_i(E_{\mathrm{kin}})$ of $N$ output files as
\begin{equation}
\label{eq:variation}
  \left<\sigma(E_{\mathrm{kin}})\right>_t = 
  \left[\frac{\sum_{i = 0}^{N} \sigma_i(E_{\mathrm{kin}}) 
              \left[ \left<E_{\mathrm{kin,i}}\right> - 
                     \overline{\left<E_{\mathrm{kin}}\right>}
  \right]^2}{N}
  \right]^{1/2},
\end{equation}
$\overline{\left<E_{\mathrm{kin}}\right>} = \frac{1}{N}
\sum_{i=0}^{N} \langle E_{\mathrm{kin,i}} \rangle$
is the mean of $\left<E_{\mathrm{kin,i}} \right>$. 

We define the boundary $R_{\mathrm{dyn}}$ (see
Fig.\,\ref{p:boundaries}) to be located at $5\%$ of the maximum value
of $\avgtime{\sigma(E_{\mathrm{kin}})}$.  In Fig.\,\ref{p:boundaries}
we also show for comparison the profiles of
$\avgtime{\avgsph{E_{\mathrm{kin}}}}$ and $\avgtime{\avgsph{U_\phi}}$.
As expected $R_{\mathrm{dyn}}$ is located at a larger radius than a
boundary that is defined at $5\%$ of the maximum value of
$\avgtime{\avgsph{E_{\mathrm{kin}}}}$.

Fig. \ref{p:boundaries} shows that our definition of $R_{\mathrm{dyn}}$
provides a radial position of the boundary that is very close to the steepest gradient in the angular
velocity component $U_\phi$, which was shown to be a good indicator
for the convective boundary by \citet{jones17}.

Another way how to determine the CZ boundary was used by
\citet{cristini17} and \citet{meakin07}, who followed the evolution of
the size of a CZ by looking at the time evolution of the chemical
composition which, in general, has an initial jump at the CZ
boundary. The radial position where the composition corresponds to the
mean between CZ and stable layer can then be defined as the boundary
$R_{\mathrm{chem}}$.

Fig.\,\ref{p:boundaries} shows the respective time-averaged profiles
obtained in our H3.5 simulation, and it also gives the corresponding
locations of the various boundaries. We set the Schwarzschild boundary
at $\nabla_{\mathrm{ex}} < - 2\,10^{-4}$  due to the
uncertainty in this quantity in \maestro{} simulations (see
\ref{s:temp_grad}). At the end of the simulation we find
$R_{\mathrm{struc}} < R_{\mathrm{dyn}} < R_{\mathrm{chem}}$.

\begin{figure} 
\resizebox{\hsize}{!}{\includegraphics{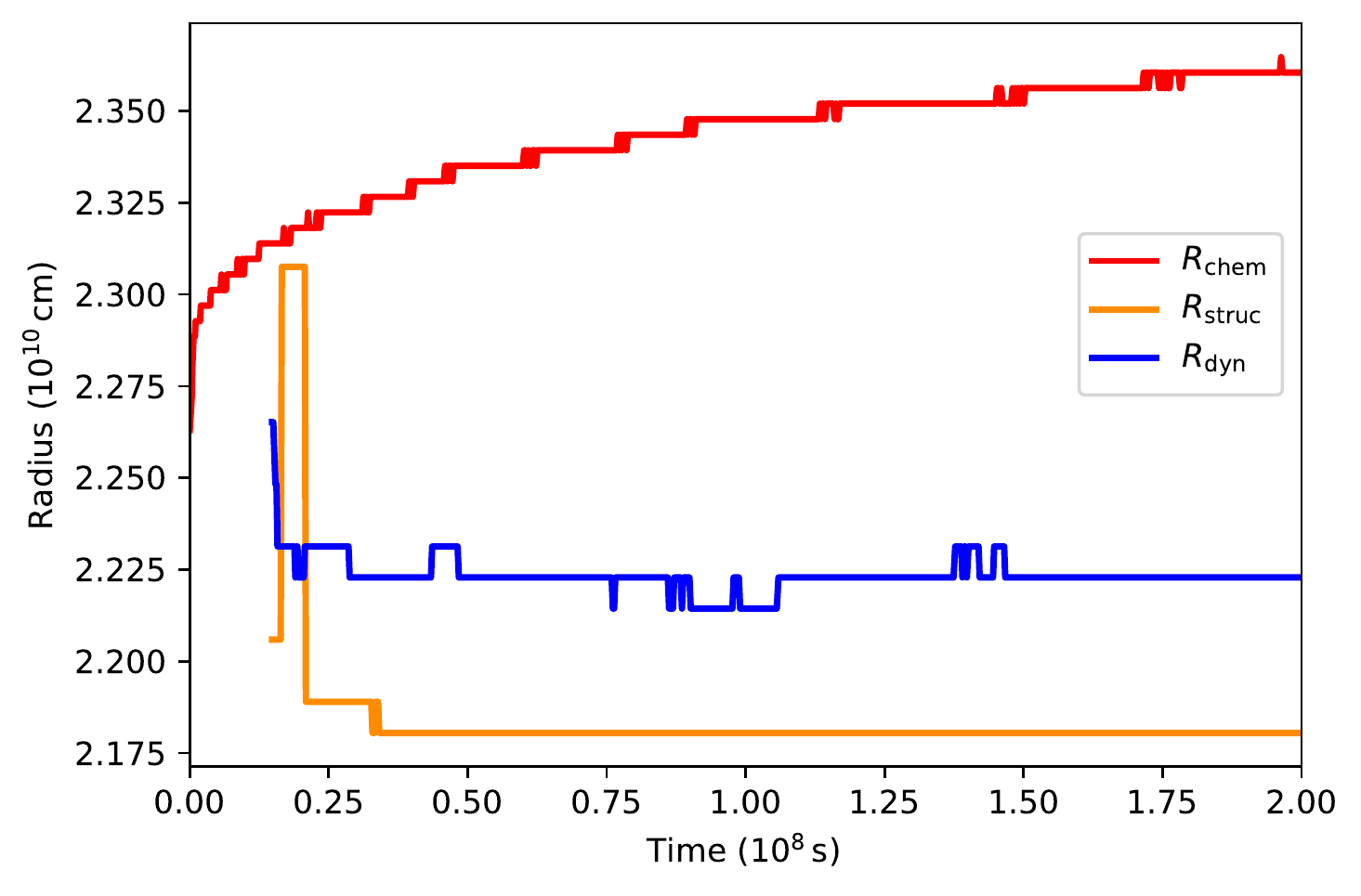}}
\caption{\label{p:boundary_evo} Temporal evolution of the various
  convective boundary locations (see text) in the H3.5 simulation.}
\end{figure}

In order to understand how mixing across the boundary progresses we
can now look at the time evolution of the different boundary
definitions.  Fig.\,\ref{p:boundary_evo} shows that the boundaries
$R_{\mathrm{dyn}}$ and $R_{\mathrm{struc}}$ are constant during the
whole simulation, except for the transient phase at the beginning.
During the transient the location of $R_{\mathrm{struc}}$ is shifted
inwards by about $7\,10^{8}\,\mathrm{cm}$, indicating a thermal
restructuring due to the release of thermal energy stored in the
temperature gradient. The released thermal energy is converted into
kinetic energy, which shifts $R_{\mathrm{dyn}}$ outwards beyond the
initial location of $R_{\mathrm{struc}}$ at
$2.25\,10^{10}\,\mathrm{cm}$.  Once $R_{\mathrm{struc}}$ retracts and
the system approaches a quasi steady state $R_{\mathrm{dyn}}$ remains
constant.

The spatial ordering of the boundaries is in agreement with the simple
picture of ballistic overshooting, which requires that mass elements
penetrate into the layer beyond the Schwarzschild boundary. The
additional motion is reflected by the fact that
$R_{\mathrm{dyn}} > R_{\mathrm{struc}}$ during most of the time. In
fact, the short period in the beginning of the simulation where the
order is reversed is due to chemical mixing, which adjusts the shape
of the temperature gradient to the more realistic Ledoux criterion.
This effect temporarily increases $\nabla$ just outside of the initial
Schwarzschild boundary.  As the chemical boundary moves further out
the change of $\nabla$ due to chemical gradients becomes negligible
for the determination of $R_{\mathrm{struc}}$, and the expected order
is re-established. We also see that the distance between
$R_{\mathrm{dyn}}$ and $R_{\mathrm{struc}}$ is only
$6\,10^{8}\,\mathrm{cm}$, corresponding to $0.04\,H_p$.

\citet{lattanzio17} give a limit of mixing around CZ during thermal
pulses of asymptotic giant branch stars. They argue that the maximum
distance a plume can mix is set by its kinetic energy and the
counteracting buoyancy force. For a general density $\rho(r)$ and
gravity $g(r)$ stratification one can then calculate the penetration
depth $d = \left|r_1 - r_0 \right|$ as
\begin{equation}
\label{eq:lattanzio}
  \frac{1}{2}v_0^2 = \int_{r_0}^{r_1} g(r) 
                     \left( \frac{\rho(r)-\rho_p}{\rho_p}\right) dr,
\end{equation}\
where $\rho_p$ is the density of the plume and $v_0$ its typical
velocity. Assuming that a plume reaches the boundary with a velocity
that corresponds to the maximum of the velocity profile in
Fig.\,\ref{p:vel_profiles} ($v_0 = 1.4\,10^5\,\mathrm{cm/s}$) and that
its density corresponds to the density at the boundary
($\rho_p = \rho(r_0)$ the maximum penetration length is
$d = 5\,10^{7}\,\mathrm{cm}$, which is considerably smaller than what
our simulations show. If we additionally account for the expansion of
the plume by requiring its density to be always
a fraction of $10^{-4}$ larger (which is a typical value we find
in our simulations)
than the surrounding density such that
$\rho_p = \rho(r) (1 + 10^{-4})$, we find $d = 5\,10^{8}\,\mathrm{cm}$
in very good agreement with our measured distance between
$R_{\mathrm{dyn}}$ and $R_{\mathrm{struc}}$.

However, we also see that the chemical boundary is evolving further
into the stable layer throughout the simulation, indicating that model
H3.5 experiences a constant mixing of hydrogen across its core
boundary. Furthermore, $R_{chem}$ becomes much larger than $R_{dyn}$,
which shows that the ballistic overshooting cannot be the main mixing
process, as it does not reach that far into the stable layer.  In the
following we will therefore look at diffusive mixing processes that
can contribute to the effects we see.

\subsection{Diffusive Mixing}
\label{s:diffusive}

\begin{figure}
\resizebox{\hsize}{!}{\includegraphics{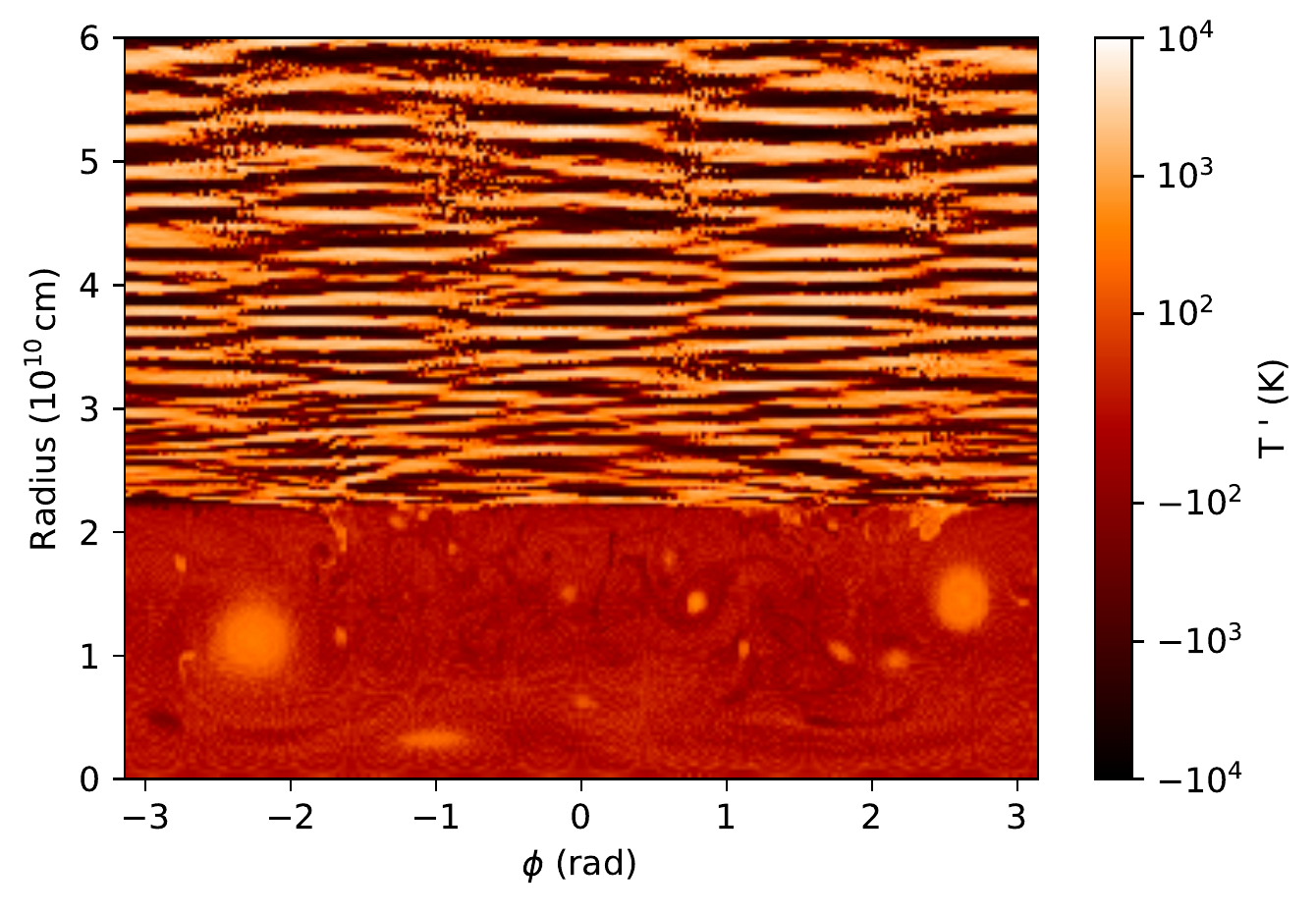}}
\caption{\label{p:temp} Temperature fluctuations around the
    angularly average value in the H3.5 simulation at
    $8\,10^7\,$s.}
\end{figure}

As was shown by \citet{rogers17}, IGW can mix the stable layer of
stars diffusively.  Even though we do not fully resolve the highest
frequency IGW in our simulations they still develop a rich IGW
spectrum, best seen by looking at temperature fluctuations
$T'(r,\phi) = T(r,\phi) - \avgsph{T}(r)$ around the angularly averaged
temperature $\avgsph{T}(r)$, where $T(r,\phi)$ is our Cartesian
temperature field interpolated onto a polar grid.

In Fig.\,\ref{p:temp}, which shows temperature fluctuations in model
H3.5 at $8\,10^7\,$s, one can recognize a clear distinction between
the CZ with an almost homogeneous temperature distribution except for
the centre of the vortices (bright spots) and the stable layer, which
shows orders of magnitude larger temperature fluctuations. The regular
flat pattern in the stable region indicates that the convective flow
creating the IGW is dominated by large vortices \citep{rogers13}.
Since this is an effect of the reduced dimensionality of our
simulations, we do not expect this wave pattern to be a good
representation of nature and therefore will not analyse it in detail.

\begin{figure}
\resizebox{\hsize}{!}{\includegraphics{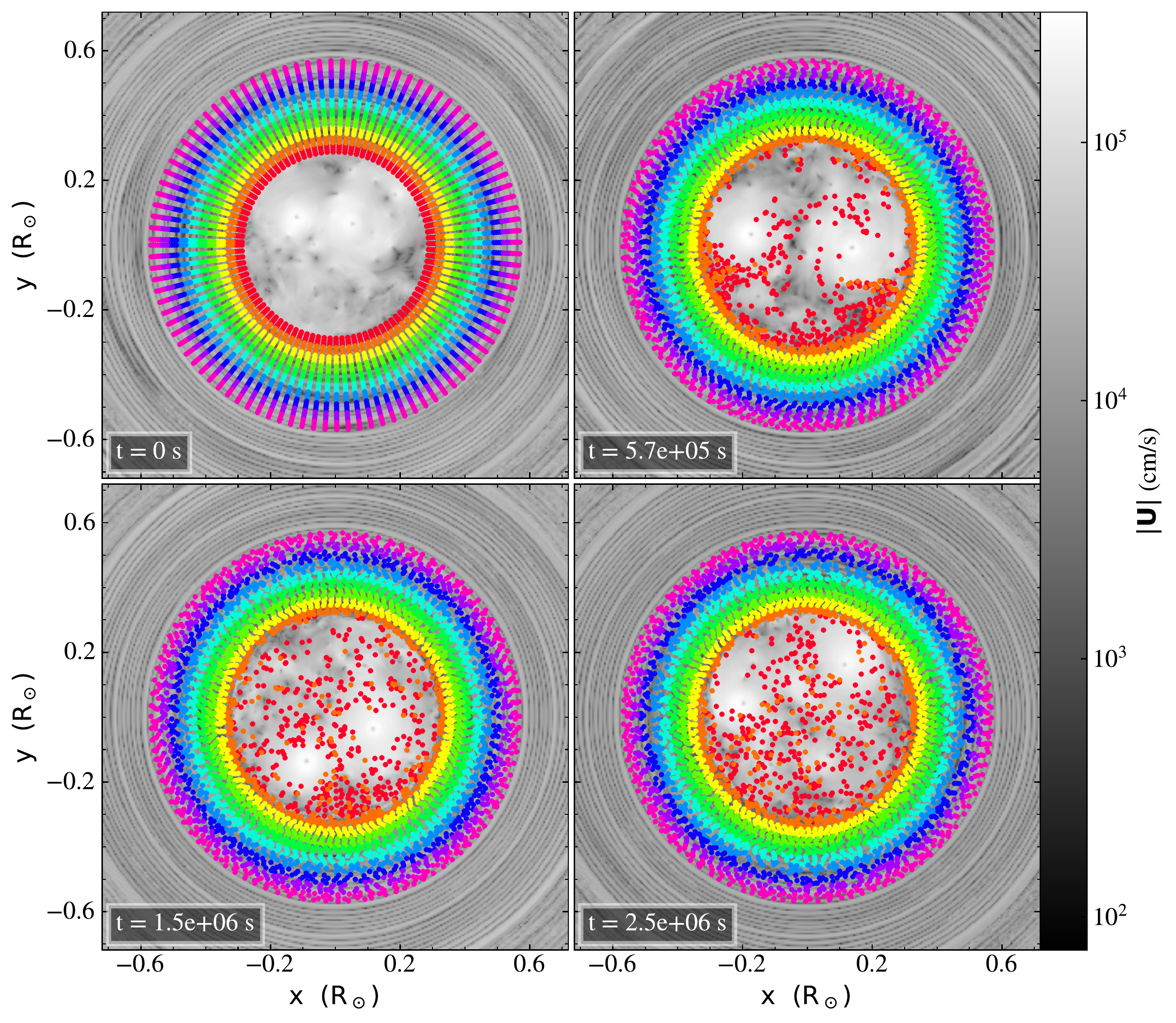}}
\caption{\label{p:tracer} Distribution of tracer particles for model
  H3.5 at several epochs.  We show only 4000 of the 40000 tracer
  particles used in the simulation, which are coloured according to
  their initial radial location given in the upper left panel.}
\end{figure}

Because the mixing caused by these waves plays a significant role
in our analysis, we estimated the amount of diffusive mixing by IGW
with tracer particles in a post processing step. Following
\citet{rogers17} we used tracer particles and tracked how the
particles were advected by the flow. For that purpose, 
the velocities of the tracer particles are 
estimated via linear interpolation on the grid. 
The positions of the tracer particles are then evolved with a
constant velocity up until the time of the next simulation output. 
To increase the accuracy of the analysis we
increased the number of simulation outputs during the analyzed timespan by a
factor of 1000, i.e., one simulation output every 1000 seconds. We
placed the tracer particles on regularly spaced spherical shells with
radii between $2.0\,10^{10}\,\mathrm{cm}$ and
$4.0\,10^{10}\,\mathrm{cm}$ to increase the resolution in the CZ
boundary region (see top left panel of Fig.\,\ref{p:tracer}). The
radial diffusion coefficient $D$ at radius $r$ is then the average of
the squared radial displacement of the tracer particles initially placed at
$r$ divided by the elapsed time.

\begin{figure}
  \resizebox{\hsize}{!}{
  \includegraphics{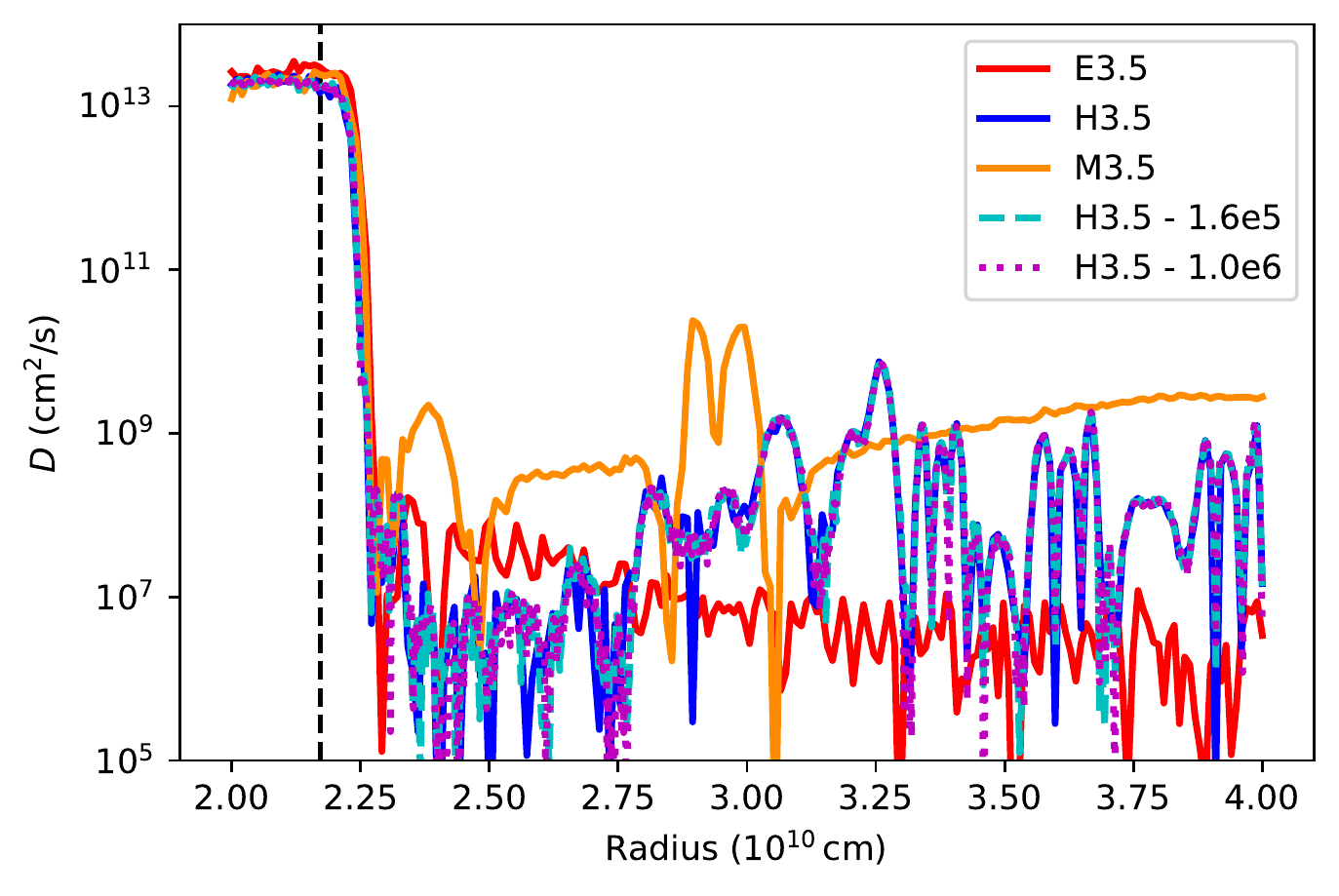}}
  \caption{\label{p:diff_resolution} Radial diffusion coefficients
    estimated from the advection of 40000 tracer particles followed for
    $1.5\,10^6\,\mathrm{s}$ for models of different grid size. The 
    dashed and dotted lines labeled "H3.5 - 1.6e5" and "H3.5 - 1.0e6" represent the 
    same quantity but evaluated with $160000$ and $10^6$ tracer particles in model H3.5, respectively. 
    The dashed vertical line indicates $R_{\mathrm{dyn}}$ of model H3.5.}
\end{figure}

\begin{figure}
  \resizebox{\hsize}{!}{
  \includegraphics{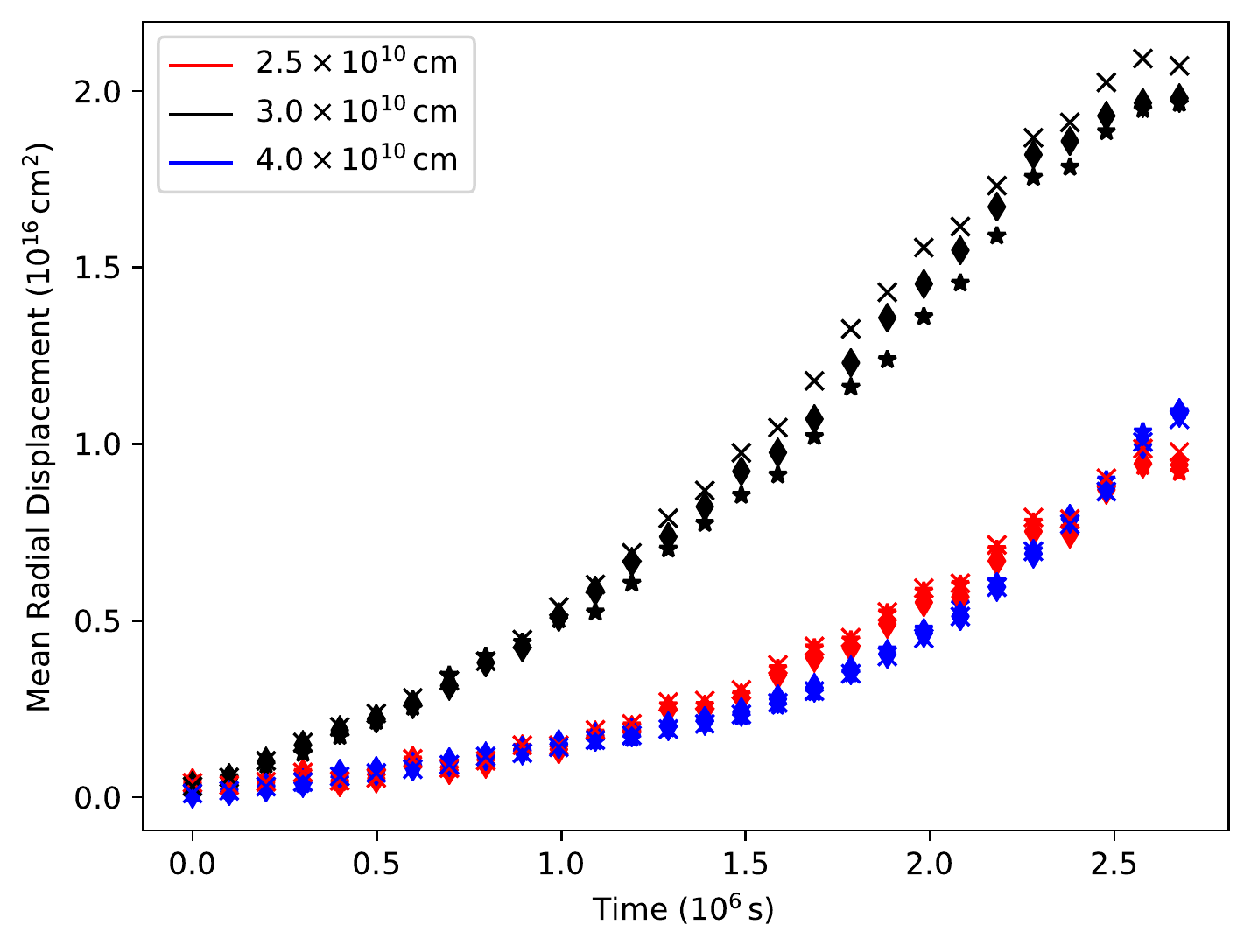}}
  \caption{\label{p:radial_displacement} Time evolution of the mean 
  squared radial displacement of tracer particles in model H3.5. The particles which where used for the analysis were taken  
  at $r=2.5\,10^6\,\mathrm{cm}$, $3.0\,10^6\,\mathrm{cm}$, and $4.0\,10^6\,\mathrm{cm}$.
  Stars, diamonds, and crosses correspond to analysis runs with a total of 
  $40000$, $160000$, and $10^6$ tracer particles, respectively. }
\end{figure}

Fig.\,\ref{p:tracer} shows snapshot of the position of the tracer
particles for model H3.5. Already early on the particles initially
located inside the CZ get transported all the way to the centre and
eventually get almost homogeneously distributed inside the CZ.  This
demonstrates the efficiency of convective mixing, corresponding to a
large radial diffusion coefficient of $D \ge 10^{13}\,$[cm$^2$/s] (see
Fig.\,\ref{p:diff_resolution}).  This value is in perfect agreement
with the value of diffusion coefficients estimated from MLT
velocities.  Even though we can compute $D$ inside the CZ we want to
note here that a diffusive treatment of convective mixing is not
correct, because the extracted value of $D$ depends on the chosen
$\Delta t$ \citep[see][]{rogers17}. In particular, one would expect
that $D \propto 1/\Delta t$ once the tracer particles are randomly
distributed inside the CZ, since the mean radial displacement cannot
increase further at that point unless particles are mixed from the CZ
into the stable layer. We also want to emphasize that the 
diffusion coefficients estimated here do not apply other stars with a different
density or temperature stratification.

Outside of the CZ the tracer particles move considerably less. At
later times there is some mixing between the particles, but mainly in
the lateral direction as can be seen from the colour coding in
Fig.\,\ref{p:tracer}. The mean of the squared radial displacement of tracer particles at 
three initial radial positions in the stable layer is given in Fig.\,\ref{p:radial_displacement}. 
A diffusive process will on average lead to a linear growth of the squared radial displacement. 
Our measurements show that a clear linear trend is maintained until $1.5\,10^{6}\,$s. 
After that point integration errors become noticeable and the curve slightly deviates from 
its linear trend. In the following we will therefore only consider the first $1.5\,10^{6}\,$s
of the tracer particle movements to estimate diffusion coefficients.
We performed this analysis with $40000$, $160000$ and $10^6$ tracer particles, 
where the latter corresponds to about 4 tracer particles per grid cell in the 
analysed radial range. In Fig.\,\ref{p:radial_displacement} the different symbols -- corresponding 
to the number of tracer particles used -- show that the number 
of tracer particles has little influence on the measured radial displacement,
indicating convergence of the analysis method. In the following we will therefore use $40000$ tracer particles. 

Measuring $D$ for all radial values, we find that $D$ is orders of magnitude smaller in the
stable layer than in the CZ (see Fig.\,\ref{p:diff_resolution}). In
the transition region between the CZ and the stable layer $D$ drops
sharply at $R \approx 2.25\,10^{10}\,\mathrm{cm}$, corresponding to
our dynamical convective boundary $R_{\mathrm{dyn}}$. Therefore, we
interpret the drop in $D$ as a very rapid decline of dynamical
convective mixing in that region.  Overall we find good agreement with
the results of \citet{rogers17}, who analysed a $3.0\,M\solar$
star at the ZAMS. This setup is very similar to the one in this work and therefore directly comparable. 
\citet{rogers17} also find $D \approx 10^{13}\,$[cm$^2$/s] in the CZ and
then a sharp drop to $D \approx 10^{8} - 10^{9}\,$[cm$^2$/s] in the
stable layer, which is approximately 1-2 orders of magnitude smaller
than what we find.  Here it should be noted that
$D = 10^{9}\,\mathrm{cm}^2/\mathrm{s}$ corresponds to an average
diffusion distance of roughly one third of a computational cell of model
H3.5 during the analysed period of $1.5\,10^6\,\mathrm{s}$. Hence, we
estimated the systematic error in the value of $D$ by repeating the
analysis with our medium resolution model M3.5 and the extremely high
resolution model E3.5 (see Fig.\,\ref{p:diff_resolution}).

We find that all three simulations give identical values of $D$ in the
CZ, showing that the convective motions are converged. The same is
true for the location and steepness of the sharp drop of $D$ at the
dynamical boundary (see Fig.\,\ref{p:diff_resolution}). However, we
find rather large differences of up to four orders of magnitude in the
stable layer. As mentioned before we attribute this to the limited
spatial resolution of our simulations.  We would expect that we will
find smaller and smaller values of $D$ if we increase the resolution
further as suggested by the small values of $D$ found in model E3.5.

The hypothesis that our values of $D$ are overestimated is also
supported by timescale arguments. Using the low values of
$D = 10^{7}\,\mathrm{cm}^2/\mathrm{s}$ found in model E3.5, we find
that the mixed core will grow by $\approx 0.4\,H_p$ during its local
thermal timescale of $160\,\mathrm{kyr}$. Considering the main-sequence 
lifetime of a $3.5\,M\solar$ star of the order of
$100\,\mathrm{Myr}$ one consequently finds that the whole star will be
fully mixed at the end of the main-sequence.  Moreover asteroseismic
observations of KIC 7760680 \citet{moravveji16} found that stellar
models can only fit the observed frequencies when an additional
diffusive mixing is added throughout the stable layer. They determined
a value of $D \approx 10\,\mathrm{cm}^2/\mathrm{s}$. A simulation that
could resolve such small diffusion coefficients is computationally
infeasible.

Overall we conclude that there is some diffusive mixing due to IGW in
our simulations, but it is very likely that the amount we find is
largely overestimating the mixing in actual stars. The most likely
reason for this is the limited resolution of our simulations, which
does not allow us to properly handle significantly smaller values of
$D$. The increased convective velocities in our 2D simulations (see
\ref{s:transient}) probably also caused values of $D$ that are
overestimated.

\subsection{Overshooting Calibration}
\label{s:calibration}

\begin{figure}
\resizebox{\hsize}{!}{\includegraphics{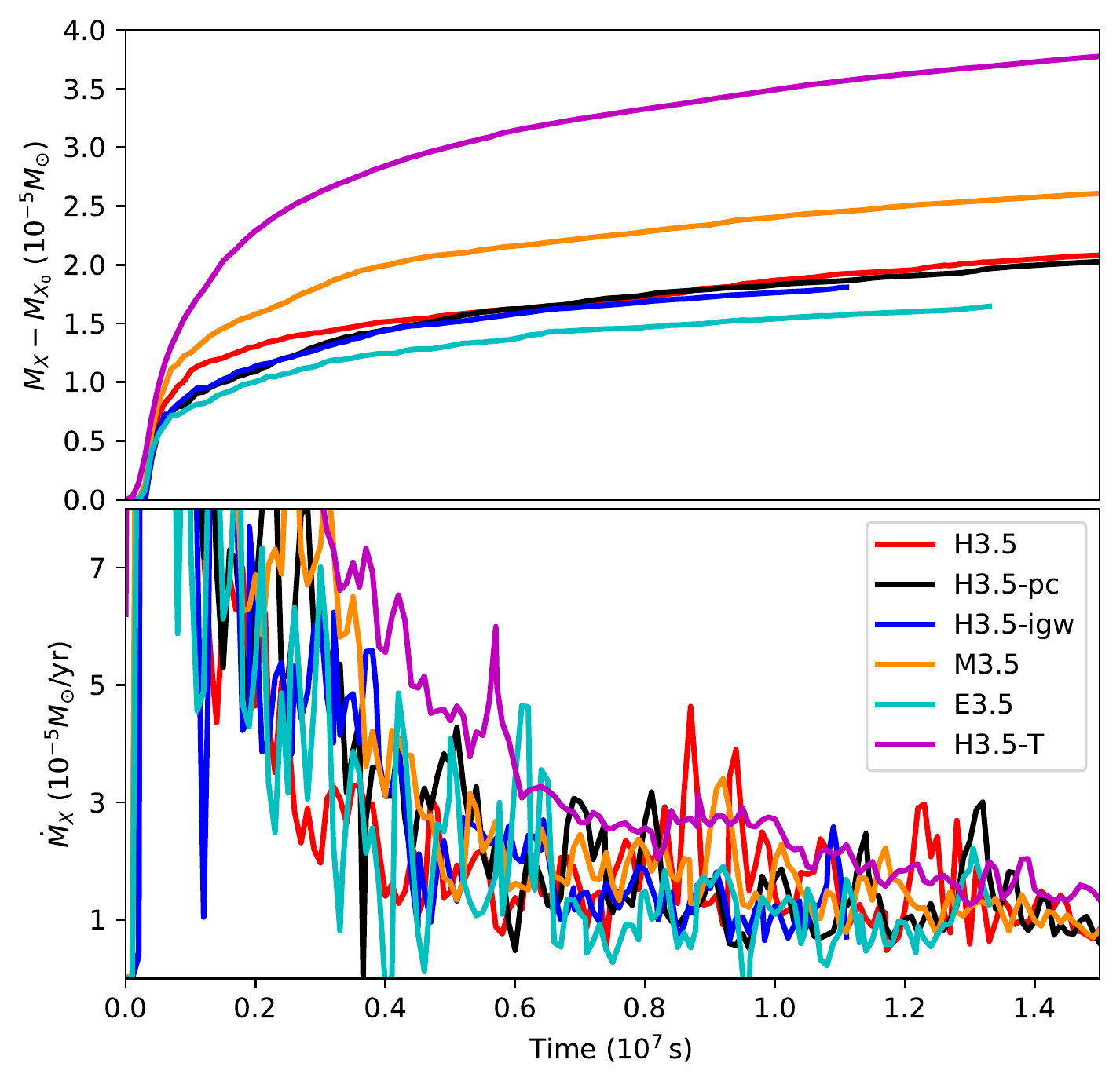}}
\caption{\label{p:mixing_details} The top panel shows the change in
  the hydrogen mass (relative to its initial value) for the initial
  convective core of our $3.5M\solar$ simulations. The bottom panel
  shows the corresponding rate of change.}
\end{figure}

The large difference between the short dynamical timescale of
convection and the long nuclear timescale of stellar evolution
prohibits the simulation of multi-D stellar evolution models with
current computers.  Therefore, 1D stellar evolution models still rely
on parametrized models of convection. A crucial parameter in this
approach is the overshooting parameter which sets the size of the
mixed region around a CZ.  In this section we will show how one can
use multi-D simulations to estimate the 1D overshooting parameter.

To this end we analysed the evolution of the hydrogen mass fraction
$X$ in the CZ in our multi-D simulations.  Initially hydrogen is
distributed homogeneously in the CZ, and the adjacent stable layer has
a larger hydrogen mass fraction than the CZ.  One way to quantify the
mixing efficiency across the convective boundary is to examine the
increase of the total hydogen mass inside the CZ $M_X$ with time.  For
this purpose we set the location of the convective boundary at the
Schwarzschild boundary of the initial model, which is not necessarily
the same as $R_{\mathrm{struc}}$ because the temperature
stratification changes slightly during the initial transient.
Furthermore, we denote $M_X$ at $t=0$ by $M_{X_0}$.

Fig.\,\ref{p:mixing_details} shows the mixing during the initial
transient and the beginning of the quasi steady state for the same
models as in Figs.\,\ref{p:vel_evolution} and \ref{p:vel_profiles}.
$M_X$ increases by $(2.0 \pm 0.5)\,10^{-5}\,M\solar$ during the first
$10^7\,\mathrm{s}$ in all simulations, except for model H3.5-T for
which the amount of mixing is about twice as high as that of, e.g.,
model H3.5.  This enhanced mixing is due to the fact that the velocity
is on average larger in the steady state of model H3.5-T, and that the
model also experienced a more extended transient phase (see
Fig.\,\ref{p:vel_evolution}) with even larger velocities.  The
duration of the transient is also the reason for the small differences
in mixing between the medium and extremely high resolution models M3.5
and E3.5.  However, once the simulations reach the quasi steady state
the mixing evolves at very similar mixing rates, which is confirmed by
looking at the time derivative $\dot{M}_X$ shown in the bottom panel
of Fig.\,\ref{p:mixing_details}. 

It is also worth noting here that
the mixing rates in models H3.5 and H3.5-igw agree almost perfectly. 
This confirms the hypothesis we made in \ref{s:maestro} that 
the unresolved high frequency IGW in model H3.5 do not contribute to 
the mixing significantly. 

\begin{figure}
\resizebox{\hsize}{!}{\includegraphics{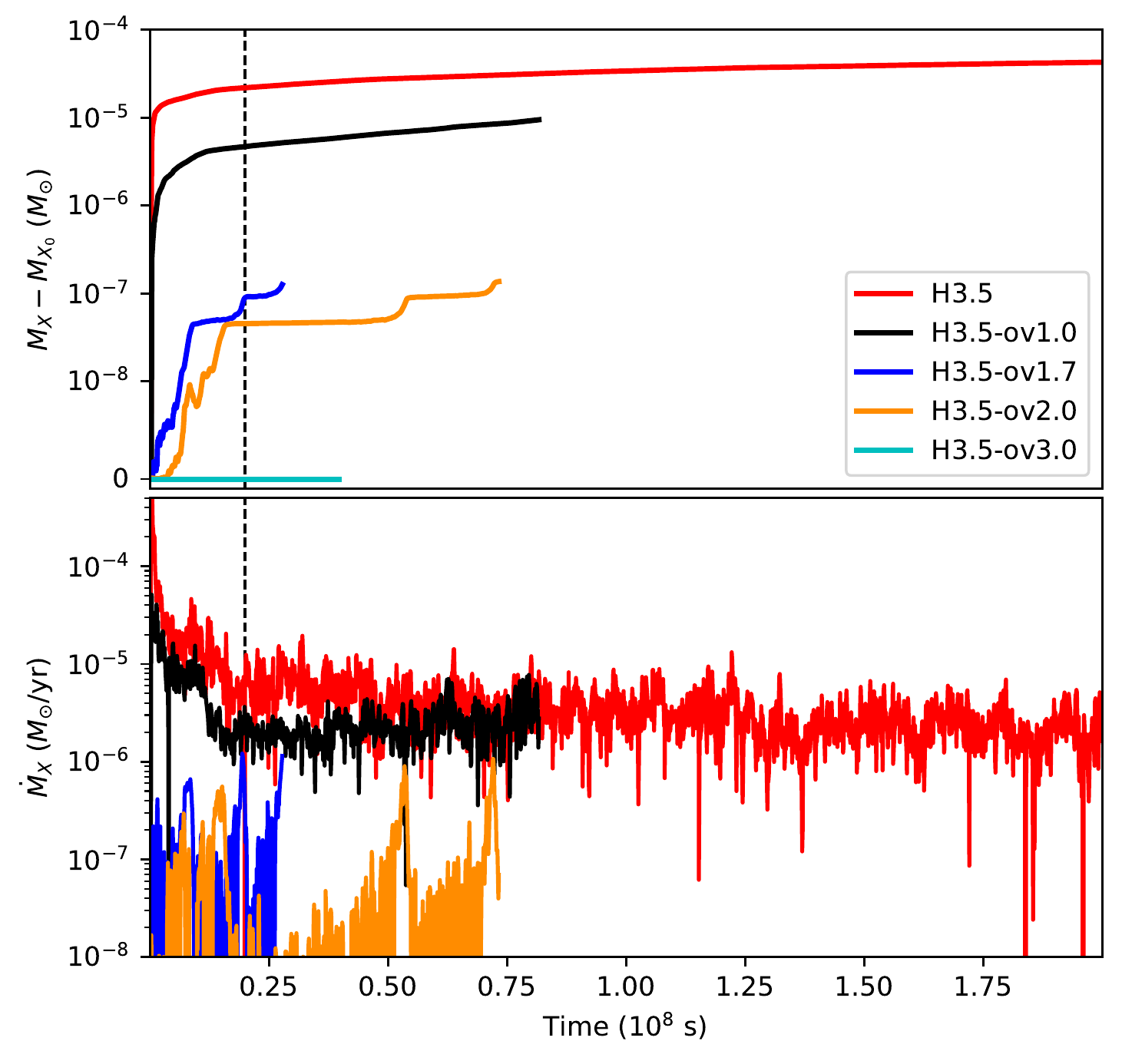}}
\caption{\label{p:mixing_3.5} Same as Fig.,\ref{p:mixing_details} but
  for simulations performed with different $f_{ov}$ values. The
  vertical dashed line indicates $t = 2\,10^7\,\mathrm{s}$ }
\end{figure}

Fig.\,\ref{p:mixing_3.5} shows the long-term evolution of mixing in
model H3.5. Averaging the mixing rate for $t>10^7\,\mathrm{s}$ we find
$\dot{M}_X = 4.0\,10^{-6}\,M\solar/\mathrm{yr}$. Over its main-sequence 
lifetime this star's overshooting region would therefore
consume the unrealistic amount of $> 400\,M\solar$. However, as the
bottom panel of Fig.\,\ref{p:mixing_3.5} shows, $\dot{M}_X$ is
slightly declining with time, and the outward motion of
$R_\mathrm{chem}$ is also slowing down with time (see
Fig.\,\ref{p:boundary_evo}). Hence, over the evolutionary timescale of
stellar models we would expect that this trend eventually leads to an
end of mixing, at which point the final location of $R_\mathrm{chem}$
would be reached.  However, as the thermal evolutionary timescale of a
$3.5\,M\solar$ star on the ZAMS is of the order of
$10^5\,\mathrm{yr}$, and as we are only able to simulate
$\approx 6\,\mathrm{yr}$ of steady convection, extrapolating our
results over five orders of magnitude is prone to large uncertainties.
Therefore, we propose a different method to estimate the maximum
extent of the mixed region by comparing simulations performed with
initial models with the same mass and evolutionary state but computed
with different values of the overshooting parameter $f_{\mathrm{ov}}$.
Increasing $f_{\mathrm{ov}}$ will increase the mass of the initially
homogeneously mixed region $M_{\mathrm{mixed},i}$, and due to our
self-consistent treatment of the evolution of the overshooting models
(see \ref{s:initial_model}) we also get slightly different masses
$M_{\mathrm{CZ},i}$ for the convective core as defined by the
Schwarzschild criterion (see Table\,\ref{t:intermediate}.

\begin{figure}
\resizebox{\hsize}{!}{\includegraphics{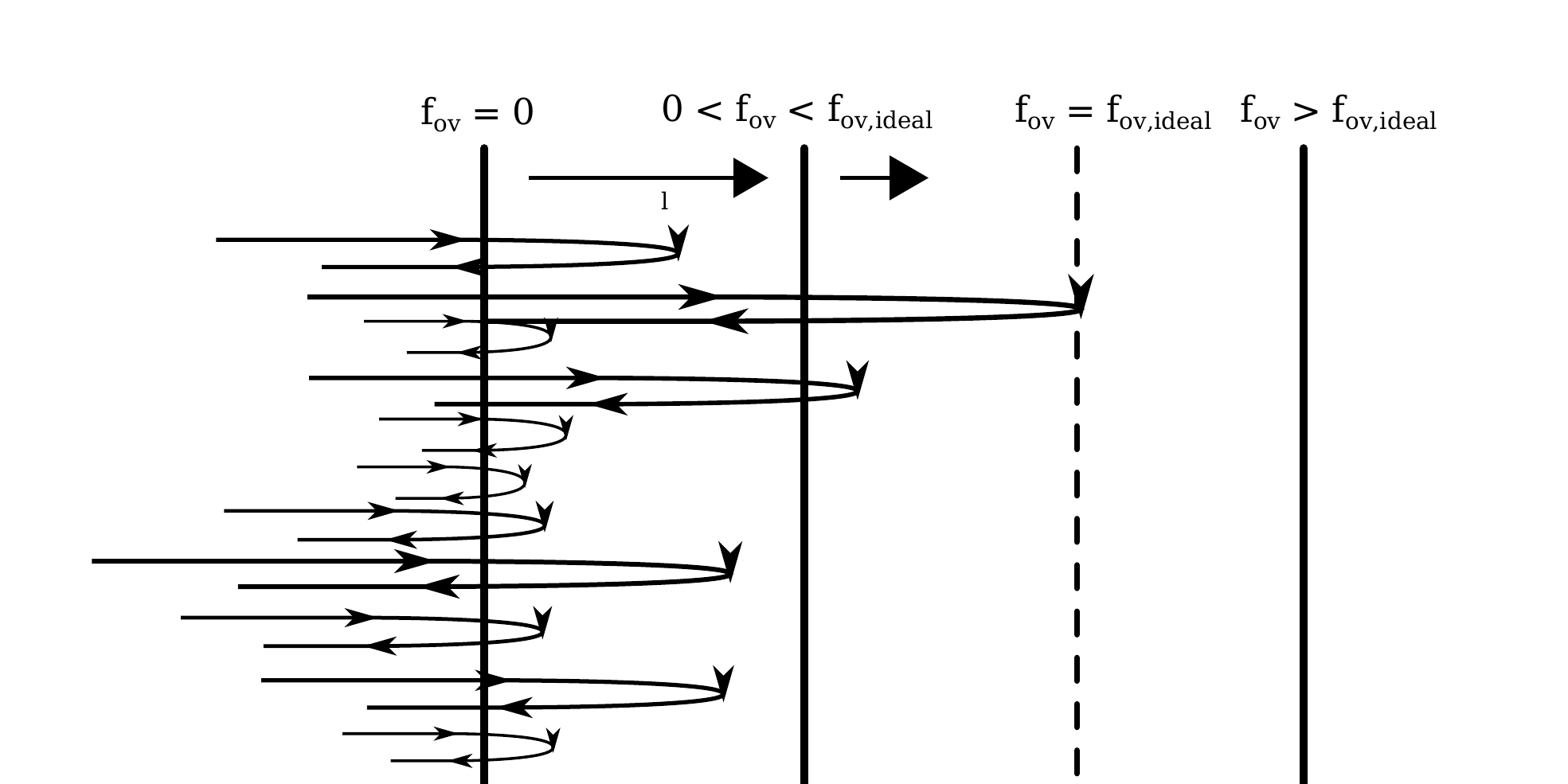}}
\caption{\label{p:calibration_scheme} Illustration of our overshooting
  calibration method (see text).}
\end{figure}

In order to explain the idea behind this method we first need to look
at the process of mixing in more detail.  As we have already discussed
in \ref{s:transient} we find that the convective flow is dominated by
two large vortices. These vortices produce shear mixing at the
convective boundary, which can be best seen by looking at the
horizontal perturbation of the hydrogen mass fraction
$X'(x,y) = X(x,y) - \avgsph{X}(r)$.  We provide snapshots of this
quantity in the right panels of Fig.\,\ref{p:flow}. The prominent
large scale inflow of hydrogen from the left into the CZ visible in
the top right panel of Fig.\,\ref{p:flow} suggests that mixing is
largely localized, and that the rather constant mixing rate one sees
in Fig.\,\ref{p:mixing_3.5} is in reality a combination of single
localized mixing events. We argue that each of these mixing events
will affect a maximum distance from the convective boundary from where
it can collect material with larger $X$.  \citet{pratt17} make a
similar argument based on plumes penetrating the boundary of a
convective envelope, and they show that the plumes with the largest
mixing depth will eventually set the maximum extent of the
overshooting region.

Our calibration of $f_{ov}$ works on the basis of this Extreme Value
approach and is illustrated in Fig.\,\ref{p:calibration_scheme}.  In a
simulation performed with an initial model without any overshooting
$R_\mathrm{chem}$ (vertical lines in Fig.\,\ref{p:calibration_scheme})
and $R_\mathrm{struc}$ will initially overlap, i.e., each single
mixing event across the convective boundary (represented by plumes in
Fig.\,\ref{p:calibration_scheme}) will reach regions with a larger $X$
and will transport some of the excess hydrogen into the CZ. As
$R_\mathrm{chem}$ moves further out into the stable layer due to
mixing, less and less mixing events will be able to reach hydrogen
rich matter.  This effect reduces the mixing as indicated by the
sequence of arrows at the top of Fig.\,\ref{p:calibration_scheme}.

By setting up simulations performed with initial models computed with
$f_{ov} > 0$ we try to find the numerical value
$f_{ov,\mathrm{ideal}}$ that results in a mixed core that exactly
reproduces the extent of the mixed region in a long-term
simulation. For values slightly smaller than $f_{ov,\mathrm{ideal}}$
we would then expect to see a very small increase of $M_X$ over time,
while we would see none for values larger than
$f_{ov,\mathrm{ideal}}$. The initial state of simulations performed
with $f_{ov} > 0$ might not fully reflect the smeared out composition
profile of a mixed model performed with $f_{ov} = 0$ for a long
time. However, we would expect that the composition profile of such a
long time mixed model steepens once the maximum extent of the mixed
region has been reached. We expect that the discrepancy to a 1D model
with $f_{ov} > 0$ eventually decreases.

We compared the average mixing rate of three simulations setting
$f_{ov}$ equal to $0.01$, $0.02$, and $0.03$, respectively. The
respective simulation names are appended by "-ov" followed by the
value of $f_{ov}$ times 100 (see Table\,\ref{t:intermediate}).  In
addition, we performed a fourth simulation with $f_{ov} = 0.017$
(model H3.5-ov1.7). This value is based on a calibration of $f_{ov}$
on observations of open clusters performed by \citet{magic10} with the
Garstec code.

After $2\,10^7\,\mathrm{s}$ all simulations have reached a quasi
steady state and the velocity fields look qualitatively the same
regardless of $f_{ov}$ (see left panels of Fig.\,\ref{p:flow}).
Comparing the right panels of Fig.\,\ref{p:flow} from top to bottom,
it is evident that increasing $f_{ov}$ decreases the perturbation
amplitude of the hydrogen mass fraction, which implies a less
efficient mixing in models simulated with a large overshooting
parameter.  

We find that model H3.5 mixes $10^{-5}\,M\solar$ of hydrogen into the
core during the whole simulation. Increasing $f_{ov}$ to $0.01$
decreases the mixing rate by a factor of $\approx 2$, which would
still corresponds to a fully mixed star at the end of the main-sequence.  
However, increasing $f_{ov}$ further leads to a massive
drop in the mixing rate by more than one order of magnitude. The
Kelvin-Helmholtz timescale of a $3.5\,M\solar$ star is
$\approx 600\,\mathrm{kyr}$, i.e., models H3.5-ov1.7 and H3.5-ov2.0
will mix much less than $0.1\,M\solar$ of hydrogen into the core over
a thermal timescale. This should allow the star to thermally adjust
its structure to the new core size, especially when we consider that
the local thermal timescale of the core is only about $150\,\mathrm{kyr}$.
Model H3.5-ov3.0 even
does not mix any hydrogen at all during the simulation time, showing
that $f_{ov} = 0.03$ is clearly too large.

\begin{figure}
\resizebox{\hsize}{!}{\includegraphics{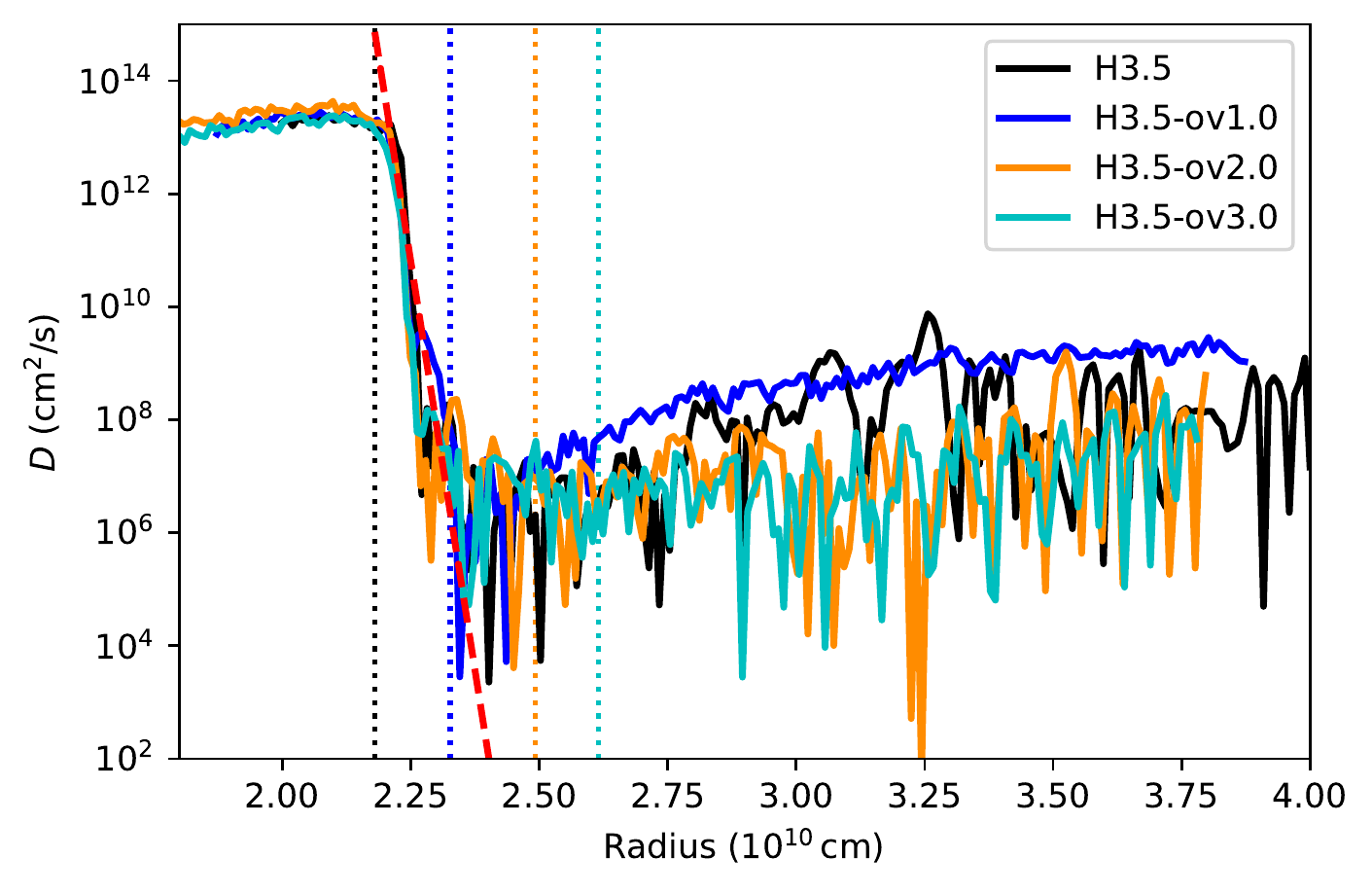}}
\caption{\label{p:diff_ov} Radial diffusion coefficients estimated
  from 40000 tracer particles over $2.5\,10^6 \,\mathrm{s}$ for the
  models with different initial values of $f_{ov}$. The lines have
  been shifted such that the radial position of the structural
  boundary matches for all the simulations. The dotted vertical lines
  indicate the position of $R_{\mathrm{chem}}$ in each run. The red
  dashed line shows $D$ according to
  Eq.\,\eqref{eq:overshooting} with $f_{ov} = 0.01$}
\end{figure}

We also find that increasing $f_{ov}$ changes the characteristic of
the time evolution of $\dot{M}_X$. While $\dot{M}_X$ is more or less
constant in models H3.5 and H3.5-ov1.0, it is dominated in models
H3.5-ov1.7 and H3.5-ov2.0 by single peaks followed by a rather
quiescent phase (see bottom panel in Fig.\,\ref{p:mixing_3.5}).  One
could interpret this behaviour as a mixing that is dominated by rare
single mixing events, i.e., only the farthest reaching mixing events
contribute to the enrichment of hydrogen. However, the situation is
more complex than that.  After a careful analysis of the velocity
field of model H3.5-ov2.0 between $4.9$ and $5.7\,10^7\,\mathrm{s}$ we
could not find any mixing event that would bridge the distance of
$0.1\,H_p$ between $R_{\mathrm{dyn}}$ and $R_{\mathrm{chem}}$. We see
at most dynamic mixing to $0.05\,H_p$ beyond $R_{\mathrm{dyn}}$.  The
missing gap can be bridged by a diffusive mixing process with
$D = 10^{11}\,\mathrm{cm}^2/\mathrm{s}$ which slowly restores the
composition profile before the next mixing event reaches the diffusion
front. This value of $D$ is larger than our estimate derived from
model H3.5-ov2.0 (see Fig.\,\ref{p:diff_ov}). However, we also have to
consider that such mixing events are highly localized in angle, i.e.,
only a small angular section of the diffusion front is mixed into the
core. Furthermore, angular diffusion is much larger than the radial
one. Hence, it can easily smooth out the composition profile between
mixing events in angular direction. Therefore, we argue that the
episodic mixing behaviour seen in our simulations can be explained as
an interplay between diffusive mixing and rare convective mixing
events. \citet{noh93} performed a series of laboratory experiments
with turbulence created by an oscillating grid. They found that the
mixing across a convective boundary in water shows a similar episodic
mixing behaviour as described above once the molecular diffusion in
the experiments becomes dominant, thereby supporting our
interpretation.

We argued in Sect.\,\ref{s:diffusive} that diffusive mixing is most
likely overestimated by several orders of magnitude in our models.
Identifying the episodic mixing as a process dominated by diffusion
then allows us to argue that models H3.5-ov1.7 and H3.5-ov2.0 would
not show any mixing in a fully realistic setup.  For our overshooting
calibration method this assumption corresponds to
$0.01 < f_{ov,\mathrm{ideal}} < 0.017$.  We can compare this estimate
to the asteroseismic constraints by \citet{moravveji16}, who
determined a value of $f_{ov} = 0.024$ for the $3.25\,M\solar$ star
KIC 7760680. This value of $f_{ov}$ is larger than the one we predict,
but one should note that the absolute values of $f_{ov}$ obtained with
different codes should not be compared directly \citep[see the
discussion in][]{angelou20}.  A more relevant evaluation of the result
is to compare the mass of the overshooting region $M_{ov}$ in both
cases. For model H3.5-ov1.7 we find $M_{ov} = 0.22\,M\solar$, which is
in perfect agreement with the $M_{ov} = 0.2239\,M\solar$ found by
\citet{moravveji16} in their grid B.  Their favoured stellar model
grid A, however, gives a larger value of $M_{ov} = 0.2642 \,M\solar$.

We note here that while we calibrate $f_{ov}$ according to
Eq.\,\eqref{eq:overshooting} it is not clear whether
Eq.\,\eqref{eq:overshooting} is indeed the best representation of the mixing
processes around a convective core.  To clarify this concern we
repeated the exercise of calculating the radial diffusion coefficients
for our calibration simulations by comparing the mixing beyond
$R_{\mathrm{struc}}$ in different runs.  In Fig.\,\ref{p:diff_ov},
which shows the outcome of this exercise, we have radially shifted
the Schwarzschild boundaries of the models such that they agree with
each other. We find that all models give remarkably similar values of
$D$ in the CZ and around the drop due to the dynamical boundary
$R_{\mathrm{dyn}}$. In Sect.\,\ref{s:diffusive} we have also seen that
the latter part of the profile of $D$ is independent of grid
size. The overshooting model defined by \citet{freytag96} (see 
Eq.\,\eqref{eq:overshooting}) is supposed to represent the run of
$D$. Indeed we can fit the drop at $R_{\mathrm{dyn}}$ perfectly with
this overshooting description and a value of $f_{ov} = 0.01$ (see red
dashed line in Fig.\,\ref{p:diff_ov}), which is smaller than our
estimated value. 
However, in contrast to this work \citet{freytag96} estimated the overshooting
in convective \emph{envelopes}, i.e., the convective plumes in \citet{freytag96} 
penetrate into regions with higher density. This has a huge impact on the behaviour of 
IGW, which are damped in \citet{freytag96} with increasing distance to the convective boundary
and are amplified in this work. The driving factor for the amplitude changes in both
cases is that amplitudes of IGW scale as $\rho^{-1/2}$. 
We can therefore assume that diffusive mixing by IGW is negligible in the simulations of \citet{freytag96} 
and that their overshooting description mainly captures the mixing caused by turbulent motions at the convective boundary.

Several groups have extended Eq. \eqref{eq:overshooting} to also account for diffusive mixing in radiative layers by means
of hydrodynamical simulations and observations. 
\citet{herwig07} fitted the mixing in simulations of helium shell flashes
on the asymptotic giant branch to two connected exponential functions with different slopes.
From their simulations they determined
$f_{ov}$ to be $0.01$ and $0.14$ for the first and second exponential,
respectively, i.e. the sharp drop of $D$ right at the convective
boundary matches perfectly with our results.  However, the second
exponential, which \citet{battino16} interpret as additional diffusive
mixing by IGW, is not present in our models. 
Fitting the asteroseismic observations of KIC 7760680 \citet{moravveji16} 
required that in addition to Eq. \eqref{eq:overshooting} a small constant diffusive mixing is present 
throughout the stable layer. 
Similarly \citet{rogers17} also predict a diffusive mixing that is active in the whole radiative region, but 
their simulations indicate that the diffusivity actually increases with increasing distance to the CZ due to the 
amplification of IGW.

Our results do seem to support that there is a constant diffusion in the stable layer as proposed by \citet{moravveji16}.
However, due to the damping of velocities in the outer regions of our simulations we can only estimate $D$ in a region relatively close to
the convective boundary. At the maximum radius considered for the tracer particle analysis ($= 4\,10^6\,$cm) the density has dropped by a factor of 2 
in respect to the convective boundary. Using the amplitude scaling of linear IGW, this corresponds to an increase of amplitude by $\approx 40\%$. 
A change that is hardly noticeable in our analysis where measured values of $D$ regularly vary by two orders of magnitude and more in the stable layer (see Fig.~\ref{p:diff_ov}).
A further increase of $D$ at larger distances to the convective boundary due to the amplification of IGW like in \citet{rogers17} can therefore not be excluded.  
 
However, in our overshooting calibration method we combine the effects of turbulent mixing and diffusive mixing by IGW into 
a single effective $f_{ov}$. We argue that due to the long evolutionary timescale on the main-sequence
we do not expect that the mass of the overshooting region does depend
on whether we use dedicated models for diffusive mixing and turbulent mixing or a
single step overshooting description with an effective $f_{ov}$ that
covers both mixing regimes. Asteroseismic properties, on the
other hand, are more sensitive to such changes
\citep{pedersen18,michielsen19}. 

\subsection{Temperature Gradients}
\label{s:temp_grad}

Another uncertainty of 1D stellar evolution is the shape of the
temperature stratification in the overshooting region. On the one
hand, the classical step overshooting method predicts a fully
adiabatic stratification. The diffusive overshooting according to
Eq.\,\eqref{eq:overshooting}, on the other hand, relies on radiative
energy transport.  We have seen in Sect.\,\ref{s:conv_boundary} that
$R_{\mathrm{dyn}} > R_{\mathrm{struc}}$ which implies that there will
be some convective motion and therefore also convective energy
transport beyond the Schwarzschild boundary.  \citet{zahn91} calls
this layer where convective energy transport is still relevant the
penetration region.  Such a penetration layer has been found in
hydrodynamic simulations of red giants \citep{viallet13} and the solar
envelope \citep{korre19} as well as in 1D stellar evolutionary
calculations where convection was modelled by one-dimensional averages
of the hydrodynamic equations instead of MLT \citep{li17}.

\begin{figure}
\resizebox{\hsize}{!}{\includegraphics{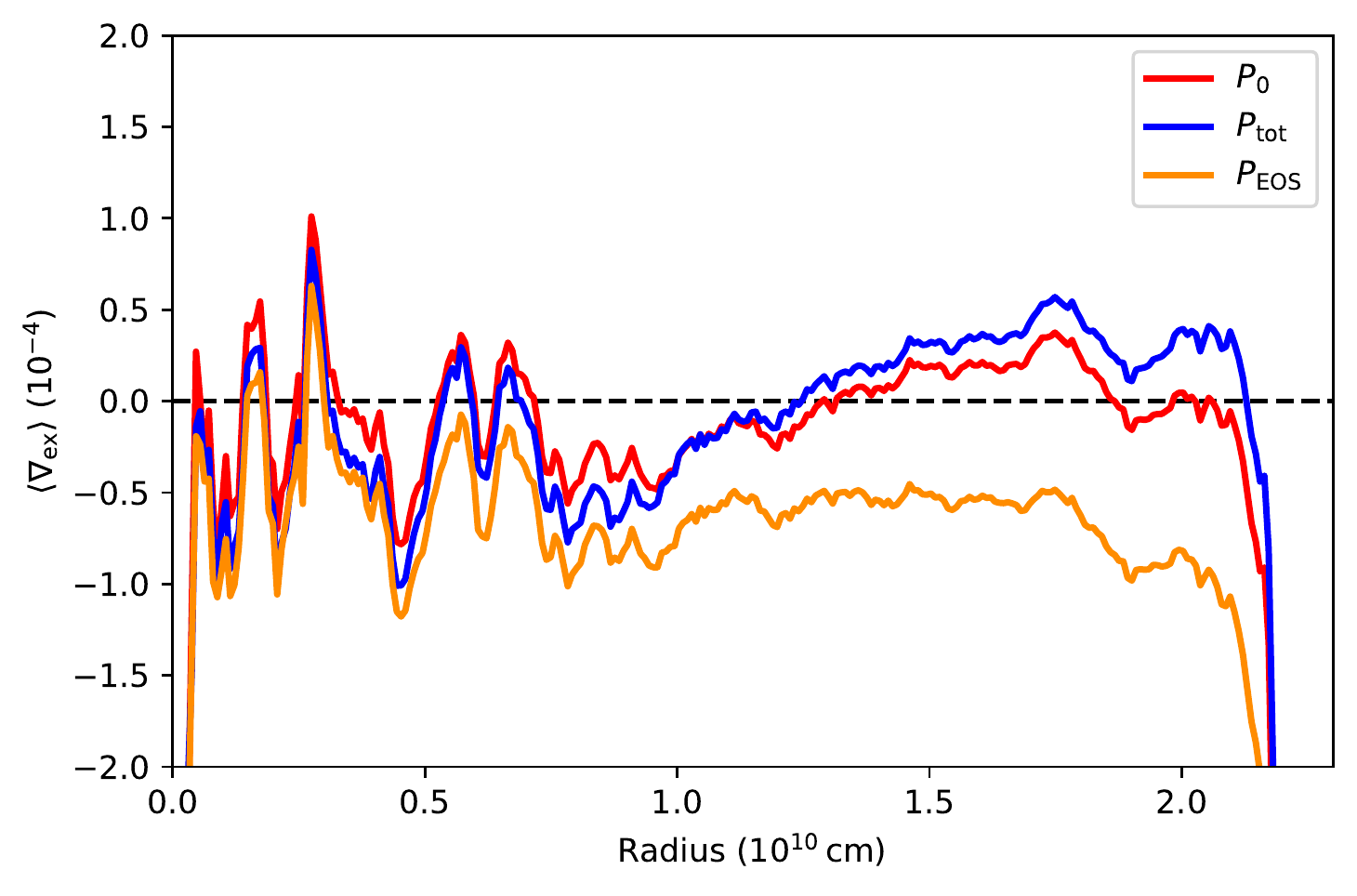}}
\caption{\label{p:nabla_compare}Comparison of the different ways to
  compute the superadiabaticity $\nabla_{\mathrm{ex}}$ in
  \maestro{} simulations (see text for more details). Shown are
  angularly averaged profiles from model H3.5 at $10^{8} \,\mathrm{s}$
}
\end{figure}

In order to see whether we also see penetration in our core convection
simulation we first have to discuss one peculiarity of \maestro{}
simulations.  Due to the fractional step approach in the time
evolution algorithm of \maestro{} the EOS functions $T(\rho,P)$
involving the pressure and $T(\rho,h)$ involving the enthalpy give
temperatures which differ relatively by about $10^{-4}$.

We used the enthalpy version both for the nuclear energy production
and the radiative energy transport.  Given the $T(\rho,h)$ temperature
stratification one has still some freedom in calculating the
temperature gradient $\nabla = \frac{d \log{T}}{d \log{P}}$. One can
compute the derivative either with respect to the total pressure
$P_{\mathrm{tot}} = P_0 + \pi$ or with respect to the
thermodynamically consistent pressure
$P_{\mathrm{EOS}} = P(\rho,T(\rho,h))$. Since the effects of pressure
perturbations $\pi$ are only considered in the momentum equation
(Eq.\,\eqref{eq:maestro_momentum}) one can also decide to compute
$\nabla$ based on $P_0$.  Fig.\,\ref{p:nabla_compare} illustrates the
effect of these different choices for $\nabla$ on the
superadiabaticity
$\nabla_{\mathrm{ex}} = \nabla - \nabla_{\mathrm{ad}}$ in the CZ,
where $\nabla_{\mathrm{ad}}$ is the adiabatic gradient provided by the
EOS. While $\nabla_{\mathrm{ex}}$ computed with $P_{\mathrm{EOS}}$
shows a mostly stably stratified CZ, using $P_0$ or $P_0 + \pi$
increases $\nabla_{\mathrm{ex}}$ by $\approx 10^{-4}$ in the outer
region of the CZ, which then classifies this region as unstable. On
the one hand, the flow field of our simulations clearly does not
correspond to a stable stratification throughout the CZ making the use
of $P_{\mathrm{EOS}}$ questionable. On the other hand,
$P_{\mathrm{EOS}}$ is the thermodynamically consistent way of
computing $\nabla_{\mathrm{ex}}$. We account for this dilemma by using
a rather loose criterion for stability in our analysis, where regions
with $|\nabla_{\mathrm{ex}}| < 10^{-4} $ are considered marginally
stable. In the stable layer, where
$\nabla = \nabla_{\mathrm{rad}} \ll \nabla_{\mathrm{ad}}$, this
modified criterion has very little influence. In the further analysis
we will use $\nabla_{\mathrm{ex}}$ as computed with $P_0$ for
simplicity, which has no effect on the results.

\begin{figure}
\resizebox{\hsize}{!}{\includegraphics{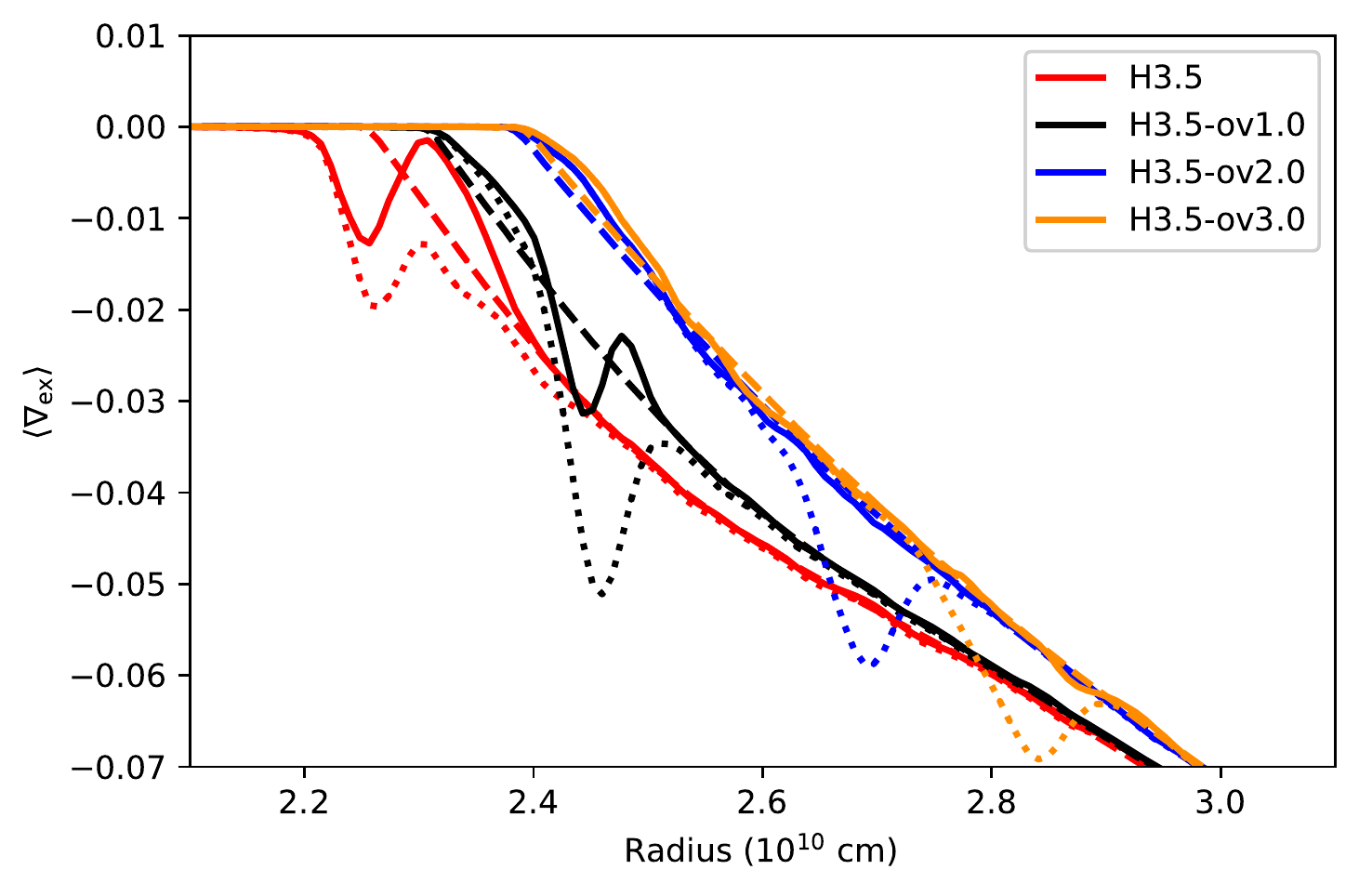}}
\caption{\label{p:nabla_profiles}Angularly averaged profiles
  of the superadiabaticity in models of a $3.5M\solar$ star with
    different amounts of overshooting. Dashed and solid lines show
  profiles according to the Schwarzschild criterion at $t=0$ and after
  $4\,10^7\,\mathrm{s}$, respectively. The dotted lines give the
  initial profiles according to the Ledoux criterion. }
\end{figure}

The presence of a penetration layer can be detected by analysing the
temporal evolution of the superadiabaticity $\nabla_{\mathrm{ex}}$. In
Fig.\,\ref{p:nabla_profiles} we compare $\nabla_{\mathrm{ex}}$ after
$4\,10^7\,\mathrm{s}$ with the initial temperature stratification.  In
model H3.5 the Schwarzschild boundary moves inward during the
evolution, but at the same time $\nabla_{\mathrm{ex}}$ increases just
outside of the initial CZ. This causes a bump in the superadiabaticity
outside of the CZ and brings the value of $\nabla_{\mathrm{ex}}$ very
close to that of the adiabatic gradient, which interferes with our
loose stability criterion and therefore with the determination of
$R_\mathrm{struc}$.  When mixing progresses, the bump moves outwards
where $\nabla_{\mathrm{ex}}$ decreases due to the increasing influence
of $\nabla_{\mathrm{rad}}$.  The wave like feature exhibited by
$\nabla_{\mathrm{ex}}$ (recognizable in Fig.\,\ref{p:nabla_profiles})
is due to the simultaneous mixing of both composition and thermal
energy in simulation H3.5, which increases $\nabla_{\mathrm{ex}}$ due
to the stabilizing effect of the molecular gradient
$\nabla_{\mu} = \frac{d \log{\mu}}{d \log{P}}$. Therefore, this
feature resembles the curve derived from the Ledoux criterion
$\nabla_{\mathrm{ex}}^* = \nabla - \nabla_{\mathrm{ad}} +
\frac{\chi_{\mu}}{\chi_t} \nabla_{\mu}$,
where $\chi_{\mu} = \frac{d \log{P}}{d \log{\mu}}$, and
$\chi_{t} = \frac{d \log{P}}{d \log{T}}$.

The same phenomenon can also be seen in model H3.5-ov1.0 but less
prominent due to the larger initial distance between
$R_\mathrm{struc}$ and $R_\mathrm{chem}$. At even larger distances
between these radii the wave like feature disappears completely as
seen in models H3.5-ov2.0 and H3.5-ov3.0
(Fig.\,\ref{p:nabla_profiles}). In these simulations we instead see
that the Schwarzschild boundary moves outwards at later
times. Adjustments to the temperature stratification on timescales
orders of magnitude shorter than the thermal timescale can be
explained when we consider the effect of penetration as a balance
between the competing processes of convective and radiative energy
transport \citep{ballegooijen82}.  Convection tries to establish an
adiabatic temperature gradient and will do so on a convective
timescale. The counteracting radiative energy transport operates on
much longer timescales so that our simulations establish a temporary
penetration layer. Since we cannot simulate long enough to establish
an equilibrium between those two processes, it is not possible to give
an estimate of the actual size of the penetration layer. However, it
is very likely that the final temperature stratification does not
correspond to the one from the 1D model and instead will show some
effects of penetration in the overshooting region.

\section{Mass Dependence}
\label{s:mass}

\begin{table*}
\center
\begin{tabular}{l|ccccccc}
Name  & $f_{ov}$ & $t_{\mathrm{max}}$& $t_{\mathrm{max}}/\tau_{\mathrm{conv}}$ & $M_{\mathrm{CZ},i}$ & $M_{\mathrm{mixed},i}$ & Mixing Rate & $Ri_{b}$\\
\hline
\hline
H2.0        & 0 & $6\cdot 10^7 $ & 90 & $0.27$ & $0.27$ & $2.0 \cdot 10^{-5}$ & $662$ \\
H2.0-ov0.5  & $0.005$ & $5\cdot 10^7 $ & 80 & $0.29$ & $0.32$ & $1.2 \cdot 10^{-5}$ & $1970$ \\
H2.0-ov1.0  & $0.01$ & $7\cdot 10^7 $ & 80 & $0.30$ & $0.37$ & $2.9 \cdot 10^{-8}$ & $5936$ \\ 
\hline
H1.5        & 0 & $1.8\cdot 10^8 $ & 180 & $0.09$ & $0.09$ & $3.0 \cdot 10^{-6}$ & $1325$ \\
H1.5-ov0.25 & 0.0025 & $1\cdot 10^8 $ & 130 & $0.10$ & $0.12$ & $1.9 \cdot 10^{-6}$ & $7407$ \\
H1.5-ov0.5  & 0.005 & $1.4\cdot 10^8 $ & 130 & $0.12$ & $0.14$ & $3.0 \cdot 10^{-7}$ & $6421$ \\ 
H1.5-ov1.0  & 0.01 & $1\cdot 10^8 $ & 110 & $0.14$ & $0.19$ & $1.4 \cdot 10^{-8}$ & $38470$ \\ 
H1.5-ov2.0  & 0.02 & $1.8\cdot 10^8 $ & 143 & $0.16$ & $0.27$ & $1.7 \cdot 10^{-8}$ & $55122$\\ 
\hline
H1.3        & 0 & $1.3\cdot 10^8 $ & 30 & $0.02$ & $0.02$ & $3.8 \cdot 10^{-7}$ & $1015$ \\
H1.3-ov0.25 & 0.0025 & $1.1\cdot 10^8 $ & 130 & $0.04$ & $0.05$ & $3.9 \cdot 10^{-7}$ & $2141$ \\
H1.3-ov0.5  & 0.005 & $1.5\cdot 10^8 $ & 150 & $0.06$ & $0.08$ & $6.0 \cdot 10^{-8}$ & $3775$ \\ 
\hline
\end{tabular}
\caption{
  \label{t:lowmass} Overview of our 2D simulations of main-sequence stars
  of different masses. The first column gives the name of the model
  and follows the naming scheme of Table\,\ref{t:intermediate}, the
  number behind the "-ov" name appendix indicating the value of 
  $f_{ov}$ used in the 1D model, which is given in the second column 
  times $100$. Columns 3 and 4 show the physical time at the 
  end of the simulation in seconds and the number of convective
  turnovers, respectively. Columns 5 and 6 give the initial
  mass of the CZ and of the homogeneously mixed region in 
  units of $M\solar$, respectively. The mixing rates averaged for
  $t>10^7\,\mathrm{s}$ in units of $M\solar/\mathrm{yr}$ are shown 
  in the second but last column, while the bulk Richardson number is 
  given in the last column.}
\end{table*}

There is currently an ongoing discussion whether the overshooting
parameter is depending on stellar mass or not.  The discussion is
mainly focusing on main-sequence stars, where core size correlates
with mass in the mass range $\approx 1.2\,M\solar$ to
$\approx 2.0\,M\solar$. \citet{claret16} proposed a linear growth of
the overshooting parameter with mass after calibrating stellar models
to fit observations of eclipsing binaries. Similarly
\citet{pietrinferni04} used a linear and \citet{vandenberg06} a
tanh-like scaling of the overshooting parameters in their stellar
evolution databases, which improves the fit of isochrones to open
clusters with turn-off masses in the range of $1.25$ to
$1.8\,M\solar$.  Other groups challenge these results based on
asteroseismic observations \citep{deheuvels16} or due to the lack of
statistical significance \citep[e.g.][]{constantino18,stancliffe15}.
With our calibration method we can contribute to this discussion with
multi-D numerical simulations. Therefore, we computed three more sets
of simulations with stellar masses of 1.3, 1.5, and $2.0 M\solar$ (see
Table\,\ref{t:lowmass}). As in the $3.5\,M\solar$ case we used ZAMS
models with a solar metallicity.

\subsection{$2.0 \,M\solar$}
\label{s:2.0}

\begin{figure}
\resizebox{\hsize}{!}{\includegraphics{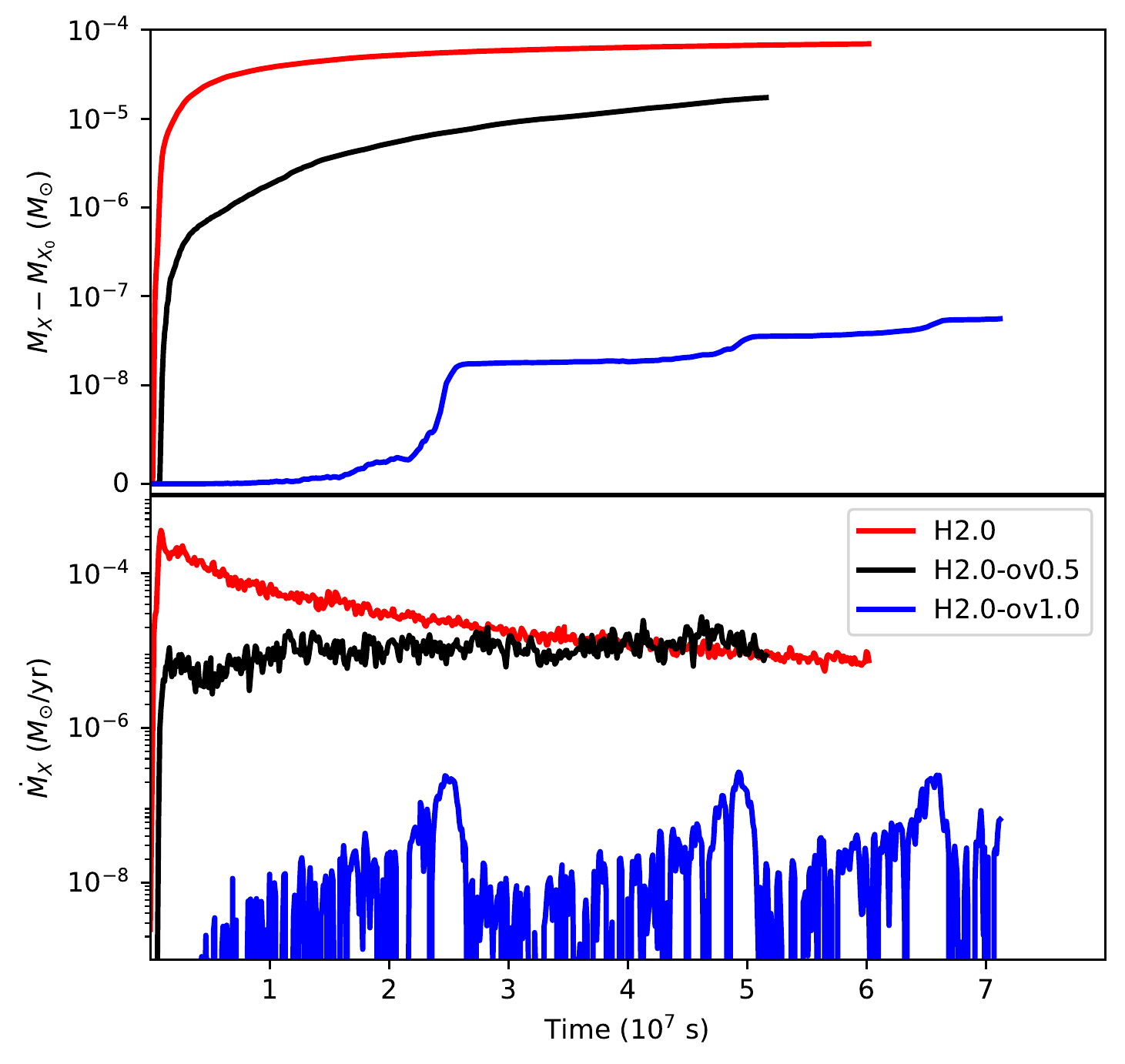}}
\caption{\label{p:mixing_2.0} Same as Fig. \ref{p:mixing_details} but
  for a $2.0\,M\solar$ star. }
\end{figure}

In our three $2.0\,M\solar$ models we used slightly smaller values of
$f_{ov}$ than in the $3.5 \,M\solar$ model, because model
  H2.0-ov1.0 with $f_{ov} = 0.01$ already showed the episodic mixing
behaviour seen in models H3.5-ov1.7 and H3.5-ov2.0 (see
Fig. \ref{p:mixing_2.0}). Following the same argumentation as before
swe find that the diffusive mixing, which is overestimated in our
simulations, implies that $f_{ov} < 0.01$. In order to establish a
lower bound to this estimate we then also computed model
H2.0-ov0.5 with $f_{ov} = 0.005$ that showed a similar
continuous mixing as model H2.0 with $f_{ov} = 0$, but at a
slightly smaller mixing rate.  Our estimate for this set of
$2.0\,M\solar$ models is therefore $0.005 < f_{ov} < 0.01$.

We can also compare this result with models of the eclipsing binary
TZ\,For, where \citet{higl18} found that this relatively close system
could not have undergone a mass transfer in its previous evolution,
and that the mass of the helium core of the $2.05\,M\solar$ primary
must have been at least $0.335\,M\solar$ (see their Table 2). This
fits perfectly to our overshooting estimate where models H2.0-ov0.5
and H2.0-ov1.0 predict He-core masses of $0.32$ and $0.37 M\solar$,
respectively, assuming that the fully mixed core of the initial model
corresponds to the He-core mass at the end of the main-sequence.

\subsection{$1.5 M\solar$}
\label{s:1.5}

\begin{figure}
\resizebox{\hsize}{!}{\includegraphics{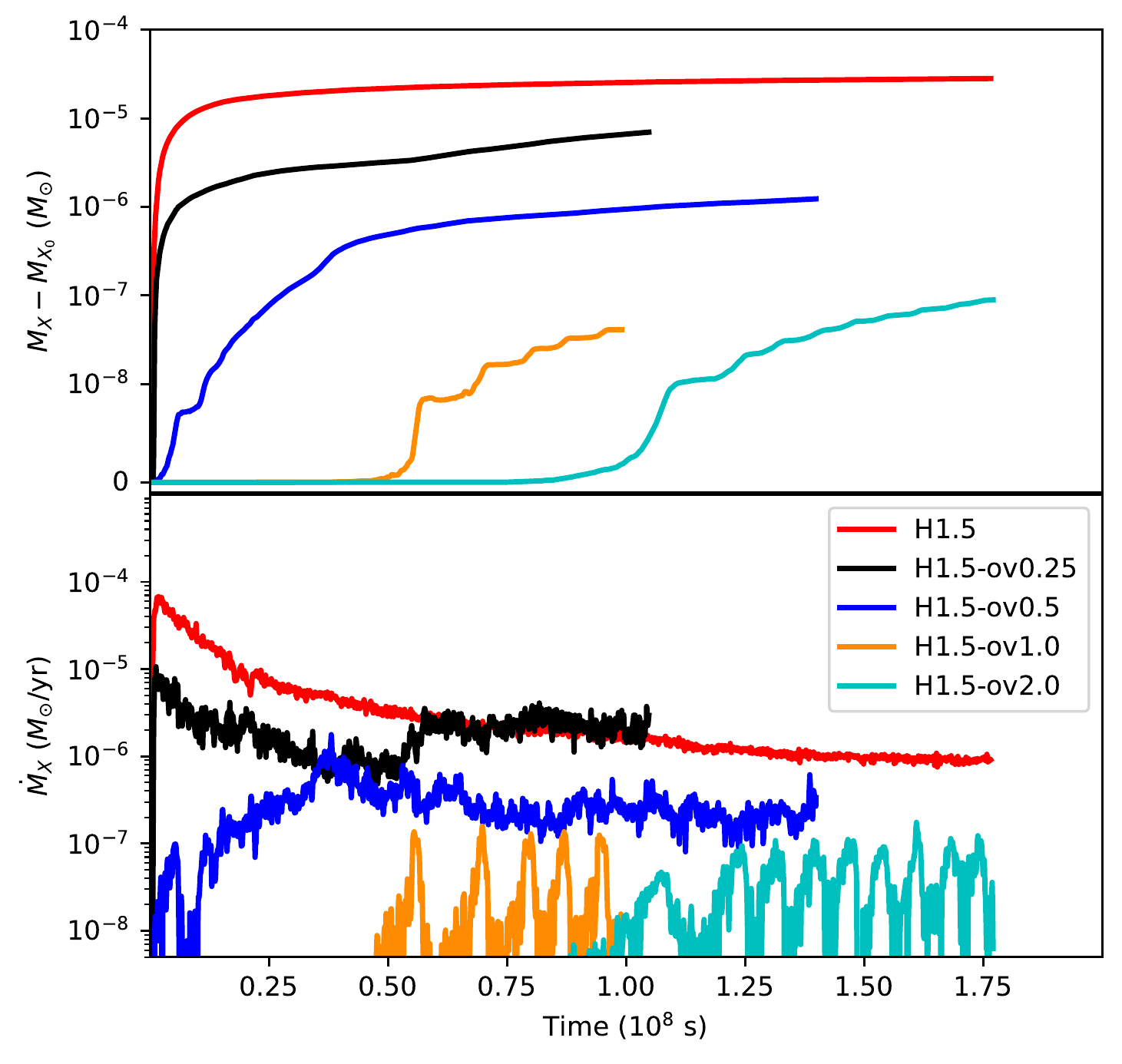}}
\caption{\label{p:mixing_1.5} Same as Fig. \ref{p:mixing_details} but
  for a $1.5\,M\solar$ star. }
\end{figure}

The pressure scale height at the boundary of the convective core of a
$1.5\,M\solar$ star is $60\%$ larger than the radial extent of the CZ
itself.  Using Eq.\,\eqref{eq:overshooting} with a moderate value of
$f_{ov} = 0.02$, increases the mass of its mixed core by a factor of
three compared to that of a no overshooting case. Therefore, we
considered in our simulations smaller values of $f_{ov}$, which
provide more reasonable core masses for a $1.5\,M\solar$ star.

Fig.\,\ref{p:mixing_1.5} shows that for this stellar mass the episodic
mixing appears first in model H1.5-ov1.0 with $f_{ov} = 0.01$, hence
providing the upper limit for our calibration.  Model H1.5-0v0.5 with
$f_{ov} = 0.005$ shows a continuous mixing behaviour resulting in an
estimate of $0.005 < f_{ov} < 0.01$.  While this is exactly the same
range of values as in the $2.0 \,M\solar$ star, we should note that
for the $1.5\,M\solar$ star $f_{ov} = 0.005$ gives a mixing rate that
is only one order of magnitude larger than in the $f_{ov} = 0.01$ run,
indicating that the ideal value of $f_{ov}$ is much closer to $0.005$
than to $0.01$ considering that the same comparison in the
$2.0\,M\solar$ star results in a difference of three orders of
magnitude in $\dot{M}_X$. Asteroseismic observations by
\citet{yang16}, on the other hand, suggest that the core of the
$\approx 1.4\,M\solar$ Kepler star KIC 9812850 has a radius 
of $0.140 \pm 0.028 \,R\solar$, which is in good agreement with 
the initial model of H1.5-ov1.0.

This $1.5\,M\solar$ models also allowed us to estimate a diffusion
coefficient without the need of tracer particles. Models H1.5-ov1.0
and H1.5-ov2.0 both show an episodic mixing behaviour but the onset in
model H1.5-ov2.0 is delayed by $\approx 3\,10^7\,\mathrm{s}$.
Considering that the distance between $R_\mathrm{struc}$ and
$R_\mathrm{chem}$ in model H1.5-ov2.0 is $1.1\,10^9 \,\mathrm{cm}$
larger than in model H1.5-ov1.0 one can estimate that
$D \approx 4\,10^{10}\,\mathrm{cm}^s/\mathrm{s}$, which is similar to
the value found for the $3.5\,M\solar$ star even though the convective
velocities in those models are twice as large as in the
$1.5 \,M\solar$ models. This again is suggesting that diffusive mixing
is numerically overestimated in our simulations.

\subsection{$1.3 \,M\solar$}
\label{s:1.3}

\begin{figure}
\resizebox{\hsize}{!}{\includegraphics{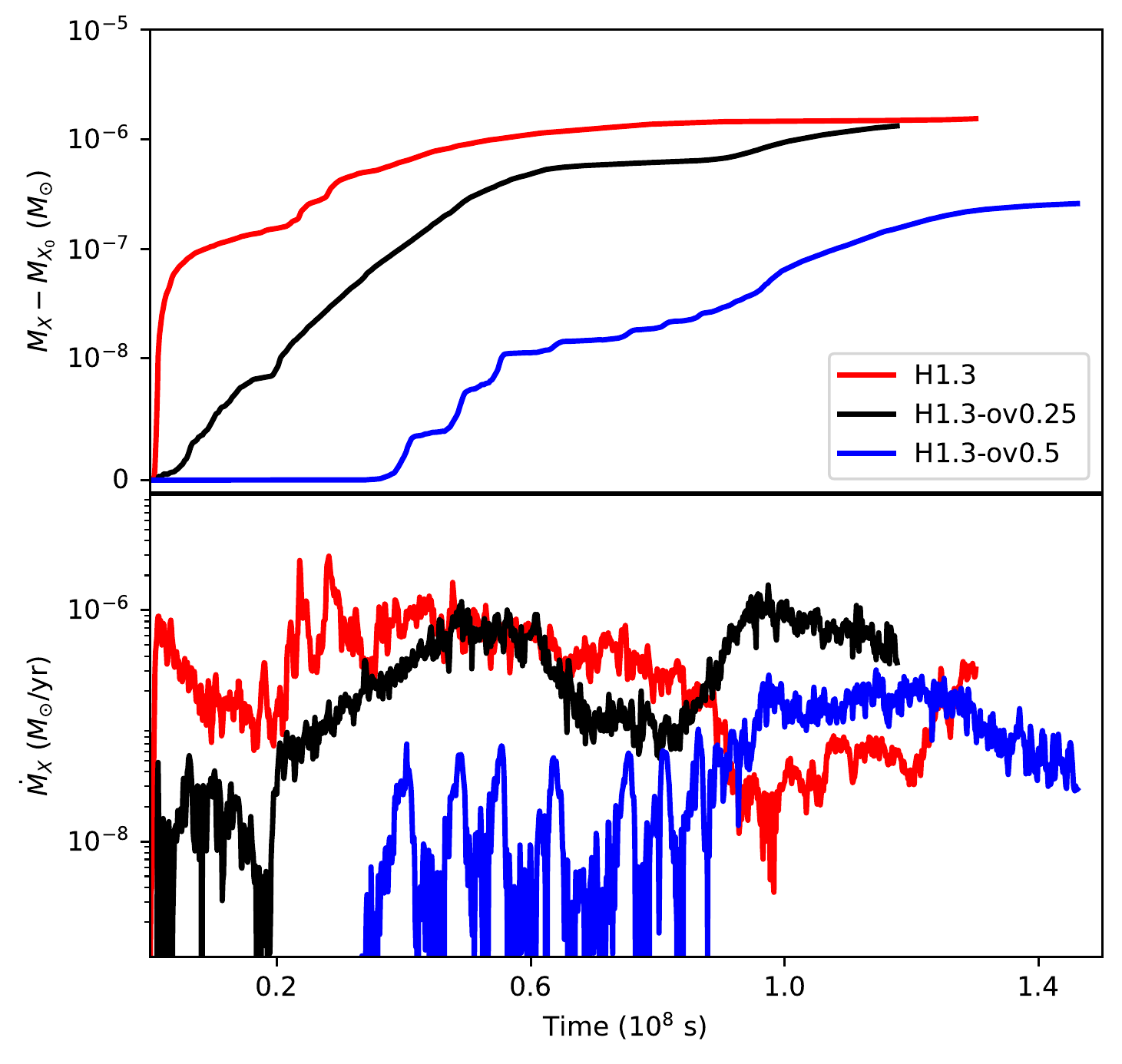}}
\caption{\label{p:mixing_1.3} Same as Fig. \ref{p:mixing_details} but
  for a $1.3\,M\solar$ star. }
\end{figure}

\begin{figure}
\resizebox{\hsize}{!}{\includegraphics{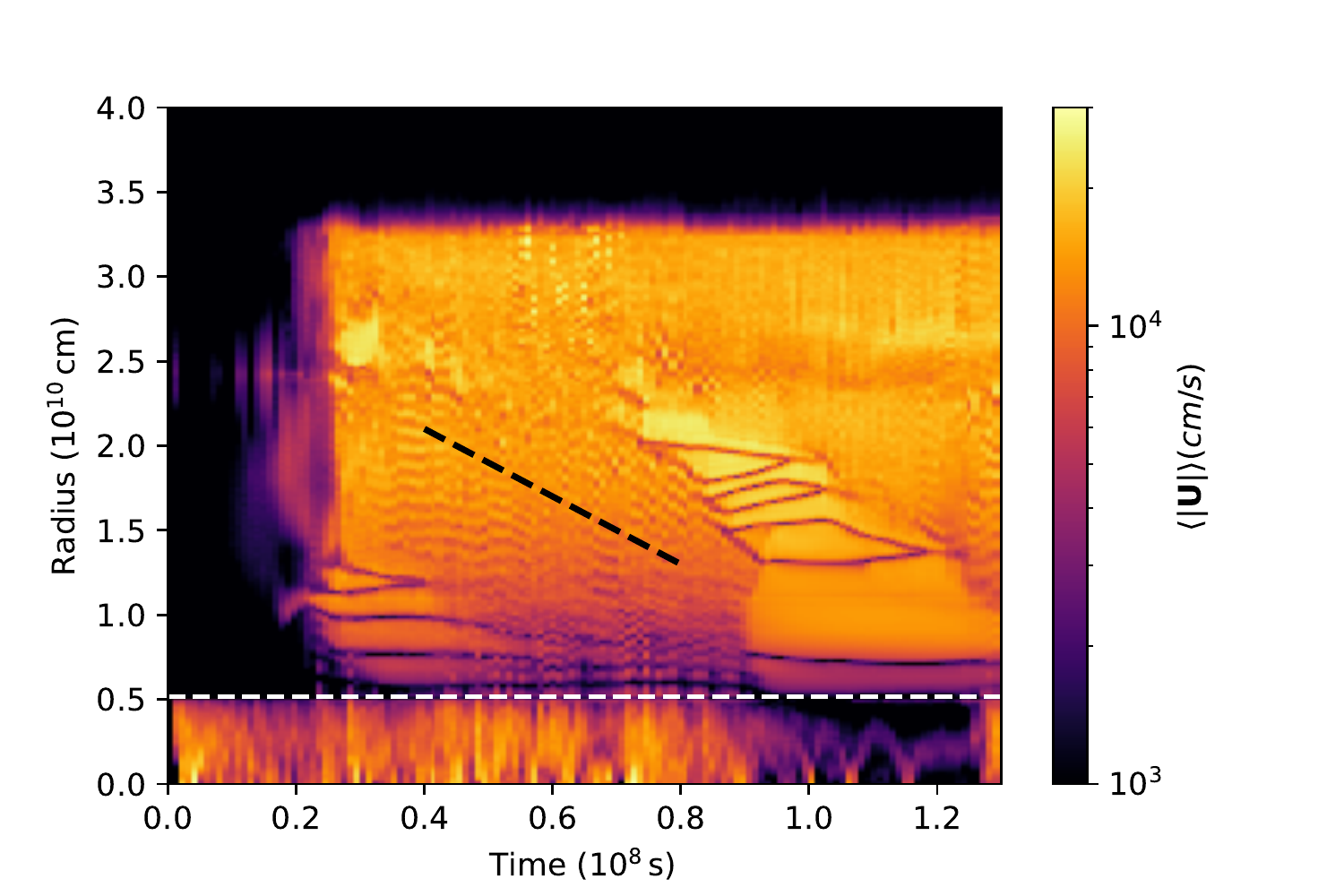}}
\caption{\label{p:1.3_velocity_evolution} Colour plot of the evolution of
  angularly averaged velocity magnitude profiles
  in model H1.3 during a timespan of $1.3\,10^8\,\mathrm{s}$. The
  radial position of the initial Schwarzschild boundary and the
  propagation direction of IGWs are marked by white and black dashed
  lines, respectively.}
\end{figure}

Since the pressure scale height diverges towards the centre of a star,
a $1.3\,M\solar$ star has an even larger reaction to an unrestricted
overshooting according to Eq.\,\eqref{eq:overshooting} than the
$1.5\,M\solar$ models.  In a $1.3\,M\solar$ star the mass of the mixed
core increases by a factor of three even for a tiny overshooting
parameter of $f_{ov}= 0.005$. Therefore, we did not compute any model
with $f_{ov} > 0.005$ for this stellar mass.

As expected model H1.3-ov0.5 does show an episodic mixing behaviour
(see Fig. \ref{p:mixing_1.3}).  However, it also starts to develop a
more continuous mixing at later times. In fact, we see that all models
experience a sudden change in $\dot{M}_X$ after
$\approx 8\,10^7 \,\mathrm{s}$, the reason being unclear. However, by
examining the time evolution of the angularly averaged velocity of
model H1.3 (Fig.\,\ref{p:1.3_velocity_evolution}) it becomes clear
that the effect must originate in the stable layer. After
$8\,10^7\,\mathrm{s}$ the velocity pattern in the stable layer changes
quickly from a wave like pattern to a homogeneous flow. This change
starts deep in the stable layer and then propagates inwards. Once it
reaches the convective boundary (marked by the white dashed line) the
convective velocities become heavily suppressed.  We see the opposite
effect in models H1.3-ov0.25 and H1.3-ov0.5 where larger velocities
lead to larger mixing rates, and in the case of model H1.3-ov0.5 to a
switch from episodic mixing to continuous mixing.

It should be noted that the origin of the strange velocity pattern is
located at around the same radial position as the $N^2$ cavity of the
stratification, suggesting that this effect is related to the
resolution of IGW. This would also explain why there is an opposite
effect in models H1.3-ov0.25 and H1.3-ov0.5 by a reflection of IGW at
the $N^2$ bump due to the composition interface.  Furthermore, the
wave pattern in model H1.3 also suggests that IGW seem to propagate
inwards in this model as indicated by the black dashed line in
Fig.\,\ref{p:1.3_velocity_evolution}.
 
It seems that simulations of a $1.3\,M\solar$ star are beyond the
proper working limit of the RK time integrator, i.e., numerically
generated IGW dominate the evolution.  It is, however, very unlikely
that the numerical effects lead to less mixing than in a properly
resolved simulation.  Hence, we argue that an upper limit
$f_{ov} < 0.005$ still holds, but we are not comfortable to give a
lower limit on $f_{ov}$ for this mass.

\subsection{Overshooting parameter recommendation for 1D models}

\begin{figure}
\resizebox{\hsize}{!}{\includegraphics{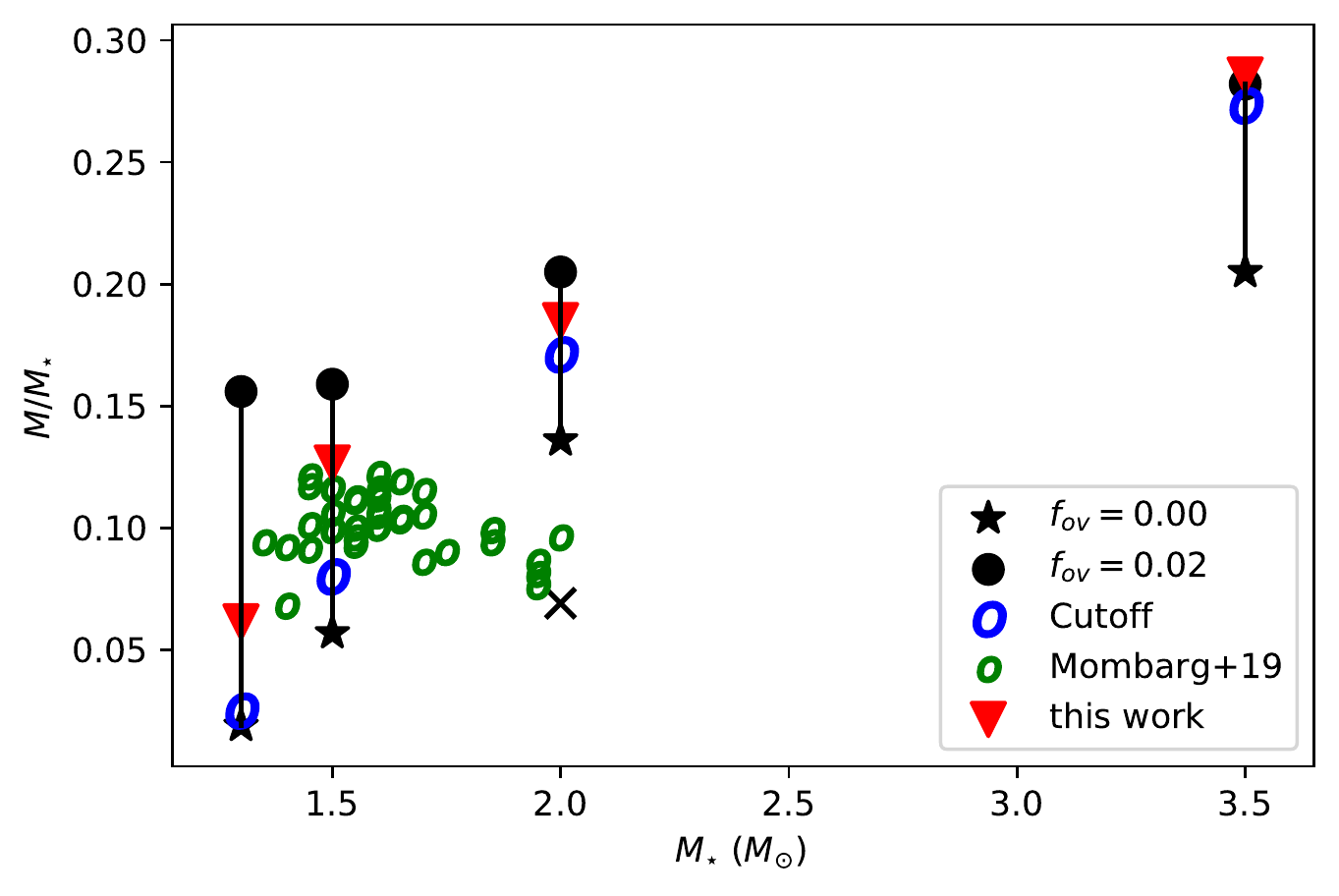}}
\caption{\label{p:masses_convcores} Convectively mixed core mass for
  models with (black dots) and without overshooting (black stars) on
  the ZAMS. Blue and green ellipses show the same quantity, but using
  a geometrical cutoff for the overshooting and observational
  constraints by \citet{mombarg19}, respectively. The red triangles
  denote the derived upper limits from our simulations. The black
  cross indicates the size of the mixed core of a $2.0\,M\solar$ star
  close to the end of its main-sequence evolution when the model is
  computed without overshooting.}
\end{figure}

Comparing our estimates for $f_{ov}$ we see a clear trend towards
smaller values for less massive stars.  In order to ease comparison
with stellar evolution models as well as with observations we provide
also core masses in addition to the value of $f_{ov}$ (see column 6 in
Tables\,\ref{t:intermediate} and \ref{t:lowmass}).  In
Fig.\,\ref{p:masses_convcores} we compare our hydrodynamic estimates
with Garstec stellar evolution models computed with and without
overshooting.  Unsurprisingly all our hydrodynamic estimates predict
core masses that are larger than those of models without overshooting.
In our most massive $3.5\,M\solar$ star we find good agreement between
our estimate and the commonly used value $f_{ov}=0.02$. We note here
that the data points of the 1D models corresponding to $f_{ov}=0.02$
were computed using Eq.\,\eqref{eq:overshooting} without any
modification. Hence, the difference between our hydrodynamic values
and those of the 1D models increases as the pressure scale height at
the convective core boundary increases due to the decreasing core size
of less massive stars.

In models with small CZ Garstec usually applies a geometrical cutoff
to prevent unrealistically large overshooting regions. The cutoff was
introduced in \citet{magic10} and is based on a comparison of the
pressure scale height at the convective boundary to the thickness of
the CZ $r_{cz}$ itself. One introduces a reduced pressure scale height
according to
\begin{equation}
\label{eq:cutoff}
  \tilde{H}_p = H_p \mathrm{max} \left(1,\left(\frac{r_{cz}}{2H_p}\right)^2\right) ,
\end{equation}
and applies the cutoff by replacing $H_p$ in
Eq.\,\eqref{eq:overshooting} with $\tilde{H}_p$. This approach was
motivated in order to reproduce the morphology of the open cluster
colour-magnitude diagram around the main-sequence turn-off of M67.
The resulting core masses for $f_{ov}=0.02$ are given in
Fig.\,\ref{p:masses_convcores} as blue ellipses.  We find that this
cutoff is, in general, much more restrictive than our hydrodynamic
estimates, which is also confirmed by \citet{higl18} who found that it
is too restrictive to model the eclipsing binary TZ\,For.

We also compare our results with recent observations of $\gamma$\,Dor
stars by \citet{mombarg19}. These stars have typical masses of
$1.5\,M\solar$ and oscillate due to the flux blocking mechanism, which
excites gravity modes that allows one to probe the core region. The
extracted core masses of 37 stars are shown in green in
Fig.\,\ref{p:masses_convcores}.  We find a good agreement with our
hydrodynamic estimates for stars around $1.5\,M\solar$ and also
confirm once more that an overshooting cutoff according to
Eq.\,\eqref{eq:cutoff} is too restrictive. A few of the observational
estimates, however, contradict our results and show core masses that
are even smaller than those in models without overshooting. This is
due to a correlation between the central hydrogen content and the
stellar mass in the analysis of \citet{mombarg19} (see their
Fig.\,7). Larger stellar masses correspond to a smaller central
hydrogen content, i.e., stars with $2\,M\solar$ are already close to
the end of their main-sequence evolution, when they have significantly
smaller convective cores than on the ZAMS as indicated by the black
cross in Fig.\,\ref{p:masses_convcores}, which shows the size of the
convective core at a central hydrogen content of $5\%$.

Overall we conclude that there is a general need for a mass-dependent
overshooting description. However, we neither have enough datapoints
to confirm the linear trend proposed by \citet{claret16} nor do we
have any other functional form of $f_{ov}(M_*)$. We also stress that
this result only holds for convective core overshooting on the main-sequence 
and does not apply to convective envelopes or any other
convective layer, as for example the intershell CZ in thermally
pulsing AGB stars \citep{wagstaff20}.

\subsection{Entrainment}

Entrainment predicts that mixing across convective boundaries can be
described as a constant growth, where the growth factor $E$ is defined
as \citep{turner86}
\begin{equation}
  E = A Ri_{b}^{-n} \label{eq:entrainment_law}
\end{equation}
where $A$ and $n$ are free parameters. The bulk Richardson number
$Ri_{b}$ is a measure of the stiffness of the boundary of a CZ,
whereby convection is characterized by a typical lengthscale $L$ and
the rms velocity $U_{\mathrm{rms}}$ of the flow.  $Ri_{b}$ can then be
written as
\begin{equation}
  Ri_{b} = \frac{\Delta b L}{U_{\mathrm{rms}}^2} , \label{eq:rib}
\end{equation} 
where $\Delta b$ describes the buoyancy jump across a boundary of
width $d_i$
\begin{equation}
  \Delta b = \int_{r_i-d_i}^{r_i+d_i} N^2 dr . \label{eq:buoyancy_jump}\\
\end{equation}

\citet{meakin07} found that an entrainment law according to
Eq.\,\eqref{eq:entrainment_law} describes the mixing in an oxygen
burning shell as well as that in a $25\,M\solar$ main-sequence star
when $A=0.027 \pm 0.38$ and $n=1.05 \pm 0.21$. \citet{cristini17} and
\citet{gilet13} find similar results for a carbon burning shell and a
$15\,M\solar$ main-sequence star.  However, using these values
inferred by hydrodynamic simulations in 1D stellar evolution on the
main-sequence leads to an unrealistically large growth rate of
  the core that allows it to engulf the complete stable layer within
a fraction of the main-sequence lifetime.

When we examine the time evolution of $\dot{M}_X$ in Figures
\ref{p:mixing_3.5}, \ref{p:mixing_2.0}, and \ref{p:mixing_1.5} we
notice that none of the simulations with $f_{ov} > 0$ surpasses the
mixing rate of the model without overshooting
significantly. Especially in Fig.,\ref{p:mixing_1.5} it seems like
model H1.5 is defining an upper limit to the mixing rate, since the
sudden increase of $\dot{M}_X$ in model H1.5-ov0.25 at
$5\,10^7\,\mathrm{s}$ stops at the same mixing rate as in model
H1.5. Subsequently, $\dot{M}_X$ evolves in both models similarly. An
upper limit of the mixing rate was proposed by \citet{linden75}, who
argued that entrainment of light material across a buoyancy jump
requires a specific amount of energy. Hence, mixing will be limited by
the available kinetic energy of convection. He further argues that
this implies $n=1$ in Eq.\,\eqref{eq:entrainment_law}.  Similarly,
\citet{jones17} and \citet{andrassy18} showed that the entrainment
rate in a convective O-burning shell is proportional to the luminosity
of the system.

\begin{figure}
\resizebox{\hsize}{!}{\includegraphics{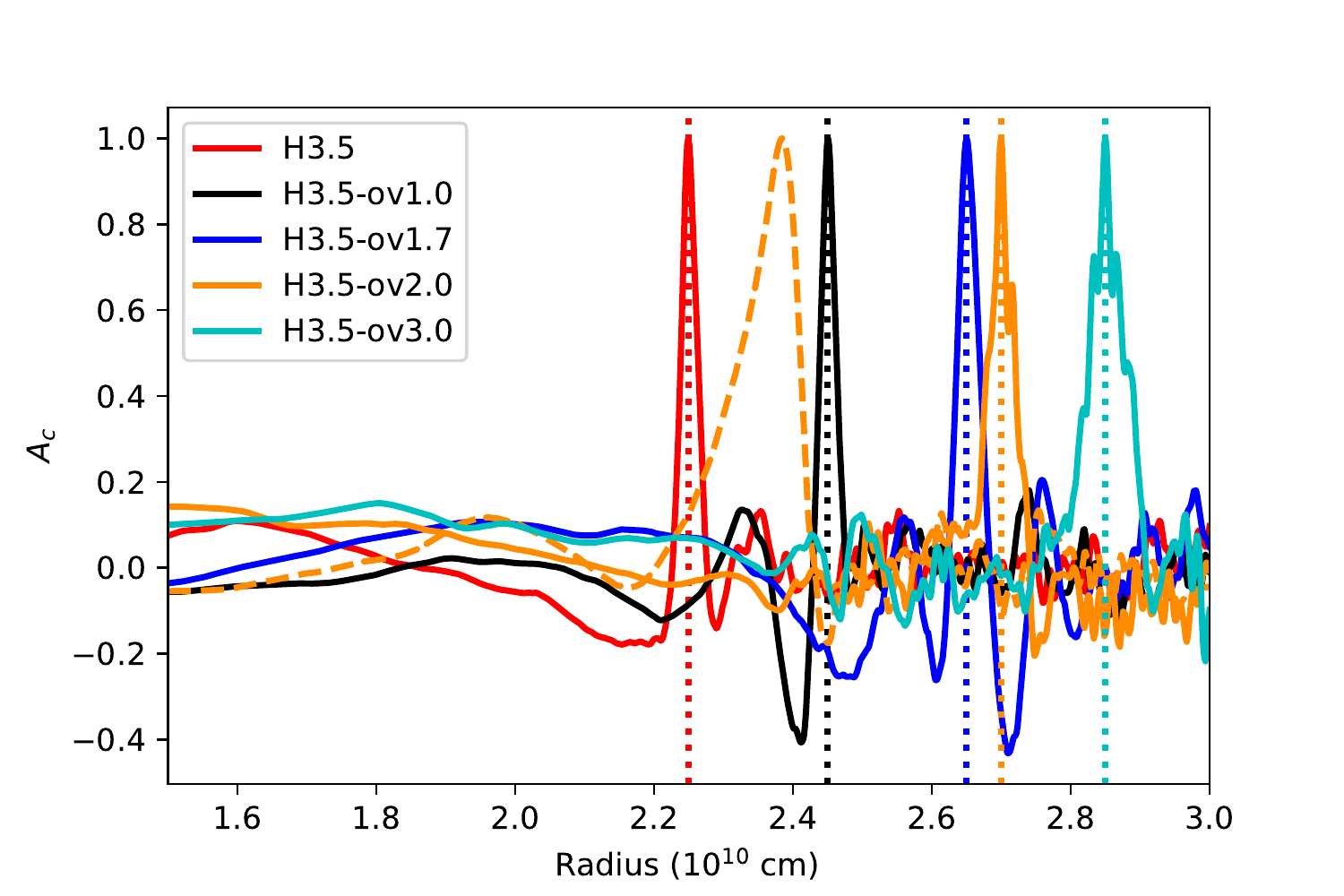}}
\caption{\label{p:correlation} Auto-correlation function (see
  Eq.\,\eqref{eq:auto-correlation}) of the radial velocities at the
  location of the composition interfaces (dotted vertical lines) at
  the end of the simulations for the $3.5\,M\solar$ models. The
  black dashed line shows the auto-correlation function at the
  Schwarzschild boundary of model H3.5-ov2.0.}
\end{figure}
 
In order to obtain estimates of $A$ and $n$ in our simulation we first
need to compute $Ri_{b}$, which requires estimates of $L$ and
$d_i$. We choose $d_i$ as the distance between $R_\mathrm{struc}$ and
$R_\mathrm{chem}$ in the model with the largest $f_{ov}$ value for
each mass. This provides comparability between different
runs of the same stellar mass, and it ensures that the integration
in Eq.\,\eqref{eq:buoyancy_jump} covers the complete interface
even when $R_\mathrm{chem}$ is propagating into the stable layer,
which is not the case for the largest $f_{ov}$ values. Estimating $L$
requires an analysis of the typical size of flow elements around the
convective boundary.  For this purpose we use an auto-correlation
function $A_c(r)$ of the angularly averaged radial velocity
$\avgsph{U_r}$ \citep{mocak09}
\begin{equation}
\label{eq:auto-correlation}
  A_c(r) = \frac{\left< U_{r}(r)U_{r}(r+dr)\right>}{\left<
           U_{r}(r)\right>^{0.5} \left< U_{r}(r+dr)\right>^{0.5}}\,.
\end{equation} 

Fig.\,\ref{p:correlation} shows $A_c(r)$ for the $3.5\,M\solar$ models
evaluated at $r=R_\mathrm{chem}$.  The peaks in each curve correspond
to $R_\mathrm{chem}$ as indicated by the dotted vertical lines. These
narrow peaks suggest that the size of connected flow elements around
$R_\mathrm{chem}$ is small, i.e., mainly small scale flows are
responsible for the mixing across the boundary. Hence, we use the
width of these peaks as our estimate for $L$.  We also show
$A_c(R_\mathrm{struc})$ for model H3.5-ov2.0 (dashed line in
Fig.\,\ref{p:correlation}), which exhibits a much broader peak
corresponding to the large scale flow inside the CZ.
   
\begin{figure}
\resizebox{\hsize}{!}{\includegraphics{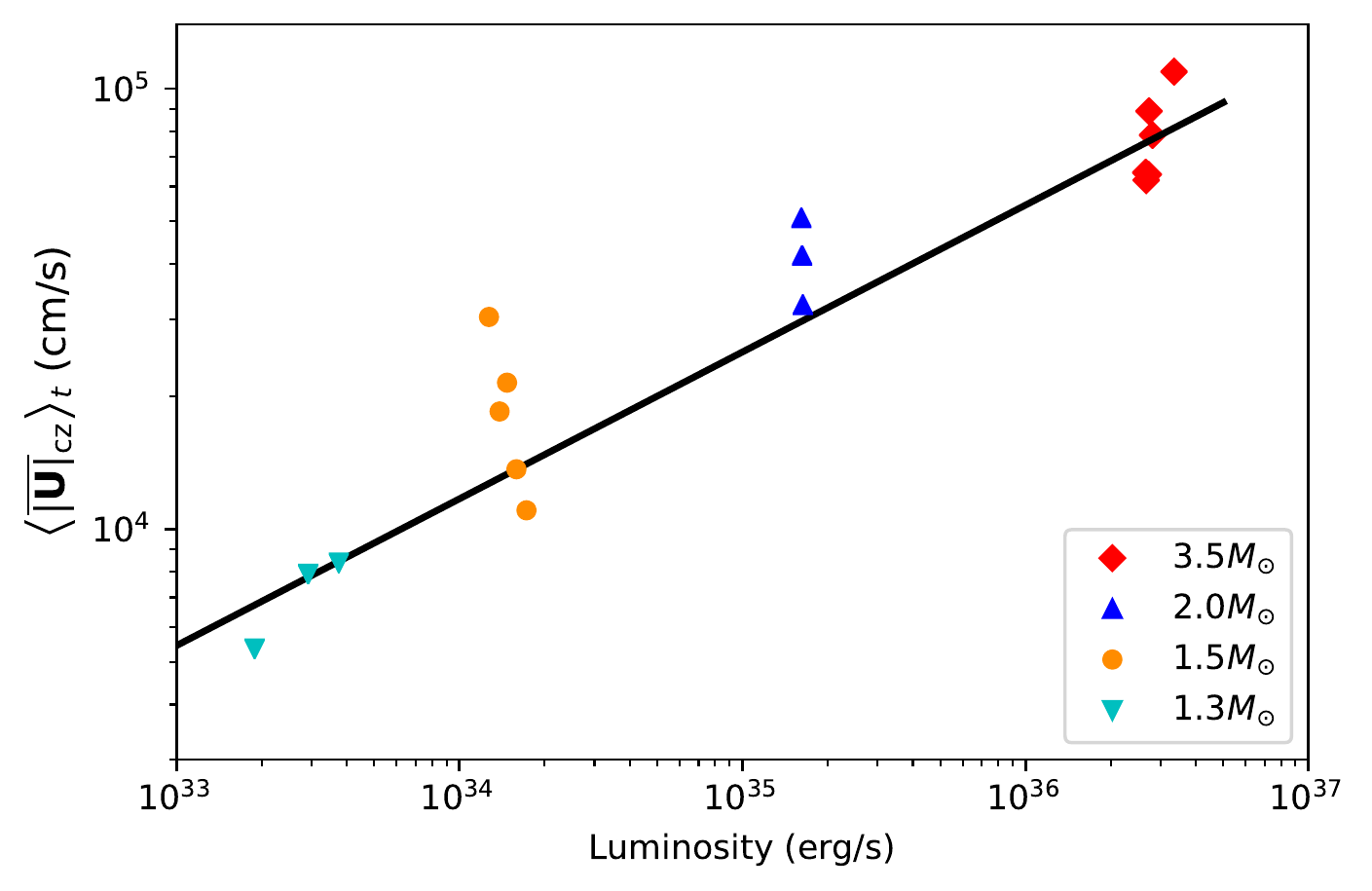}}
\caption{\label{p:mlt_velocity}Time-averaged convective velocities as
  a function of stellar luminosity. The predicted MLT scaling is shown
  by the black line.}
\end{figure}

At this point we can also check whether our rms velocities fulfill the
scaling relation
$\avgtime{\avgvol{|\vec{U}|}_\mathrm{CZ}}^3 \propto F$ predicted by
MLT.  We find that our simulations in the analysed mass range agree
fairly well with this scaling relation (see
Fig.\,\ref{p:mlt_velocity}). However, the spread in
$\avgtime{\avgvol{|\vec{U}|}_\mathrm{CZ}}$ for a given luminosity is
much larger than expected. Models with the same stellar mass have a
similar luminosity, but the mass of the convective core is 
quite different due to the self-consistent 
evolution of the 1D models including overshooting (see column 5 in
Tables\,\ref{t:intermediate} and \ref{t:lowmass}).  This suggests that
second order effects like the mass of a CZ, or respectively its size,
should also be considered in velocity estimates.

\begin{figure}
\resizebox{\hsize}{!}{\includegraphics{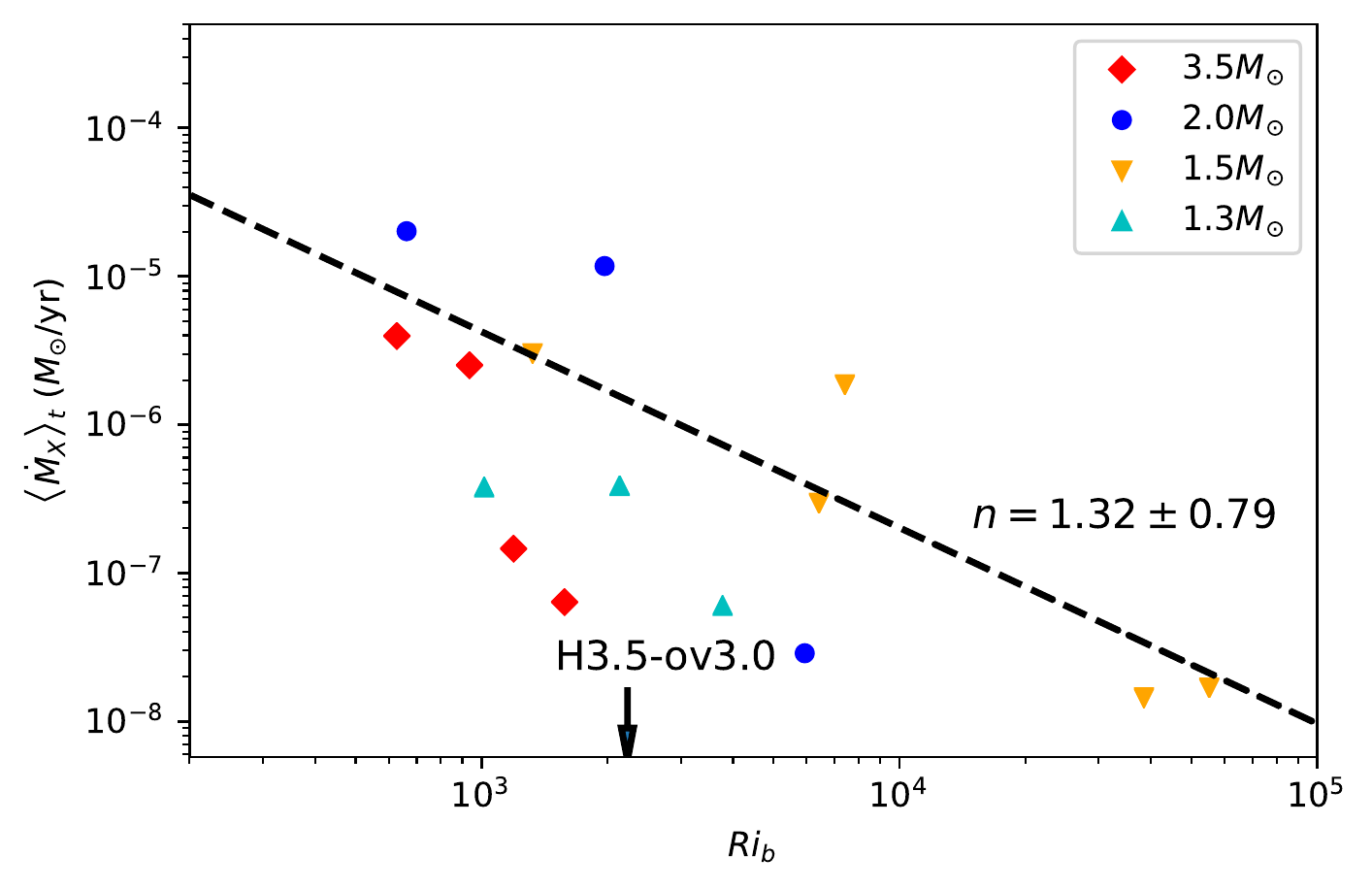}}
\caption{\label{p:rib_fit_all} Time-averaged mass entrainment rate as
  a function of the bulk Richardson number for all our models. Each
  symbol corresponds to a different model. The black dashed line
  represents an entrainment law according to
  Eq.\,\eqref{eq:entrainment_law} with n=1.32.}
\end{figure}

We find that the convective boundaries of our models have bulk
  Richardson numbers in the range $600 < Ri_{b} < 55000$, which is
comparable to the unenhanced luminosity simulation in
\citet{cristini19}. Combining these values with the mixing rates in
Table\,\ref{t:intermediate} and \ref{t:lowmass} we are able to fit an
entrainment law to our simulations.  This fit gives $n=1.32 \pm 0.79$
and $A = 4 \cdot 10^{-2}$. Even though this is in agreement with the
theoretical prediction $n=1$ by \citet{linden75} we would argue that
the large error in the fit actually suggests that (1) our simulations
should not be fitted by a single universal entrainment law, and that
(2) they indicate that the entrainment law does not only depend on
$Ri_{b}$. One possible extension to the entrainment law would be to
include the Peclet number in the analysis as it was proposed by
\citet{noh93}. Obviously, further research is required to analyse the
different influences on the entrainment process.

\section{Conclusion}\label{s:conclusion}

We presented a total of 21 two-dimensional simulations of convective
cores in ZAMS stars ranging from $1.3$ to $3.5\,M\solar$.  The
simulation domain covers the convective core and a large fraction of
the convectively stable layer on top of it.  Due to the pseudo
incompressible approximation of \maestro{} we were able to follow the
convective flow for many convective turnover times at the nominal
stellar luminosity, which allowed us to study the time evolution of
the very low Mach number flows and their mixing across the convective
core boundary in detail. By comparing a simulation that resolves 
high frequency IGW in time with one that only resolves the advection timescale 
we determined that high frequency IGW do not
play a significant role in the mixing and therefore decided to not
fully resolve them in time. This allowed us to increase the numerical
timestep further and hence to simulate more than 100 convective
turnover times in most of our simulations. In order to guarantee the 
accuracy of simulations using large timesteps we replaced the advection scheme in the time
integrator of \maestro{} with a 4th order Runge-Kutta scheme, which reduces 
the amplitude of numerical artifacts in the convectively stable layer. 

We used Garstec 1D stellar evolution models with a solar-like
composition as input for our hydrodynamic simulations. During the
mapping procedure from 1D to 2D we preserved the thermal
stratification of the 1D model following \citet{edelmann17}, which
reduced the influence of the transient phase due to thermal
readjustments at the beginning of the simulations.  We analysed the
time evolution of the convective boundary in detail and identified
three fundamental types of boundary definitions.  We showed that for
main-sequence convection dynamic (defined by the flow velocity) and
structural (defined by the temperature stratification) boundaries
remain mostly constant for the simulated timescales but at different
radial locations. In contrast, the chemical boundary (defined by the
composition) evolves in time, demonstrating the effects of
mixing. While the time evolution of the chemical boundary indicates
that mixing is slowing down with time, we were not able to simulate
long enough to establish an equilibrium state. In order to estimate
the maximum extent of the mixed region around a convective core we
therefore used a series of 1D models computed with increasing
values of the overshooting parameter $f_{ov}$.

We found that increasing $f_{ov}$ in the initial model reduces the
mixing rate of hydrogen into the convective core. Furthermore,
increasing $f_{ov}$ beyond a certain limiting value leads to an
abrupt change in the mixing characteristics, namely from a continuous
entrainment process to an episodic mixing behaviour. We attribute this
change to an increasing influence of diffusive mixing due to IGW and
of numerics as the distance between the composition interface and the
Schwarzschild boundary of the convective core increases.

We measured the diffusion coefficients of the flow with tracer
particles, and we found that the results within the CZ and in the
immediate surrounding of it are well converged in our
simulations. However, this is not the case further away from the
convective boundary, where the estimated diffusion coefficients are
orders of magnitude smaller than in the CZ. On the one hand, 
the clear separation of these two regions hints that the
overshooting description itself has to be altered to account for
different mixing processes. On the other hand, comparing our results with
asteroseismic constraints we argue that the contribution of diffusive
mixing is largely overestimated in our simulations, i.e., the
simulations that are dominated by diffusion would show even smaller
mixing rates in a higher resolved simulation. This allows us to
constrain $f_{ov}$ to values where no episodic mixing can be seen in
our simulations.

With this procedure we determined $f_{ov}$ to be in the range
$0.01 < f_{ov} < 0.017$ in a $3.5\,M\solar$ star, which is in rough
agreement with empirical estimates.  Reducing the simulated stellar
mass and therefore the size of the convective core shows that the
tight connection between $f_{ov}$ and the pressure scale height
requires a reduction of $f_{ov}$ used on the main-sequence towards
lower stellar masses, e.g., we could limit $f_{ov}$ in a $1.3\,M\solar$
star with a tiny convective core to $f_{ov} < 0.005$.  This
result confirms findings when comparing isochrones with open
cluster observations \citep[e.g.][]{pietrinferni04,magic10} and it
agrees with results using eclipsing binaries \citep{claret16}. In
particular, we find that our overshooting estimates are in good
agreement with asteroseismic observations of $\gamma$\,Dor stars by
\citet{mombarg19}.

Practically limiting the overshooting can be achieved by a mass
dependent overshooting parameter or by geometrically limiting the
overshooting region based on the size of the convective region itself,
where the latter is the physically more relevant property. However,
the exact functional form remains to be determined.

Moreover, we find that models with large values of $f_{ov}$ develop a
thin penetration region where the temperature gradient lies between
the radiative and the adiabatic gradient as it was predicted by, e.g.,
\citet{ballegooijen82} and \citet{zahn91}.  In models with small
$f_{ov}$ values this effect is probably masked by changes in the
molecular gradient due to the strong chemical mixing in these
simulations. Nevertheless, it is not possible to estimate the maximum
extent of the penetration region in any of our simulations, since we
are not able to simulate the relevant thermal timescales.

We also investigated the possibility to describe chemical mixing in
the form of an entrainment law as proposed by \citet{meakin07}. We
were not able to find an acceptable universal fit that covers all
stellar masses in this work, indicating that entrainment laws need to
include more parameters as they currently do.

Three-dimensional simulations are needed to confirm these results.
Initial tests in \citet{higl19} point in this direction, but
computational limits regarding the numerical stability of the
convectively stable layer in these simulations currently prevent us
from making conclusive statements. Proposed well-balancing methods
\citep[see e.g.][]{berberich19} could potentially help to provide
these answers in the future.

\begin{acknowledgements} 
The JH would like to thank M. Zingale for fruitful discussion on the numerics of \maestro{}. The E3.5 simulation and most of the diffusion coefficient analysis has been computed on the Hydra and Draco clusters at the Max Planck Computing and Data Facility (MPCDF). JH acknowledges support by the Klaus Tschira Foundation. This work made use of the Matplotlib \citep{hunter07} package and used MPI4Py \citep{dalcin11} for parallel data processing. 
\end{acknowledgements}

-------------------------------------------------------------------

\bibliographystyle{aa}
\bibliography{OVCalibration}

\begin{thebibliography}{90}
\expandafter\ifx\csname natexlab\endcsname\relax\def\natexlab#1{#1}\fi

\bibitem[{{Aerts} {et~al.}(2010){Aerts}, {Christensen-Dalsgaard}, \&
  {Kurtz}}]{Aerts10}
{Aerts}, C., {Christensen-Dalsgaard}, J., \& {Kurtz}, D.~W. 2010,
  {Asteroseismology}

\bibitem[{{Aerts} {et~al.}(2018){Aerts}, {Mathis}, \& {Rogers}}]{Aerts18}
{Aerts}, C., {Mathis}, S., \& {Rogers}, T. 2018, arXiv e-prints,
  arXiv:1809.07779

\bibitem[{{Almgren} {et~al.}(2008){Almgren}, {Bell}, {Nonaka}, \&
  {Zingale}}]{almgren08}
{Almgren}, A.~S., {Bell}, J.~B., {Nonaka}, A., \& {Zingale}, M. 2008, \apj,
  684, 449

\bibitem[{{Almgren} {et~al.}(2006{\natexlab{a}}){Almgren}, {Bell}, {Rendleman},
  \& {Zingale}}]{almgren06a}
{Almgren}, A.~S., {Bell}, J.~B., {Rendleman}, C.~A., \& {Zingale}, M.
  2006{\natexlab{a}}, \apj, 637, 922

\bibitem[{{Almgren} {et~al.}(2006{\natexlab{b}}){Almgren}, {Bell}, {Rendleman},
  \& {Zingale}}]{almgren06b}
{Almgren}, A.~S., {Bell}, J.~B., {Rendleman}, C.~A., \& {Zingale}, M.
  2006{\natexlab{b}}, \apj, 649, 927

\bibitem[{{Andrassy} {et~al.}(2018){Andrassy}, {Herwig}, {Woodward}, \&
  {Ritter}}]{andrassy18}
{Andrassy}, R., {Herwig}, F., {Woodward}, P., \& {Ritter}, C. 2018, arXiv
  e-prints, arXiv:1808.04014

\bibitem[{{Angelou} {et~al.}(2020){Angelou}, {Bellinger}, {Hekker}, {Mints},
  {Elsworth}, {Basu}, \& {Weiss}}]{angelou20}
{Angelou}, G.~C., {Bellinger}, E.~P., {Hekker}, S., {et~al.} 2020, \mnras, 493,
  4987

\bibitem[{{Aparicio} {et~al.}(1990){Aparicio}, {Bertelli}, {Chiosi}, \&
  {Garcia-Pelayo}}]{aparicio90}
{Aparicio}, A., {Bertelli}, G., {Chiosi}, C., \& {Garcia-Pelayo}, J.~M. 1990,
  \aap, 240, 262

\bibitem[{{Batchelor}(1969)}]{batchelor69}
{Batchelor}, G.~K. 1969, Physics of Fluids, 12, II

\bibitem[{{Battino} {et~al.}(2016){Battino}, {Pignatari}, {Ritter}, {Herwig},
  {Denisenkov}, {Den Hartogh}, {Trappitsch}, {Hirschi}, {Freytag},
  {Thielemann}, \& {Paxton}}]{battino16}
{Battino}, U., {Pignatari}, M., {Ritter}, C., {et~al.} 2016, \apj, 827, 30

\bibitem[{{Bell} {et~al.}(2002){Bell}, {Day}, {Almgren}, {Lijewski}, \&
  {Rendleman}}]{bell02}
{Bell}, J.~B., {Day}, M.~S., {Almgren}, A.~S., {Lijewski}, M.~J., \&
  {Rendleman}, C.~A. 2002, International Journal for Numerical Methods in
  Fluids, 40, 209

\bibitem[{Berberich {et~al.}(2019)Berberich, Chandrashekar, \&
  Klingenberg}]{berberich19}
Berberich, J.~P., Chandrashekar, P., \& Klingenberg, C. 2019, arXiv preprint
  arXiv:1903.05154

\bibitem[{{Bertelli} {et~al.}(1992){Bertelli}, {Bressan}, \&
  {Chiosi}}]{bertelli92}
{Bertelli}, G., {Bressan}, A., \& {Chiosi}, C. 1992, \apj, 392, 522

\bibitem[{{B{\"o}hm-Vitense}(1958)}]{boehmvitense58}
{B{\"o}hm-Vitense}, E. 1958, \zap, 46, 108

\bibitem[{{Bowman} {et~al.}(2019){Bowman}, {Aerts}, {Johnston}, {Pedersen},
  {Rogers}, {Edelmann}, {Sim{\'o}n-D{\'\i}az}, {Van Reeth}, {Buysschaert},
  {Tkachenko}, \& {Triana}}]{bowman19}
{Bowman}, D.~M., {Aerts}, C., {Johnston}, C., {et~al.} 2019, \aap, 621, A135

\bibitem[{{Brummell} {et~al.}(2002){Brummell}, {Clune}, \&
  {Toomre}}]{brummell02}
{Brummell}, N.~H., {Clune}, T.~L., \& {Toomre}, J. 2002, \apj, 570, 825

\bibitem[{{Christy}(1966)}]{christy66}
{Christy}, R.~F. 1966, \apj, 144, 108

\bibitem[{{Claret} \& {Torres}(2016)}]{claret16}
{Claret}, A. \& {Torres}, G. 2016, \aap, 592, A15

\bibitem[{{Constantino} \& {Baraffe}(2018)}]{constantino18}
{Constantino}, T. \& {Baraffe}, I. 2018, \aap, 618, A177

\bibitem[{{Couston} {et~al.}(2018){Couston}, {Lecoanet}, {Favier}, \& {Le
  Bars}}]{couston18}
{Couston}, L.-A., {Lecoanet}, D., {Favier}, B., \& {Le Bars}, M. 2018, \prl,
  120, 244505

\bibitem[{{Cristini} {et~al.}(2019){Cristini}, {Hirschi}, {Meakin}, {Arnett},
  {Georgy}, \& {Walkington}}]{cristini19}
{Cristini}, A., {Hirschi}, R., {Meakin}, C., {et~al.} 2019, \mnras, 484, 4645

\bibitem[{{Cristini} {et~al.}(2017){Cristini}, {Meakin}, {Hirschi}, {Arnett},
  {Georgy}, {Viallet}, \& {Walkington}}]{cristini17}
{Cristini}, A., {Meakin}, C., {Hirschi}, R., {et~al.} 2017, \mnras, 471, 279

\bibitem[{Dalcin {et~al.}(2011)Dalcin, Paz, Kler, \& Cosimo}]{dalcin11}
Dalcin, L.~D., Paz, R.~R., Kler, P.~A., \& Cosimo, A. 2011, Advances in Water
  Resources, 34, 1124 , new Computational Methods and Software Tools

\bibitem[{{Deheuvels} {et~al.}(2016){Deheuvels}, {Brand{\~a}o}, {Silva
  Aguirre}, {Ballot}, {Michel}, {Cunha}, {Lebreton}, \&
  {Appourchaux}}]{deheuvels16}
{Deheuvels}, S., {Brand{\~a}o}, I., {Silva Aguirre}, V., {et~al.} 2016, \aap,
  589, A93

\bibitem[{{Durran}(1989)}]{durran89}
{Durran}, D.~R. 1989, Journal of Atmospheric Sciences, 46, 1453

\bibitem[{{Edelmann} {et~al.}(2019){Edelmann}, {Ratnasingam}, {Pedersen},
  {Bowman}, {Prat}, \& {Rogers}}]{edelmann19}
{Edelmann}, P.~V.~F., {Ratnasingam}, R.~P., {Pedersen}, M.~G., {et~al.} 2019,
  \apj, 876, 4

\bibitem[{{Edelmann} {et~al.}(2017){Edelmann}, {R{\"o}pke}, {Hirschi},
  {Georgy}, \& {Jones}}]{edelmann17}
{Edelmann}, P.~V.~F., {R{\"o}pke}, F.~K., {Hirschi}, R., {Georgy}, C., \&
  {Jones}, S. 2017, \aap, 604, A25

\bibitem[{{Freytag} {et~al.}(1996){Freytag}, {Ludwig}, \&
  {Steffen}}]{freytag96}
{Freytag}, B., {Ludwig}, H.-G., \& {Steffen}, M. 1996, \aap, 313, 497

\bibitem[{{Gilet} {et~al.}(2013){Gilet}, {Almgren}, {Bell}, {Nonaka},
  {Woosley}, \& {Zingale}}]{gilet13}
{Gilet}, C., {Almgren}, A.~S., {Bell}, J.~B., {et~al.} 2013, \apj, 773, 137

\bibitem[{{Gilet}(2012)}]{gilet12}
{Gilet}, C.~E. 2012, PhD thesis, University of California, Berkeley

\bibitem[{{Herwig} {et~al.}(2007){Herwig}, {Freytag}, {Fuchs}, {Hansen},
  {Hueckstaedt}, {Porter}, {Timmes}, \& {Woodward}}]{herwig07}
{Herwig}, F., {Freytag}, B., {Fuchs}, T., {et~al.} 2007, in Astronomical
  Society of the Pacific Conference Series, Vol. 378, Why Galaxies Care About
  AGB Stars: Their Importance as Actors and Probes, ed. F.~{Kerschbaum},
  C.~{Charbonnel}, \& R.~F. {Wing}, 43

\bibitem[{Higl(2019)}]{higl19}
Higl, J. 2019, PhD thesis, Technische Universit{\"a}t M{\"u}nchen

\bibitem[{{Higl} {et~al.}(2018){Higl}, {Siess}, {Weiss}, \& {Ritter}}]{higl18}
{Higl}, J., {Siess}, L., {Weiss}, A., \& {Ritter}, H. 2018, \aap, 617, A36

\bibitem[{{Higl} \& {Weiss}(2017)}]{higl17}
{Higl}, J. \& {Weiss}, A. 2017, \aap, 608, A62

\bibitem[{{Horst} {et~al.}(2020){Horst}, {Edelmann}, {Andrassy}, {Roepke},
  {Bowman}, {Aerts}, \& {Ratnasingam}}]{horst20}
{Horst}, L., {Edelmann}, P.~V.~F., {Andrassy}, R., {et~al.} 2020, arXiv
  e-prints, arXiv:2006.03011

\bibitem[{Hunter(2007)}]{hunter07}
Hunter, J.~D. 2007, Computing in Science \& Engineering, 9, 90

\bibitem[{{Iben}(1975)}]{iben75}
{Iben}, I., J. 1975, \apj, 196, 525

\bibitem[{{Jacobs} {et~al.}(2016){Jacobs}, {Zingale}, {Nonaka}, {Almgren}, \&
  {Bell}}]{jacobs16}
{Jacobs}, A.~M., {Zingale}, M., {Nonaka}, A., {Almgren}, A.~S., \& {Bell},
  J.~B. 2016, \apj, 827, 84

\bibitem[{{Jones} {et~al.}(2017){Jones}, {Andrassy}, {Sandalski}, {Davis},
  {Woodward}, \& {Herwig}}]{jones17}
{Jones}, S., {Andrassy}, R., {Sandalski}, S., {et~al.} 2017, \mnras, 465, 2991

\bibitem[{{J{\o}rgensen} {et~al.}(2018){J{\o}rgensen}, {Mosumgaard}, {Weiss},
  {Silva Aguirre}, \& {Christensen-Dalsgaard}}]{jorgensen18}
{J{\o}rgensen}, A. C.~S., {Mosumgaard}, J.~R., {Weiss}, A., {Silva Aguirre},
  V., \& {Christensen-Dalsgaard}, J. 2018, \mnras, 481, L35

\bibitem[{{Kercek} {et~al.}(1998){Kercek}, {Hillebrandt}, \&
  {Truran}}]{kercek98}
{Kercek}, A., {Hillebrandt}, W., \& {Truran}, J.~W. 1998, \aap, 337, 379

\bibitem[{{Kippenhahn} {et~al.}(2012){Kippenhahn}, {Weigert}, \&
  {Weiss}}]{kippenhahn12}
{Kippenhahn}, R., {Weigert}, A., \& {Weiss}, A. 2012, {Stellar Structure and
  Evolution}

\bibitem[{{Korre} {et~al.}(2019){Korre}, {Garaud}, \& {Brummell}}]{korre19}
{Korre}, L., {Garaud}, P., \& {Brummell}, N.~H. 2019, \mnras, 484, 1220

\bibitem[{{Kraichnan}(1967)}]{kraichnan67}
{Kraichnan}, R.~H. 1967, Physics of Fluids, 10, 1417

\bibitem[{{Lattanzio} {et~al.}(2017){Lattanzio}, {Tout}, {Neumerzhitckii},
  {Karakas}, \& {Lesaffre}}]{lattanzio17}
{Lattanzio}, J.~C., {Tout}, C.~A., {Neumerzhitckii}, E.~V., {Karakas}, A.~I.,
  \& {Lesaffre}, P. 2017, \memsai, 88, 248

\bibitem[{{Lecoanet} {et~al.}(2014){Lecoanet}, {Brown}, {Zweibel}, {Burns},
  {Oishi}, \& {Vasil}}]{lecoanet14}
{Lecoanet}, D., {Brown}, B.~P., {Zweibel}, E.~G., {et~al.} 2014, \apj, 797, 94

\bibitem[{{Li}(2017)}]{li17}
{Li}, Y. 2017, \apj, 841, 10

\bibitem[{{Linden}(1975)}]{linden75}
{Linden}, P.~F. 1975, Journal of Fluid Mechanics, 71, 385

\bibitem[{{Magic} {et~al.}(2013){Magic}, {Collet}, {Asplund}, {Trampedach},
  {Hayek}, {Chiavassa}, {Stein}, \& {Nordlund}}]{magic13}
{Magic}, Z., {Collet}, R., {Asplund}, M., {et~al.} 2013, \aap, 557, A26

\bibitem[{{Magic} {et~al.}(2010){Magic}, {Serenelli}, {Weiss}, \&
  {Chaboyer}}]{magic10}
{Magic}, Z., {Serenelli}, A., {Weiss}, A., \& {Chaboyer}, B. 2010, \apj, 718,
  1378

\bibitem[{{Meakin} \& {Arnett}(2007)}]{meakin07}
{Meakin}, C.~A. \& {Arnett}, D. 2007, \apj, 667, 448

\bibitem[{{Mellado}(2017)}]{mellado17}
{Mellado}, J.~P. 2017, Annual Review of Fluid Mechanics, 49, 145

\bibitem[{{Michielsen} {et~al.}(2019){Michielsen}, {Pedersen}, {Augustson},
  {Mathis}, \& {Aerts}}]{michielsen19}
{Michielsen}, M., {Pedersen}, M.~G., {Augustson}, K.~C., {Mathis}, S., \&
  {Aerts}, C. 2019, \aap, 628, A76

\bibitem[{{Miczek} {et~al.}(2015){Miczek}, {R{\"o}pke}, \&
  {Edelmann}}]{miczek15}
{Miczek}, F., {R{\"o}pke}, F.~K., \& {Edelmann}, P.~V.~F. 2015, \aap, 576, A50

\bibitem[{{Moc{\'a}k} {et~al.}(2009){Moc{\'a}k}, {M{\"u}ller}, {Weiss}, \&
  {Kifonidis}}]{mocak09}
{Moc{\'a}k}, M., {M{\"u}ller}, E., {Weiss}, A., \& {Kifonidis}, K. 2009, \aap,
  501, 659

\bibitem[{{Mombarg} {et~al.}(2019){Mombarg}, {Van Reeth}, {Pedersen},
  {Molenberghs}, {Bowman}, {Johnston}, {Tkachenko}, \& {Aerts}}]{mombarg19}
{Mombarg}, J.~S.~G., {Van Reeth}, T., {Pedersen}, M.~G., {et~al.} 2019, \mnras,
  485, 3248

\bibitem[{{Moravveji} {et~al.}(2016){Moravveji}, {Townsend}, {Aerts}, \&
  {Mathis}}]{moravveji16}
{Moravveji}, E., {Townsend}, R.~H.~D., {Aerts}, C., \& {Mathis}, S. 2016, \apj,
  823, 130

\bibitem[{{Mosumgaard} {et~al.}(2020){Mosumgaard}, {J{\o}rgensen}, {Weiss},
  {Silva Aguirre}, \& {Christensen-Dalsgaard}}]{mosumgaard20}
{Mosumgaard}, J.~R., {J{\o}rgensen}, A. C.~S., {Weiss}, A., {Silva Aguirre},
  V., \& {Christensen-Dalsgaard}, J. 2020, \mnras, 491, 1160

\bibitem[{Noh \& Fernando(1993)}]{noh93}
Noh, Y. \& Fernando, H.~J. 1993, Dynamics of atmospheres and oceans, 17, 187

\bibitem[{{Nonaka} {et~al.}(2010){Nonaka}, {Almgren}, {Bell}, {Lijewski},
  {Malone}, \& {Zingale}}]{nonaka10}
{Nonaka}, A., {Almgren}, A.~S., {Bell}, J.~B., {et~al.} 2010, \apjs, 188, 358

\bibitem[{{Pedersen} {et~al.}(2018){Pedersen}, {Aerts}, {P{\'a}pics}, \&
  {Rogers}}]{pedersen18}
{Pedersen}, M.~G., {Aerts}, C., {P{\'a}pics}, P.~I., \& {Rogers}, T.~M. 2018,
  \aap, 614, A128

\bibitem[{{Pietrinferni} {et~al.}(2004){Pietrinferni}, {Cassisi}, {Salaris}, \&
  {Castelli}}]{pietrinferni04}
{Pietrinferni}, A., {Cassisi}, S., {Salaris}, M., \& {Castelli}, F. 2004, \apj,
  612, 168

\bibitem[{{Pols} {et~al.}(1997){Pols}, {Tout}, {Schroder}, {Eggleton}, \&
  {Manners}}]{pols97}
{Pols}, O.~R., {Tout}, C.~A., {Schroder}, K.-P., {Eggleton}, P.~P., \&
  {Manners}, J. 1997, \mnras, 289, 869

\bibitem[{{Pratt} {et~al.}(2017){Pratt}, {Baraffe}, {Goffrey}, {Constantino},
  {Viallet}, {Popov}, {Walder}, \& {Folini}}]{pratt17}
{Pratt}, J., {Baraffe}, I., {Goffrey}, T., {et~al.} 2017, \aap, 604, A125

\bibitem[{{Pratt} {et~al.}(2020){Pratt}, {Baraffe}, {Goffrey}, {Geroux},
  {Constantino}, {Folini}, \& {Walder}}]{pratt20}
{Pratt}, J., {Baraffe}, I., {Goffrey}, T., {et~al.} 2020, arXiv e-prints,
  arXiv:2003.04643

\bibitem[{{Pratt} {et~al.}(2016){Pratt}, {Baraffe}, {Goffrey}, {Geroux},
  {Viallet}, {Folini}, {Constantino}, {Popov}, \& {Walder}}]{pratt16}
{Pratt}, J., {Baraffe}, I., {Goffrey}, T., {et~al.} 2016, \aap, 593, A121

\bibitem[{{Ribas} {et~al.}(2000){Ribas}, {Jordi}, \& {Gim{\'e}nez}}]{ribas00}
{Ribas}, I., {Jordi}, C., \& {Gim{\'e}nez}, {\'A}. 2000, \mnras, 318, L55

\bibitem[{{Robinson} {et~al.}(2003){Robinson}, {Demarque}, {Li}, {Sofia},
  {Kim}, {Chan}, \& {Guenther}}]{Robinson03}
{Robinson}, F.~J., {Demarque}, P., {Li}, L.~H., {et~al.} 2003, \mnras, 340, 923

\bibitem[{{Rogers}(2015)}]{rogers15}
{Rogers}, T.~M. 2015, \apjl, 815, L30

\bibitem[{{Rogers} {et~al.}(2013){Rogers}, {Lin}, {McElwaine}, \&
  {Lau}}]{rogers13}
{Rogers}, T.~M., {Lin}, D.~N.~C., {McElwaine}, J.~N., \& {Lau}, H.~H.~B. 2013,
  \apj, 772, 21

\bibitem[{{Rogers} \& {McElwaine}(2017)}]{rogers17}
{Rogers}, T.~M. \& {McElwaine}, J.~N. 2017, \apjl, 848, L1

\bibitem[{{Stancliffe} {et~al.}(2015){Stancliffe}, {Fossati}, {Passy}, \&
  {Schneider}}]{stancliffe15}
{Stancliffe}, R.~J., {Fossati}, L., {Passy}, J.~C., \& {Schneider}, F.~R.~N.
  2015, \aap, 575, A117

\bibitem[{{Staritsin}(2013)}]{staritsin13}
{Staritsin}, E.~I. 2013, Astronomy Reports, 57, 380

\bibitem[{Stevens(2002)}]{stevens02}
Stevens, B. 2002, Quarterly Journal of the Royal Meteorological Society: A
  journal of the atmospheric sciences, applied meteorology and physical
  oceanography, 128, 2663

\bibitem[{Sutherland(2010)}]{sutherland10}
Sutherland, B.~R. 2010, Internal gravity waves (Cambridge university press)

\bibitem[{{Timmes}(2000)}]{timmes00b}
{Timmes}, F.~X. 2000, \apj, 528, 913

\bibitem[{{Timmes} \& {Swesty}(2000)}]{timmes00}
{Timmes}, F.~X. \& {Swesty}, F.~D. 2000, \apjs, 126, 501

\bibitem[{{Turk} {et~al.}(2011){Turk}, {Smith}, {Oishi}, {Skory}, {Skillman},
  {Abel}, \& {Norman}}]{turk11}
{Turk}, M.~J., {Smith}, B.~D., {Oishi}, J.~S., {et~al.} 2011, The Astrophysical
  Journal Supplement Series, 192, 9

\bibitem[{{Turner}(1986)}]{turner86}
{Turner}, J.~S. 1986, Journal of Fluid Mechanics, 173, 431

\bibitem[{{Valle} {et~al.}(2016){Valle}, {Dell'Omodarme}, {Prada Moroni}, \&
  {Degl'Innocenti}}]{valle16}
{Valle}, G., {Dell'Omodarme}, M., {Prada Moroni}, P.~G., \& {Degl'Innocenti},
  S. 2016, \aap, 587, A16

\bibitem[{{van Ballegooijen}(1982)}]{ballegooijen82}
{van Ballegooijen}, A.~A. 1982, \aap, 113, 99

\bibitem[{{VandenBerg} {et~al.}(2006){VandenBerg}, {Bergbusch}, \&
  {Dowler}}]{vandenberg06}
{VandenBerg}, D.~A., {Bergbusch}, P.~A., \& {Dowler}, P.~D. 2006, \apjs, 162,
  375

\bibitem[{{Vasil} {et~al.}(2013){Vasil}, {Lecoanet}, {Brown}, {Wood}, \&
  {Zweibel}}]{vasil13}
{Vasil}, G.~M., {Lecoanet}, D., {Brown}, B.~P., {Wood}, T.~S., \& {Zweibel},
  E.~G. 2013, \apj, 773, 169

\bibitem[{{Viallet} {et~al.}(2013){Viallet}, {Meakin}, {Arnett}, \&
  {Moc{\'a}k}}]{viallet13}
{Viallet}, M., {Meakin}, C., {Arnett}, D., \& {Moc{\'a}k}, M. 2013, \apj, 769,
  1

\bibitem[{{Wagstaff} {et~al.}(2020){Wagstaff}, {Miller Bertolami}, \&
  {Weiss}}]{wagstaff20}
{Wagstaff}, G., {Miller Bertolami}, M.~M., \& {Weiss}, A. 2020, \mnras, 493,
  4748

\bibitem[{{Weaver} {et~al.}(1978){Weaver}, {Zimmerman}, \&
  {Woosley}}]{weaver78}
{Weaver}, T.~A., {Zimmerman}, G.~B., \& {Woosley}, S.~E. 1978, \apj, 225, 1021

\bibitem[{{Weiss} \& {Schlattl}(2008)}]{weiss08}
{Weiss}, A. \& {Schlattl}, H. 2008, \apss, 316, 99

\bibitem[{{Yang}(2016)}]{yang16}
{Yang}, W. 2016, \apj, 829, 68

\bibitem[{{Zahn}(1991)}]{zahn91}
{Zahn}, J.~P. 1991, \aap, 252, 179

\bibitem[{{Zingale} {et~al.}(2009){Zingale}, {Almgren}, {Bell}, {Nonaka}, \&
  {Woosley}}]{zingale09}
{Zingale}, M., {Almgren}, A.~S., {Bell}, J.~B., {Nonaka}, A., \& {Woosley},
  S.~E. 2009, \apj, 704, 196

\end{thebibliography}

\end{document}